\documentclass[10pt,twocolumn,english,american,prx,amsmath,amssymb,superscriptaddress]{revtex4-2}
\usepackage[T1]{fontenc}
\UseRawInputEncoding
\usepackage{xcolor}
\usepackage{babel}
\usepackage{verbatim}
\usepackage{units}
\usepackage{mathtools}
\usepackage{bm}
\usepackage{amsmath}
\usepackage{amsthm}
\usepackage{amssymb}
\usepackage{graphicx}
\usepackage{microtype}
\usepackage[pdfusetitle,
 bookmarks=true,bookmarksnumbered=false,bookmarksopen=false,
 breaklinks=false,pdfborder={0 0 0},pdfborderstyle={},backref=false,colorlinks=false]
 {hyperref}
\hypersetup{
 colorlinks=true,linkcolor=black,citecolor=black,urlcolor=black,filecolor=black}

\makeatletter
\DeclarePairedDelimiter{\ev}{\langle}{\rangle}
\DeclarePairedDelimiter{\cum}{\llangle}{\rrangle}

\DeclarePairedDelimiter\abs{\lvert}{\rvert}%

\newcommand{\T}{T}  %

\DeclareMathOperator*{\argmax}{arg\,max}%
\DeclareMathOperator*{\argmin}{arg\,min}%

\newcommand{\bw}{\bm{w}}
\newcommand{\bphi}{\bm{\phi}}
\newcommand{\bx}{\bm{x}}

\usepackage{tikz}
\usepackage{mathtools}
\usepackage{bm}
\usepackage{MnSymbol}
\usepackage{graphicx}
\usepackage[misc]{ifsym}

\usepackage{notoccite}

\usepackage{cleveref}
\crefname{equation}{Eq.}{Eqs.}\Crefname{equation}{Eq.}{Eqs.}
\crefname{figure}{Fig.}{Figs.}\Crefname{figure}{Fig.}{Figs.}
\crefname{table}{Table}{Tables}\Crefname{table}{Table}{Tables}
\crefname{section}{Sec.}{Secs.}\Crefname{section}{Sec.}{Secs.}
\crefname{appendix}{Appendix}{Appendices}\Crefname{appendix}{Appendix}{Appendices}
\makeatletter
\let\bc@oldappendix\appendix
\renewcommand\appendix{\bc@oldappendix\crefalias{section}{appendix}}
\makeatother

\usepackage{siunitx}[=v2]

\wlog{LaTeX Warning:Rerun to get cross-references}

\makeatother

\usepackage{cleveref}
\begin{document}
\selectlanguage{english}%
\global\long\def\cN{\mathcal{N}}%
\global\long\def\T{\intercal}%
\global\long\def\bh{\mathbf{h}}%
\global\long\def\bth{\tilde{\mathbf{h}}}%
\global\long\def\bx{\bm{x}}%
\global\long\def\bw{\bm{w}}%
\global\long\def\bv{\mathbf{v}}%
\global\long\def\ty{\tilde{y}}%
\global\long\def\D{\mathcal{D}}%
\foreignlanguage{american}{
\global\long\def\smallmath#1{\scalebox{.75}{\ensuremath{#1}}}%
\global\long\def\av#1#2{\left\langle #1\right\rangle _{#2}}%
\global\long\def\ev#1#2{\left\langle #1\right\rangle _{#2}}%
\global\long\def\D{\mathcal{D}}%
\global\long\def\bphi{\bm{\phi}}%
\global\long\def\bl{\boldsymbol{l}}%
\global\long\def\bxi{\boldsymbol{\xi}}%
\global\long\def\bh{\boldsymbol{h}}%
\global\long\def\bz{\boldsymbol{z}}%
\global\long\def\bJ{\boldsymbol{J}}%
\global\long\def\N{\mathcal{N}}%
\global\long\def\hh{\hat{h}}%
\global\long\def\bhh{\hat{\boldsymbol{h}}}%
\global\long\def\T{\intercal}%
\global\long\def\Tbi{\mathrm{T}_{\mathrm{p}}}%
\global\long\def\TI{\mathrm{T}_{\mathrm{I}}}%
\global\long\def\by{\mathrm{\mathbf{y}}}%
\global\long\def\diag{\mathrm{diag}}%
\global\long\def\th{\tilde{h}}%
\global\long\def\argmax{\text{argmax}}%
\global\long\def\tj{\tilde{j}}%
}

\selectlanguage{american}%
\global\long\def\Gammafl{\Gamma_{\mathrm{fl}}}%
\global\long\def\gammafl{\gamma_{\mathrm{fl}}}%
\global\long\def\R{\mathbb{R}}%
\global\long\def\Z{\mathcal{Z}}%
\global\long\def\jt{j^{\mathrm{T}}}%
\global\long\def\Tb{T}%
\global\long\def\MR{\boldsymbol{R}}%
\global\long\def\MQ{\boldsymbol{Q}}%
\global\long\def\Qi{{\cal Q}}%
\global\long\def\Qs{Q}%
\global\long\def\Qr{Q_{12}}%
\global\long\def\Qri{Q^{\mathrm{I}}_{12}}%
\global\long\def\Qrb{Q^{\mathrm{b}}_{12}}%
\global\long\def\Qzb{Q^{\mathrm{b}}_{0}}%
\global\long\def\Qzi{Q^{\mathrm{I}}_{0}}%
\global\long\def\qr{q_{12}}%
\global\long\def\qrb{q^{\mathrm{b}}_{12}}%
\global\long\def\qri{q^{\mathrm{I}}_{12}}%
\global\long\def\xe{x^{1}}%
\global\long\def\xz{x^{2}}%
\global\long\def\he{h^{1}}%
\global\long\def\hz{h^{2}}%
\global\long\def\Qrd{\epsilon\left(s\right)}%
\global\long\def\Qrid{\epsilon^{\mathrm{I}}\left(s\right)}%

\global\long\def\b#1{\boldsymbol{#1}}%
 
\global\long\def\av#1#2{\left\langle #1\right\rangle _{#2}}%
\global\long\def\cum#1#2{{\left\llangle #1\right\rrangle }_{#2}}%
 
\global\long\def\D{\mathcal{D}}%
 
\global\long\def\d{\mathrm{d}}%
 
\global\long\def\at#1#2{\left.#1\right|_{#2}}%
 
\global\long\def\abs#1{\left|#1\right|}%

\global\long\def\bR{\mathbb{R}}%
\global\long\def\Var{\mathrm{Var}}%

\title{Discrete signaling mediates chaotic regularization in recurrent neural
networks}
\author{Jan Bauer}
\email{ja.bauer@fz-juelich.de, jan.bauer@ucl.ac.uk}

\affiliation{ELSC, The Hebrew University, Jerusalem, Israel}
\affiliation{Institute for Advanced Simulation (IAS-6), J\"ulich Research Centre,
Germany}
\affiliation{Gatsby Computational Neuroscience Unit, University College London,
United Kingdom}
\author{Christian Keup}
\affiliation{Institute for Advanced Simulation (IAS-6), J\"ulich Research Centre,
Germany}
\affiliation{Statistical physics of computation laboratory, EPFL, Lausanne, Switzerland}
\affiliation{INFN Gruppo Collegato di Parma, Parma, Italy}
\author{Jonathan Kadmon}
\affiliation{ELSC, The Hebrew University, Jerusalem, Israel}
\author{Moritz Helias}
\affiliation{Institute for Advanced Simulation (IAS-6), J\"ulich Research Centre,
Germany}
\affiliation{Department of Physics, Faculty 1, RWTH Aachen University, Aachen,
Germany}
\date{\today}
\begin{abstract}
Cortical circuits operate in a regime of intrinsic chaos, where even
tiny changes in input can lead to divergent neural responses. Yet,
remarkably, population codes in the brain vary smoothly with sensory
stimuli, forming coherent representational manifolds. How can chaotic
networks sustain such stable coding? Here, we develop a theoretical
framework that links the microscopic chaos of recurrent networks to
the macroscopic geometry of neural representations. Combining kernel
methods with dynamical mean-field theory, we show that chaotic dynamics
induce local roughness (introducing sharp distortions at small scales)
while preserving global smoothness across larger stimulus variations.
This structural property acts as an intrinsic regularizer, enhancing
generalization while maintaining expressivity. Moreover, we show how
chaotic networks naturally produce power-law spectral signatures,
closely matching experimental observations in cortical recordings.
These results explain how chaotic spiking networks can sustain smooth,
differentiable population codes and establish a theoretical framework
linking network dynamics, computational structure, and recorded neural
activity.

\end{abstract}
\maketitle

\section{Introduction\label{sec:Introduction}}

Neuronal circuits in the cortex present a paradox. Experimental and
theoretical studies reveal that spiking neural networks, which communicate
through discrete all-or-nothing events, are intrinsically chaotic
and highly sensitive to small changes in input \citep{VanVreeswijk96Chaosneuronalnetworks,London10_123,Kadmon15_041030}.
Yet, despite this apparent instability, the brain encodes and processes
sensory stimuli with remarkable reliability, likely by constructing
stable and coherent representations of the external world. Indeed,
large-scale recordings show that population codes in the visual cortex
remain remarkably smooth, with neural representations varying continuously
and predictably as a function of the presented sensory stimuli \citep{Stringer19_361,Munoz17_954}.
In a chaotic system, one would instead expect representations to be
rough and non-differentiable, with small differences in input leading
to abrupt, unpredictable shifts in activity. How can such smooth and
functionally useful representations emerge from the underlying volatile
and chaotic dynamics?

A key to understanding this puzzle lies in recurrent connectivity,
a defining feature of cortical circuits. In recurrent networks, inputs
are continuously reshaped by internal feedback. This recurrent processing
alters the geometry of stimulus representations, embedding them into
a higher-dimensional activity manifold shaped by the network\textquoteright s
own dynamics. Such transformations can enhance expressivity and flexibility,
supporting complex computations and improving generalization across
both biological and artificial systems \citep{Maass02_2531,Jaeger04_87}.

Previous studies have investigated how disordered recurrent connectivity
shapes representations, highlighting principles of expressivity and
inductive bias \citep{biswas2022}. However, much less is known about
the representational consequences of internally generated chaos\textemdash the
spontaneous fluctuations arising in recurrent networks. While prior
work has characterized how chaos affects temporal predictability,
its impact on the \textit{spatial} structure of representations (i.e.,
its stimulus dependence) remains poorly understood. In particular,
it is unclear how chaotic dynamics deform the activity manifolds that
encode continuous variations in sensory input. In stable regimes,
nearby stimuli are mapped to nearby points on a smooth activity manifold.
In contrast, one might expect that chaos renders the manifold rough
or fragmented, with small input differences leading to discontinuous
jumps in the neural code. Although chaos has been linked to enhanced
expressivity in deep feedforward networks \citep{Poole16_3360,Schoenholz17_01232},
the consequences for the smoothness or roughness of internal spatial
representations in recurrent systems have not been systematically
explored.

Here, we address these questions by developing a theoretical framework
that links chaotic network dynamics to the geometry of neural representations.
The theory shows how internally generated activity reshapes the structure
of stimulus representations.  To expose the implications for computation
and coding, we adopt the perspective of kernel methods. Kernels provide
a geometric lens on learning and generalization, by characterizing
how networks transform similarity between inputs into similarity between
outputs \citep{yang2019,Segadlo22_accepted}. They are powerful especially
in large systems, where the network can be treated as a Gaussian process.
Here, we extend this approach to recurrent networks operating in the
chaotic regime. Using tools from statistical physics (a two-replica
dynamical mean-field theory (DMFT) \citep{Keup21_021064}), we derive
an effective, time-dependent kernel that captures how chaotic activity
shapes internal representations. This framework bridges the microscopic
dynamics of recurrent chaos with the macroscopic structure of neural
codes, offering a new perspective on how cortical circuits maintain
stable representations despite their intrinsic instability.

The remainder of this work is organized as follows. In \cref{sec:Background}
we introduce our framework that links reservoir computing, two-replica
DMFT, and kernel regression: \cref{subsec:Signal-transmission-RNNs}
introduces the description of the recurrent network dynamics for continuous
and discrete signaling paradigms. To operationalize computation in
these models, \cref{subsec:RC} introduces the notion of reservoir
computing, and \cref{subsec:GP} recasts this computational paradigm
into the language of kernel regression. Finally, \cref{subsec:Spectral-analysis}
introduces the kernel spectrum to analyze generalization to unseen
inputs.

We then put this theoretical framework to use in \cref{subsec:chaos-can-compute}:
through the lens of kernel regression we expose how disordered chaotic
networks can perform demanding computations, even though their neural
code is rough. \cref{subsec:Binary-networks-are-strong-regularizers}
then focuses on intrinsic regularization arising in the specific case
of discrete signaling. Subsequently, we show in \cref{subsec:rate-nets}
how strong synaptic connectivity can emulate a similar effect in continuous
networks. Finally, we show how this framework can be related to experimentally
observed power spectra.

\section{Background\label{sec:Background}}

\begin{figure*}[!t]
\centering{}\includegraphics[width=1\textwidth]{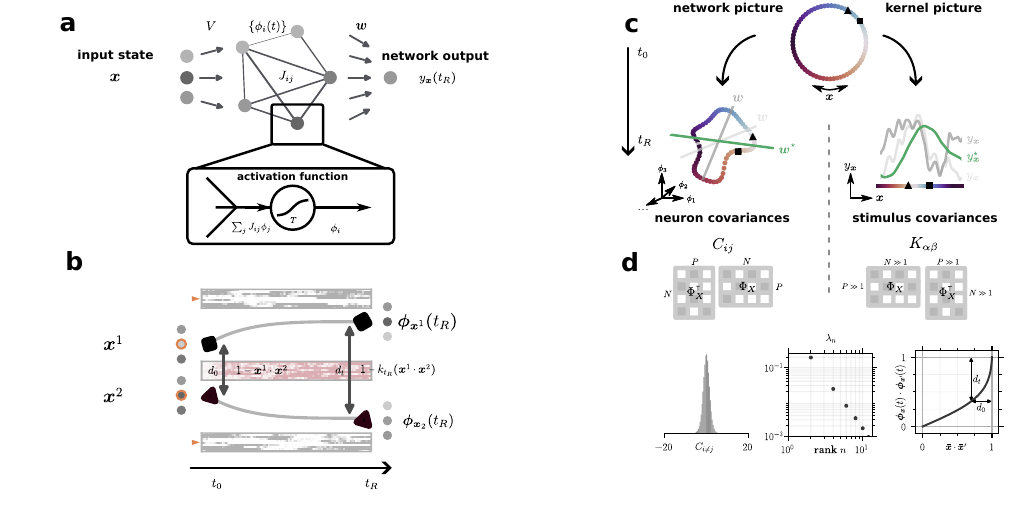}\caption{\textbf{\label{fig:neuron-models}Describing computation in RNNs in
kernel space. a}~A recurrent neural network (RNN) receives an input
stimulus $\bm{x}$ at an initial time $t_{0}$. The stimulus is propagated
by disordered random synapses $J_{ij}$ of strength $g$ to produce
a linear readout $y_{\bm{x}}(t_{R})=\bm{w}\cdot\bm{\phi}^{J}_{\bm{x}}(t_{R})$
at a readout time $t_{R}$. \textbf{b}~Trajectories of two stimuli
$\bm{x}^{1}$ (\emph{square marker}) and $\bm{x}^{2}$ (\emph{triangle
marker}) in a chaotic network are characterized by an increase of
their separation with time. \textbf{c}~Propagation of a collection
of $P$ stimuli $X$ that gets deformed to the neural code manifold
$\Phi^{J}_{X}(t_{R})$ as a result of the chaotic network activity.
\textbf{c, left}~\emph{Network picture}: A function on the input
stimuli $X$ is implemented by projecting the propagated stimuli $\Phi^{J}_{X}(t_{R})$
on a readout vector $\bm{w}$. The empirical correlation between \emph{neural}
activities can be obtained as $C_{ij}=\frac{1}{P}\sum^{P}_{\alpha}\phi^{J}_{\protect\bx^{\alpha},i}\,\phi^{J}_{\protect\bx^{\alpha},j}$.
\textbf{c, right}~\emph{Kernel picture}: The set of all possible
readouts forms an ensemble of functions (\emph{shades of gray}). Fitting
to observations corresponds to selecting a compatible function in
terms of a set of readout weights $\bm{w}^{\star}$ from this ensemble.
The pairwise \emph{stimulus} correlations are obtained as $K_{\alpha\beta}=\frac{1}{N}\sum^{N}_{i}\phi^{J}_{\protect\bx^{\alpha},i}\,\phi^{J}_{\protect\bx^{\beta},i}$.
For large networks, the kernel function $k_{t}(\protect\bx^{\alpha}\cdot\protect\bx^{\beta})\simeq K_{\alpha\beta}$
is the key component that summarizes how the transformation of correlations
impacts the output function $y_{\protect\bx}\sim\mathcal{N}\left(0,\,k_{t}\right)$.
Both matrices $C$ and $K$ have identical spectra $\{\lambda_{n}\}$
for $n=\min(N,\,P)$ (up to the multiplicative constant $N/P$).}
\end{figure*}

\subsection{Archetypes for discrete and continuous signaling in recurrent networks\label{subsec:Signal-transmission-RNNs}}

In building theoretical models of the brain, the transition from spikes
as discrete all-or-nothing events to a continuous-valued averaged
rate signal appears as a particularly strong simplification. In this
work, we therefore focus on\emph{ discrete dynamics} between two states
\citep{McCulloch43,Hebb49,Amit1985,Vreeswijk96_1724,Renart10_587},
capturing that neurons are either actively firing or quiescent. In
\cref{app:transmission_laws}, we give a more detailed description
of this model and contrast it to \emph{a matched continuous-rate model},
which has an identical dynamical-mean-field theory (DMFT) for the
first- and second-order statistics of the activity when presented
a single input. We will find that matching the cross-trajectory statistics
between two different inputs, however, is impossible between the two
models, thus exposing a qualitative difference between the discrete
and the continuous model.

Both models are based on two basic assumptions about recurrent networks.
First, the recurrence is realized by modeling the input $h_{i}$ to
each neuron $i$ as a weighted sum of upstream neurons\textquoteright{}
activities $\phi_{j}$, 
\begin{equation}
h_{i}(t)=\sum^{N}_{j=1}J_{ij}\,\phi_{j}(t),\quad J_{ij}\;\overset{\mathrm{i.i.d.}}{\sim}\;\mathcal{N}\bigl(0,\;g^{2}/N\bigr),\label{eq:recurrence}
\end{equation}
where $\lvert J_{ij}\rvert\propto g/\sqrt{N}$ ensures $\Var[h_{i}]$
remains $O(1)$ as $N\to\infty$.

Second, for the discrete model, we restrict each neuron\textquoteright s
output to a binary state $\phi_{i}(t)\in\{-1,1\}$ and update its
state via asynchronous \textit{Glauber dynamics} \citep{Glauber63_294}
with Poisson updates: in any infinitesimal interval $dt$, an update
variable $\theta_{\mathrm{up}}\in\{0,1\}$ is set to 1 with probability
$dt/\tau$, and if $\theta_{\mathrm{up}}=1$ the new state $\mathcal{T}_{i}\in\{-1,1\}$
is realized with probability 
\begin{equation}
\Pr\bigl[\mathcal{T}_{i}(h)=1\bigr]=\tfrac{1}{2}\bigl(T(h)+1\bigr)\,.\label{eq:transmission_law_main}
\end{equation}
This yields the discrete-time update rule 
\begin{equation}
\phi^{\text{disc}}_{i}(t+dt)=\bigl(1-\theta_{\mathrm{up}}\bigr)\,\phi^{\text{disc}}_{i}(t)\;+\;\theta_{\mathrm{up}}\,\mathcal{T}_{i}\bigl(h_{i}(t)\bigr).\label{eq:transmission_dynamics_glauber}
\end{equation}
On average, updates occur every $\tau$, reflecting the neuronal membrane
time constant.

The matched continuous-rate model can be obtained in the form of an
ODE by averaging over both the update process and the stochastic activation,
$\langle\mathcal{T}_{i}(h_{i})\rangle=T(h_{i})$. Expanding to first
order in $dt$ gives \citep{Amari77,Sompolinsky88_259} 
\begin{equation}
\phi^{\text{cont}}_{i}(t+dt)=\Bigl(1-\tfrac{dt}{\tau}\Bigr)\,\phi^{\text{cont}}_{i}(t)\;+\;\tfrac{dt}{\tau}\,T\bigl(h_{i}(t)\bigr)\;+\;\xi_{i}(t),\label{eq:rate_from_avg}
\end{equation}
where $\xi_{i}(t)$ is a zero-mean noise term capturing residual fluctuations
from the discrete updates.

Despite these different treatments of activation range and update
timing, Eqs.~\cref{eq:transmission_dynamics_glauber} and \cref{eq:rate_from_avg}
yield identical DMFT for the mean $\langle h_{i}(t)\rangle$ and covariance
$\langle h_{i}(t)\,h_{i}(s)\rangle$, while exhibiting distinct cross-trajectory
statistics \cref{app:DMFT}.

\subsection{Neural computation via reservoir computing\label{subsec:RC}}

The brain processes inputs through a large network of nonlinear units,
two aspects that have been found crucial to expressivity of neural
systems \citep{Hornik1989}, and which have served as a successful
paradigm for artificial systems \citep{McCulloch43}. Within this
class, reservoir computing is one of the simplest frameworks to make
use of a recurrent architecture \citep{Jaeger04_87}.

A canonical model of this approach has been put forward in form of
the Echo State Network \citep{Jaeger04_87}: By imprinting a stimulus
to the network as an external input, replacing \cref{eq:recurrence}
by $h_{i}(t)\rightarrow h_{i}(t)+h_{\text{ext},i}(t)$, it becomes
possible to perform computations with the network. In this work, we
consider inputs of the form 
\begin{align}
h_{\text{ext},i}(t) & =a(t)\,\sum^{N_{\text{in}}}_{j=1}V_{ij}x_{j}\,,\label{eq:input_projection}
\end{align}
where $V_{ij}\sim\mathcal{N}(0,g^{2}_{V}/N_{\text{in}})$ is an unstructured
input projection to ensure matching dimensions. Its variance scales
$\propto N^{-1}_{\text{in}}$ to preserve an $\mathcal{O}(1)$ norm
of the input, assuming that $x_{j}=\mathcal{O}(1)$, and $a(t)$ is
a scalar envelope that represents the time over which the stimulus
is presented to the network. We denote the resulting activity of recurrent
network \cref{eq:transmission_dynamics_glauber} or \cref{eq:rate_from_avg}
as $\bm{\phi}^{J}_{\bm{x}}(t)$.

Reservoir computing considers a dataset of $P$ observed examples
$\mathcal{D}=\left(X,Y_{X}\right)=\left\{ (\bx^{\alpha},y_{\bm{x}^{\alpha}})\right\} _{\alpha=1,\ldots,P}$
and tries to learn the mapping from inputs $\bm{x}$ to outputs $y$
with the help of a network as a linear readout $y_{\bm{x}}(t_{R})=\bm{w}\cdot\bm{\phi}^{J}_{\bm{x}}(t_{R})$.
It depends on the time of readout $t_{R}$, the recurrent connectivity
$J$, the readout vector $\bm{w}$, the input projection $V$, and
the synaptic noise $\xi$ in the system. . The rationale is that the
network activity $\bm{\phi}^{J}_{\bm{x}^{\alpha}}$ on each sample
$\bm{x}^{\alpha}$ will form an expressive basis, so that an optimized
readout $\bm{w}$ is able to read out the desired mapping. The optimal
readout $\bm{w}^{\ast}$ is found by minimizing the loss via a regularized
ridge regression objective \citep{Hoerl70Ridgeregressionapplications}

\begin{align}
\mathcal{L}(\bm{w};J)=\sum^{P}_{\alpha=1}\;\left(y_{\bm{x}^{\alpha}}-\bm{w}\cdot\bm{\phi}^{J}_{\bm{x}^{\alpha}}\right)^{2} & \:+\:N\Lambda\,\bigl\Vert\bm{w}\bigr\Vert^{2}.\label{eq:loss-rr}
\end{align}
For an underdetermined problem, it will converge to the unique solution
of minimum norm, which is enforced by the additional ridge-term $N\Lambda\,\bigl\Vert\bm{w}\bigr\Vert^{2}$.

Treating the recurrent weights as fixed, the optimization of $\bm{w}$
in \cref{eq:loss-rr} becomes a linear problem. Hence, its solution
can be obtained in closed form in terms of the features $\Phi^{J}_{X}\in\bR^{P\times N}$
which yields the predictor $\bar{y}_{\bm{x}^{*}}$ on an unseen test
point $\bm{x}^{*}$ as
\begin{align}
\bar{y}_{\bm{x}^{*}} & =\bm{w}^{\star}\cdot\bm{\phi}^{J}_{\bm{x}^{*}}\label{eq:y_pred_RR}\\
 & =Y^{\T}_{X}\Phi^{J}_{X}\left(\Phi^{J\,\T}_{X}\Phi^{J}_{X}+N\Lambda\right)^{-1}\cdot\bm{\phi}^{J}_{\bm{x}^{*}}\nonumber \\
 & =K^{J}_{x^{*}X}\,\left(K^{J}_{XX}+\Lambda\right)^{-1}\,Y_{X}.\label{eq:y_pred_GP}
\end{align}
In the last line, we rephrased this predictive equation from neuronal
correlations $C\coloneqq\Phi^{J\,\T}_{X}\Phi^{J}_{X}\in\mathbb{R}^{N\times N}$
to correlations between samples, $\frac{1}{N}\Phi^{J}_{X}\Phi^{J}_{X}{}^{\T}\eqqcolon K^{J}_{XX}\in\bR^{P\times P}$
(\cref{app:Kernel-trick}): In this formulation, the downstream predictions
of the network at a time $t$ depend only on the \emph{kernel matrix}
$K^{J}_{XX}\in\mathbb{R}^{P\times P}$, which trades operations in
a high dimensional neuronal space for operations that act directly
on the data. This transformation is known as the kernel trick \citep{Schoelkopf02Learningkernelssupport}
and allows us to abstract away microscopic details while preserving
computational properties of the networks, here in form of the predictor
$\bar{y}_{\bm{x}^{*}}$.

\subsection{Equivalence between optimal readout and maximum a posteriori predictor\label{subsec:GP}}

The kernel trick can be equivalently seen from the perspective of
Bayesian inference, which builds a connection to how a Bayes-optimal
observer would arrive at predictions (see \cref{app:Two-replica}
for details): One here assumes a prior distribution for the trainable
weights $V$, $J$, and $\bm{w}$ of the network and, subsequently,
computes their posterior by conditioning on the training data. For
networks that comprise a sufficiently large number of units $N\rightarrow\infty$,
and for a prior that is i.i.d. in the readout weights $w_{i}$, the
higher-order statistics of $y$ vanish and the network output $y_{\bm{x}}+\eta_{\bm{x}}$,
including a potential measurement noise $\eta_{\bm{x}}\stackrel{\text{i.i.d. over \ensuremath{\bm{x}}}}{\sim}\N(0,\,\Lambda)$,
converges to a Gaussian process $y_{\bm{x}}+\eta_{\bm{x}}\sim\mathcal{N}(0,\,k_{t})$
\citep{Neal96,lee2019wide,Yang19,segadlo2022} over the space of
inputs $\bm{x}$. Its statistics are then completely characterized
by a time-dependent covariance function, in this context called the
\emph{kernel function} $k_{t}(\bm{x},\bm{x}{}^{\prime})$. For a prior
that is i.i.d. in the read-in weights $V_{ij}$, we show that $k_{t}$
only depends on the dot product $\bm{x}\cdot\bm{x}{}^{\prime}$

\begin{align}
k_{t}(\bm{x}\cdot\bm{x}{}^{\prime}) & =\frac{1}{N}\langle\bm{\phi}^{J}_{\bm{x}}(t)\cdot\bm{\phi}^{J}_{\bm{x}^{\prime}}(t)\rangle_{J}+\Lambda\delta_{\bm{x}\bm{x}^{\prime}}\label{eq:kernel-def}\\
 & =\langle\phi_{\bm{x}}(t)\,\phi_{\bm{x}^{\prime}}(t)\rangle_{\N(0,\mathcal{Q}(\bm{x}\cdot\bm{x}{}^{\prime}))}+\Lambda\delta_{\bm{x}\bm{x}^{\prime}}\,,\label{eq:kernel-from-mft-1}
\end{align}
which captures the transformation of input overlaps, relating the
similarity of the input patters $\bm{x}\cdot\bm{x}{}^{\prime}$ to
correlations between outputs of the network $\bm{\phi}^{J}_{\bm{x}}(t)\cdot\bm{\phi}^{J}_{\bm{x}^{\prime}}(t)$
(see \cref{app:Two-replica}). We here scale the readout as $w_{i}\stackrel{\text{i.i.d.}}{\sim}\N(0,\,1/N)$.
Taking the expectation in \cref{eq:kernel-def} over the network's
recurrent weights $J$ renders the expression analytically tractable
while still being exact for large networks.

This operation reduces the calculation of the kernel function $k_{t}$
to the calculation of the cross-replica Gaussian statistics $\N\big(0,\mathcal{Q}(\bm{x}\cdot\bm{x}{}^{\prime})\big)$
of scalar synaptic input fields $h_{\bx}(t)$ for two different external
input vectors $\bm{x}$ and $\bm{x}^{\prime}$ \cref{eq:def_exp_replica},
while higher-order fluctuations vanish. The corresponding measure
can be obtained by solving a scalar differential equation in time
which describes the propagation of correlations in the network. Thus,
the correlation structure in large networks in the limit $N\rightarrow\infty$
only depends on the average connection strength $g$ \cref{eq:recurrence}
and the transmission law $\mathcal{T}$ \cref{eq:transmission_law_main}.

\subsubsection{Reservoir computing in large networks implements Bayesian inference}

Moving to kernel space opens up a complementary perspective on learning
in the reservoir computing paradigm. Suggestively, \cref{eq:y_pred_GP}
takes the form of the Gaussian process posterior mean \citep{WilliamsRasmussen06}.
Indeed, it is possible to rephrase the regression objective \cref{eq:loss-rr}
in a probabilistic setting, where the likelihood $\mathcal{L}(\bm{w},J)$
from \cref{eq:loss-rr} controls the probability $p(y_{\bm{x}}|\bm{w},J)=\N(y_{\bm{x}}\,|\,\bm{w}\cdot\bm{\phi}^{J}_{\bm{x}},\,\Lambda)\propto\exp\big(-\nicefrac{1}{2\Lambda}\left(y_{\bm{x}}-\bm{w}\cdot\bm{\phi}^{J}_{\bm{x}}\right){}^{2}\big)$.
Details are given in \cref{app:RC-yields-GP}.

In the limit of large networks, we show in \cref{app:RC-yields-GP}
that averaging under a Gaussian model for the disordered connectivities
$w_{i}\overset{\text{i.i.d.}}{\sim}\N(0,\,\Lambda)$ and $J_{ij}\overset{\text{i.i.d.}}{\sim}\N(0,\,g^{2}/N)$
results in an ensemble over functions that follows a Gaussian distribution,
$\ev{p(y|\bm{w},J)}{\bm{w},J}\propto\exp\left(-\nicefrac{1}{2}\,y^{\T}\,k^{-1}_{t}\,y\right)$.
Bayesian inference \citep{WilliamsRasmussen06} then provides a paradigm
to incorporate the evidence from the observations $\mathcal{D}=\left(X,Y_{X}\right)$
that is equivalent to the posterior of both $\bm{w}$ and $J$ if
the networks are large. Equivalently, this is the solution produced
by Langevin gradient descent with a weight decay \citep{Cohen21_023034}.
In particular, downstream predictions are obtained as means over the
posterior distribution
\begin{align}
\bar{y}_{x^{*}} & =\ev{y_{x^{*}}}{p(y_{x^{*}}|Y_{X})}\label{eq:gp-post}\\
 & =k_{t}(\bm{x}^{*}X^{\T})\,k_{t}(XX^{\T})^{-1}\,Y_{X}.\nonumber 
\end{align}
Here, the action of $k_{t}(\circ)$ is element-wise. Intuitively,
this selects all functions expressible by the network that match the
training data $Y_{X}$, as illustrated in \cref{fig:neuron-models}c.

\subsubsection{Spectral analysis of generalization in kernel regression\label{subsec:Spectral-analysis}}

So far, we have analyzed learning as fitting a set of data points
via regression. For sufficiently large networks, it is always possible
to fit the data perfectly \citep{Cybenko1989,Barron1994}. More generally,
however, a learned representation should also be useful on stimuli
that haven't been seen at training time. Inferring such representations
from finite data is the problem of induction \citep{Hume1896} and
requires prior assumptions about the ground truth.

The perspective of kernel regression makes these biases inherent to
neural networks explicit: Broadly, neural networks converge to Gaussian
processes \citep{Neal96,Lee17_00165}. More specifically, the spectral
Mercer decomposition of the kernel \citep{Williams06}
\begin{equation}
k(\bm{x}\cdot\bm{x}{}^{\prime})=\sum_{n}\lambda_{n}\psi_{n}(\bm{x})\psi_{n}(\bm{x}{}^{\prime})\label{eq:mercer}
\end{equation}
gives information of how this class is comprised: For the case of
dot-product kernels, the associated eigenvalues $\lambda_{n}$ to
its eigenfunctions $\psi_{n}$ determine which primitives will get
learned first \citep{LeCun91_2396}. Formally, this can be seen from
rephrasing the network prediction \cref{eq:y_pred_GP} in terms of
the eigenmodes $\psi_{n}$ \citep{canatar2021}. As the $\psi_{n}$
for dot-product kernels correspond to spherical harmonics of different
frequencies \citep{dutordoir2020}, the spectrum of the kernel may
either emphasize fast or slow modes in the data depending on the eigenvalues
of these modes.

\section{Results\label{sec:Results}}

In the preceding section, we have introduced a framework to connect
statistical properties such as the typical amplitude of a synaptic
weight $g$ to computational properties in terms of the kernel. We
introduced learning through readout weight optimization, abstracted
the network using Gaussian process regression, and finally examined
generalization through the lens of the kernel spectrum. Here, we put
this framework to use to understand the effect of chaos in neural
networks, in particular regarding discrete signaling. To this end,
we will derive the effective equations resulting from the $N\rightarrow\infty$
limit of the mean-field theory and leverage this framework to understand
the effect of chaos in neural networks, in particular regarding discrete
signaling.

\subsection{Computing the kernel via dynamical mean-field theory\label{subsec:kernel-dmft}}

To derive $k_{t}$ analytically, we employ dynamical mean-field theory
(DMFT) in the large-$N$ limit. We are interested in the case of two
\textit{different} stimuli, which we thus explicitly label $\bx=\bx^{1},\,\bx'=\bx^{2}$
in the following, and the induced correlation of the network $\ev{\bphi_{\bx^{1}}\cdot\bphi_{\bx^{2}}}{}$.
This requires a two-replica formulation: two identical networks with
the same connectivity $J$ that differ only in the input they receive,
$x^{1}$ or $x^{2}$, respectively.

Crucially, discrete and continuous networks can be matched to have
identical marginal (i.e., within-replica) statistics: their mean activity
and autocorrelation evolve identically when appropriate noise is added
to the continuous model. However, their two-replica cross-correlations
$Q^{12}_{ts}\coloneqq g^{2}\langle\phi^{1}(t)\,\phi^{2}(s)\rangle$,
which determine the kernel, evolve fundamentally differently:

For continuous networks, the two-replica dynamics remain smooth, governed
by 
\begin{multline}
(\tau\partial_{t}+1)(\tau\partial_{s}+1)Q^{12}_{ts}\\
=g^{2}\langle T(h^{1}(t))T(h^{2}(s))\rangle_{\N(0,\mathcal{Q}^{12})}\:+\:\langle\xi^{1}(t)\,\xi^{2}(s)\rangle\,,\label{eq:dmft_continuous}
\end{multline}
where $\mathcal{Q}^{12}$ denotes the cross-replica correlation of
the mean-field synaptic inputs $h^{1}_{t},\,h^{2}_{s}$, and the cross-noise
term describes the stochasticity in neuronal inputs due to the dynamics
in \cref{eq:transmission_dynamics_glauber,eq:rate_from_avg}.

For discrete networks, the stochastic Poisson update process introduces
qualitatively different dynamics. Most strikingly, the equal-time
cross-correlation evolves with a non-analytic dependence, 
\begin{multline}
(\tau\partial_{t}+1)(\tau\partial_{s}+1)Q^{12}_{ts}\\
=g^{2}\langle1-|T(h^{1}(t))-T(h^{2}(s))|\rangle_{\N(0,\mathcal{Q}^{12})}.\label{eq:dmft_discrete_diagonal}
\end{multline}
The absolute value term\textemdash absent in continuous dynamics\textemdash arises
from the discrete nature of neuronal states: It expresses the probability
of $h^{1}_{t}$ and $h^{2}_{t}$ both being above the activity threshold
$h_{t}=0$ simultaneously. Its effect is that two nearly identical
replicas can instantaneously decorrelate when one neuron changes state
while the other remains at its past state. A related non-analyticity
has been shown to entail a distinctly stronger kind of chaos due to
discrete signaling \citep{Kadmon15_041030}.

Together with the single-replica autocorrelations, which evolve according
to standard DMFT (\cref{app:DMFT}, see also \citep{Sompolinsky88_259,Helias19_10416}),
these equations close so that the PDE can be integrated forward in
time with an Euler-scheme (see \cref{app:kernel_mft} for details).
The equal-time kernel at time $t$ is then given by:
\begin{equation}
k_{t}(\bx^{1}\cdot\bx^{2})=Q^{12}_{tt}/g^{2}.\label{eq:kernel_from_q12}
\end{equation}
The distinction is fundamental: continuous networks undergo a smooth
evolution in their two-replica correlations, while discrete networks
exhibit discontinuous dynamics due to their all-or-nothing state changes,
manifested in the non-analytic right-hand side of \cref{eq:dmft_discrete_diagonal}.
As we will demonstrate, this microscopic difference manifests as a
sharp discontinuity in the kernel at $\bx^{1}=\bx^{2}$: a signature
of a qualitatively different, strong local chaos \citep{keup2021}.
The full technical derivation and additional cases (such as correlated
vs. independent update clocks) are detailed in~\cref{app:DMFT}.

\subsection{Computational capabilities are determined by global geometry of the
neural code\label{subsec:chaos-can-compute}}

\begin{figure}[!tbp]
\centering{}\includegraphics[width=1\columnwidth]{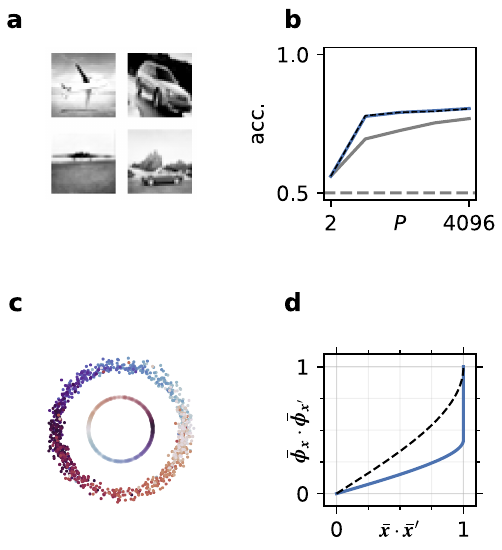}\caption{\textbf{\label{fig:chaos_can_compute}Chaotic network models can perform
reliable computation despite rough neural code. a}~Four examples
from the image dataset CIFAR10. \textbf{b}~Accuracy of the binary
classification between cars and planes as a function of the number
of presented training samples $P$. \emph{Blue: }Heaviside-activated
(i.e., $T=H$) recurrent network with Glauber dynamics as in \cref{eq:transmission_dynamics_glauber}
with $g=1.1$. \emph{Gray}: Linear classifier. \emph{Dashed black:}
Single-layer Heaviside-activated feedforward network. Readout time
for the recurrent network, and regularization for the feedforward
network were optimized in the networks to give best performance for
each number of training patterns. \emph{Horizontal line} indicates
baseline of random classification. \textbf{c}~Visualization of the
distortion of the neural code. \emph{Inner circle:} Cross-section
along the first two principal components (PCs) of the image data,
sorted and colored by angle. \emph{Outer circle}: Same sorting as
for inner circle, but for responses $\Phi=\text{Mat}\bigl(\protect\bphi^{\alpha}\bigr)_{\alpha=1,\ldots,P}$
propagated through the discrete recurrent network. While local structure
is distorted, global structure (coloring) is preserved. The cross-section
is obtained by measuring angle in the PC-space and setting the radial
components to $r^{\alpha}=1+\beta\left(\left\Vert \protect\bphi^{\alpha}\right\Vert -\langle\left\Vert \protect\bphi^{\alpha}\right\Vert \rangle_{\alpha}\right)$
according to the norms of the full high-dimensional vectors. $\beta=10$
is chosen as an amplification factor for clear visualization. \textbf{d}~Kernel
function describing the transmission of correlations $\protect\bphi_{\protect\bx}(t)\cdot\protect\bphi_{\protect\bx^{\prime}}(t)=k_{t}(\protect\bx\cdot\protect\bx^{\prime})$
for the discrete network that gives rise to the code in \textbf{c}.
Shown is the classification-optimal readout time of $t_{R}=3.3\tau$
(\emph{blue}). \emph{Dashed black}: Kernel for the feedforward network.}
\end{figure}

As discussed in \cref{subsec:Spectral-analysis}, suitable extrapolation
from the training data is necessary to perform reliable generalization
on unseen data. However, this a priori stands at odds with the implementation
through biological neural networks: If they operate in the chaotic
regime, slight perturbations $\epsilon$ from the training data diverge
exponentially. In the language of dynamical systems, this property
manifests as a positive Lyapunov exponent. Along the same lines, prior
work has argued that such networks form an unreliable substrate for
computation. For example, \citet{Poole16_3360} argue that initial
conditions will eventually be forgotten, \citet{Stringer19_361} point
to a diverging derivative of the neural code preventing extrapolation,
and \citet{London10_123} have argued that in chaotic networks the
change of a single spike may have a dramatic effect on all subsequent
spikes. Conversely, chaos has been argued to benefit expressivity\foreignlanguage{english}{
or memory \citep{Bertschinger04_1413,Toyoizumi11_051908}}. Here,
we use the kernel to argue that the divergence of trajectories in
response to a perturbation is an effect that only asymptotically impairs
coding: Firstly, on short timescales with moderate synaptic strength,
initial similarities in the data are not completely forgotten. This
preservation of similarity can hence be seen as a partial fulfillment
of the Echo State property \citep{Jaeger01_echo}. Secondly, while
two very similar patterns may show strong relative divergence, two
states that are initially less similar face a much smaller relative
decorrelation.

We exemplify these properties in \cref{fig:chaos_can_compute}: We
there consider classification of natural images (\cref{fig:chaos_can_compute}a),
a task that has been considered in \citep{Stringer19_361,bordelon2022population}
and noted to be challenging. The computation is implemented with a
chaotic discrete network: We consider the period of computation after
the stimuli had been presented and dominate the network state, i.e.,
$\Phi(t=0)\coloneqq VX$ with a projection matrix $V$ that linearly
transforms the data to an amplitude $(V\bm{x})_{i}=\mathcal{O}(1)$
that lies in the dynamic range of the activation function. In general,
the initial condition will be a combination of ongoing network activity
and the input. To isolate effects of the recurrent connectivity, we
here consider explicitly \textit{setting }the initial condition to
$VX$, which is only valid for strong input. The network's kernel
function exhibits a sharp peak, corresponding to a locally diverging
derivative, similar to a white noise process (\cref{fig:chaos_can_compute}d).
Yet, the base of the kernel function is decaying smoothly, reflecting
finite correlation length at all but the shortest scales. As a result,
the neural code from propagating the images is locally rough, but
maintains correlations on longer scales, as illustrated in \cref{fig:chaos_can_compute}c.
This allows the network to extrapolate from its training data, achieving
a test accuracy of 80\% on about 4000 training samples, which is well
above the chance level of 50\% (\cref{fig:chaos_can_compute}b). For
comparison, we show the accuracy and kernel of a Heaviside-activated
feedforward network where the ridge regularization has been optimized
to maximize accuracy. While accuracy is identical, only the recurrent
network's kernel has a sharp falloff, acting as intrinsic regularization.

This reconciles the paradox posed by \citet{Stringer19_361}: neural
codes can approach the edge of non-differentiability while maintaining
computational utility, because roughness remains spatially confined.

\subsection{Deterministic chaos in discrete networks mediates effective regularization\label{subsec:Binary-networks-are-strong-regularizers}}

\begin{figure}[!tbp]
\centering{}\includegraphics[width=1\columnwidth]{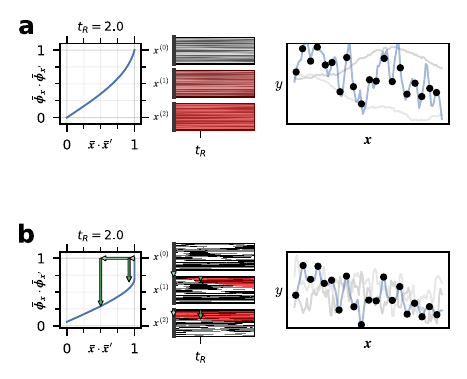}\caption{\textbf{\label{fig:bnry_rate_regularization}Strong local chaos acts
as effective regularization in discretely-coupled networks. a }Inference
for a continuously signaling network \cref{eq:rate_from_avg}, \textbf{b
}for discrete signaling \cref{eq:transmission_dynamics_glauber}.
\textbf{Left panels }show the kernel functions for each network. \textbf{Center
panels }show a subset of the network activity in simulation of networks
with $N=6000$ units for a base stimulus $\bm{x}^{(0)}$, a mild perturbation
thereof $\bm{x}^{(1)}=\bm{x}^{(0)}+\bm{\epsilon}$ realized through
changing the state of a single neuron, and a noticeably distinct stimulus
$\bm{x}^{(2)}$ in which $1/4$ of the neurons are changed relative
to $\bm{x}^{(0)}$; changes are applied at initial time $t=0$ (\emph{gray
bars}). \emph{Light green arrows} indicate the size of the initial
perturbation, \emph{dark green arrows} the induced change at a later
time $t_{R}$. \emph{Red shading} marks the fraction of neurons that
have changed relative to the base stimulus $\protect\bx^{(0)}$. The
neurons have been sorted by time of first change. \textbf{Right panels
}show regression on a synthetic task: \emph{Gray curves }show network
prior before fitting a readout $\bm{w}$. \emph{Black markers }are
training observations \textbf{$X$},\textbf{ $Y_{X}$ }that are shared
between the two signaling paradigms. \emph{Solid curves} show the
inference from the training data under the kernel. \emph{Blue }is
the regression result using the respective kernel. Parameters for
both networks: Recurrent coupling strength $\langle J^{2}_{ij}\rangle=g^{2}/N=3.0^{2}/N$,
activation function $T=\mathtt{erf}$. The variance of the independent
noise input for the continuous network is $D=0.01$ at every neuron.}
\end{figure}

The two-replica dynamics derived in \cref{app:Kernel-trick} reveal
a fundamental distinction between discrete and continuous signaling:
discrete networks exhibit \emph{strong local chaos}, manifesting itself
as a sharp discontinuity in the kernel function $k_{t}(\bx\cdot\bx^{\prime})$
at the origin, when $\bx\rightarrow\bx^{\prime}$. Here, we examine
the computational consequences of this microscopic instability. Compared
to inputs that are identical, the kernel abruptly drops from unity
by an amount $\Delta$ when considering a pair of stimuli that are
infinitesimally separated: 
\begin{equation}
k_{t}(1^{-})=1-\Delta,\quad\text{where}\quad\Delta=g^{2}\left(1-e^{-t/\tau}\right)^{2}>0.\label{eq:kernel_step}
\end{equation}
This discontinuity, derived in~\cref{app:Small-decorrelation-Deltapeak},
arises from the non-analytic term in \cref{eq:dmft_discrete_diagonal}:
the absolute value function captures the discrete nature of neuronal
state changes, where even infinitesimal input differences can trigger
macroscopic decorrelation when one neuron flips while its counterpart
in the other replicon maintains its state.

The step corresponds to infinite local sensitivity\textemdash formally,
a diverging Lyapunov exponent (\cref{app:microscopic-analysis-of-chaos-discrete}).
Yet, remarkably, the kernel remains smooth away from the origin, preserving
finite correlations for non-identical inputs (\cref{fig:chaos_can_compute}d).
This structure implies that the neural code is \emph{locally non-differentiable}
but \emph{globally coherent}.

The effect on regression of this discontinuity is displayed in \cref{fig:bnry_rate_regularization},
where we contrast a binary network with a network of continuous signaling
neurons when presented with noisy observations. It can be understood
in both the network and the kernel picture:

\paragraph{Network picture}

When a test observation $\bm{x}^{*}$ that closely resembles one of
the training points $\bm{x}\in X$ is propagated through the trained
network, the final state $\bm{\phi}_{\bm{x}^{*}}$ will show a similarity
with $\bm{\phi}_{\bm{x}}$ that has been reduced by $\Delta$. When
projected onto the trained linear readout $\bm{w}^{\star}$, the divergence
is reflected as a likewise reduction in the posterior output $\bar{y}_{\bm{x}^{\ast}}$.
A dual perspective to this phenomenon can be obtained via the kernel
picture, as outlined in \cref{app:Local-extrapolation-shape}: the
steep decay of the kernel function $k_{t}$ introduced in \cref{eq:kernel-def}
at small decorrelations controls the local extrapolation of the predictive
mean around an isolated training point $x\in X$ as $\overline{y}^{*}_{\bm{x}+\delta\bm{x}}\simeq k(1-|\delta\bm{x}|^{2})\,y_{\bx}=(1-\Delta)y_{\bx}\ll y_{\bx}$,
which shows that the predictor is drawn towards the mean of the prior,
which is zero here, as soon as one moves away from the training point.
This robust handling of outliers has been termed \emph{benign overfitting}
in the context of linear regression \citep{Bartlett20Benignoverfittinglinear}.

\paragraph{Kernel picture}

More abstractly, Bayesian inference explains this difference through
task-model alignment: The data that we expose the network to appears
noisy. The discrete network is a model that has a matching inductive
bias: Any realization of the neural code parametrized in terms of
the connectivities $\bm{w},\,J$ is locally discontinuous (\cref{fig:bnry_rate_regularization},
\emph{gray} \emph{curves}). High-variance observations appear likewise
discontinuous and are hence natural to the network. Viewed through
the readout-optimization picture in \cref{eq:y_pred_RR}, it is straightforward
to find a readout direction $\bm{w}^{\star}$ that matches the observations.

\paragraph{Deterministic versus stochastic regularization}

The effect of the jump in the kernel is identical to the Bayesian
assumption that the labels used for training were corrupted by a noise
$\eta$ of strength $\sigma^{2}_{\eta}=\Delta$ (\cref{app:RC-yields-GP}).
Its origin, however, is the strong local chaos that is a property
of the \textit{deterministic} circuit dynamics of discrete networks:
a single neuron's state change leads to divergence in the global network
state at an unbounded rate (formally, the Lyapunov exponent $\lambda$
is infinite \citep{Keup21_021064}). In contrast, continuous networks
require a strong, decorrelated external drive to yield a qualitatively
similar effect (the bare dynamics have $\lambda<\infty$).. We detail
this connection in \cref{app:RC-yields-GP}.

\subsection{Moderate chaos promotes multi-scale computation\label{subsec:rate-nets}}

\begin{figure}[!tbp]
\centering{}\includegraphics[width=1\columnwidth]{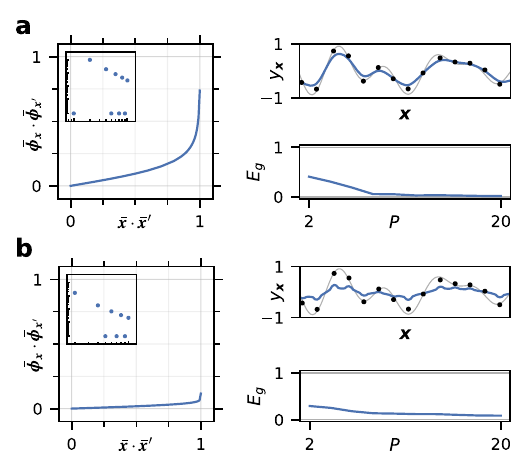}\caption{\textbf{\label{fig:chaos_noise_Eg}Chaotic rate networks have a rich
spectral repertoire. }Computation in the regular (\textbf{top}) and
chaotic regime (\textbf{bottom}). \textbf{Left panels} show the kernel
functions in either regime after stimulus propagation for $t=20\tau$
and white noise variance $D=0.01$. Inset shows the eigenvalues $\lambda_{n}$
of the kernel as in \cref{eq:mercer}, with identical $y$-axis between
panels. \textbf{Right upper panels }show a 1D regression task that
is composed of different length scales with respect to $\protect\bx$
and constitutes the regression target, with training data indicated
as \emph{black markers, }with the\emph{ blue line} being the kernel
predictor \cref{eq:gp-post}. \textbf{Right lower panels} show the
generalization error for this task as a function of the number of
presented data samples $P$.}
\end{figure}

While local chaos provides intrinsic regularization through kernel
discontinuities, its global effects on computation remain to be explored.
Does chaos confer computational advantages beyond regularization?
To address this question, we must compare chaotic and non-chaotic
regimes\textemdash a comparison only possible in continuous networks,
where synaptic strength $g$ controls the transition to chaos. Discrete
networks, by contrast, are always chaotic due to their all-or-nothing
signaling \citep{Vreeswijk96_1724,Kadmon15_041030,Keup21_021064}.

\subsubsection{The chaos transition shapes computational repertoires}

For continuous systems \cref{eq:recurrence}, the general impact of
synaptic strength on the collective dynamics has been studied: It
has been shown that continuous networks with many neurons exhibit
a transition to chaos at large synaptic strengths. For hyperbolic
tangent transfer function $T(h)=\tanh(h)$ in particular, the critical
transition is at $g=1$ \citep{Sompolinsky88_259}. More generally,
driven networks are chaotic when the neural output variance exceeds
the variance on the input \citep{Schuecker18_041029} 
\[
g^{2}\ev{T(h)\,T(h)}{h\sim\mathcal{N}(0,\,Q_{0})}>Q_{0}.
\]
This result from mean-field theory describes an amplification condition
due to the recurrence in the network: Below this threshold, networks
operate in a regular regime where perturbations decay exponentially.
Above it, chaos emerges and trajectories diverge. We provide a mechanistic
explanation of this condition in \cref{app:microscopic-analysis-of-chaos-discrete}.

Here, we again adopt the model-independent mean-field theory of \citet{Keup21_021064}
for these continuous networks described by \cref{eq:rate_from_avg}
to analyze the influence of the synaptic strength $g$ using \cref{eq:kernel-from-mft-1}.
The details of this calculation can be found in \cref{app:kernel_mft}.

\paragraph*{Deeply chaotic networks suffer from limited expressivity}

\cref{fig:chaos_noise_Eg} shows the kernels of a network just above
the transition to chaos $g\gtrsim1$ and one deeper into the chaotic
regime, $g\gg1$. Here the kernel functions do not reach $1$ due
to white noise added to the network with strength $D=0.01$ which
accumulates over time to a $g$-dependent decorrelation at $\bm{\bar{x}}\cdot\bm{\bar{x}}{}^{\prime}=1$,
as seen comparing the left panels a and b. The deterministic part
of the kernel in addition develops a steep section, the slope of which
approaches infinity in the limit $g\to\infty$. Deeply within the
chaotic regime, the kernel decays to low output similarities between
states for $t=20\tau$ (\cref{fig:chaos_noise_Eg}b). The initial
correlations are forgotten.

\paragraph*{Moderate chaos enables multi-scale computation}

In contrast, the moderately chaotic regime allows computations for
transient periods well beyond the neuronal time constant $\tau$ \citep{Toyoizumi11_051908}.
Leveraging the full dynamic range of the nonlinearity, the kernel
exhibits a spectrum with multiple nonzero modes (\cref{fig:chaos_noise_Eg}a),
enabling the network to model data that varies across multiple length-scales
(\cref{fig:chaos_noise_Eg}, \emph{right panels}) and thereby endowing
it with a richer functional repertoire. The kernel spectrum displays
several significant eigenvalues, each tied to a distinct spatial frequency,
arising from nonlinear mixing in chaotic dynamics: 
\begin{equation}
k_{t}(\bx\cdot\bx^{\prime})=\sum^{\infty}_{\ell=0}\lambda_{\ell}(t)P_{\ell}(\bx\cdot\bx^{\prime})\,,\label{eq:rich_spectrum}
\end{equation}
where $P_{\ell}$ are Legendre polynomials and the time-dependent
eigenvalues $\lambda_{\ell}(t)$ remain substantial across multiple
modes. This multi-mode structure supports simultaneous representation
of features from smooth, slowly varying components (low $\ell$) to
sharp, rapidly changing ones (high $\ell$). On regression tasks demanding
multi-scale structure, chaotic networks thus achieve lower generalization
error with fewer samples (\cref{fig:chaos_noise_Eg}a).

Crucially, moderate chaos ($g\approx1$\textendash $2$) preserves
this expressivity over extended timescales. Whereas regular networks
relax to a fixed point, so that the cross-replica correlations become
independent of the input, chaotic dynamics sustain input-dependent
correlations 
\begin{equation}
Q^{12}_{t}=g^{2}\langle\phi^{1}(t)\,\phi^{2}(t)\rangle\sim\mathcal{O}(1)\quad\text{for}\quad t\gg\tau\,,\label{eq:sustained_correlation}
\end{equation}
even when $t$ exceeds $\tau$. This fulfills a key requirement for
reservoir computing: maintaining a fading memory of past inputs while
avoiding both trivial forgetting and the complete decorrelation of
strong chaos \citep{Jaeger04_87}.

Thus, the moderately chaotic regime ($g\approx1$\textendash $2$)
optimally balances these extremes. The kernel maintains the following
properties:
\begin{itemize}
\item \textbf{Local roughness}: The derivative of the kernel remains steep,
providing regularization.
\item \textbf{Global smoothness}: The kernel decays gradually with decreasing
similarity $\bar{\bx}\cdot\bar{\bx}^{\prime}<1$, preserving information
about the similarity of inputs.
\item \textbf{Spectral richness}: Multiple eigenvalues $\lambda_{\ell}$
remain $\mathcal{O}(1)$, supporting diverse features.
\end{itemize}

\subsection{Kernel mean-field theory can explain experimentally observed neural
spectra\label{subsec:spectral_data}}

\begin{figure}[!tbp]
\centering{}\includegraphics[width=1\columnwidth]{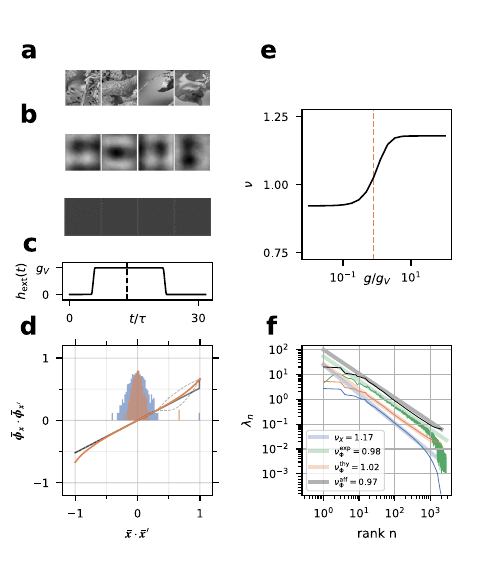}\caption{\textbf{\label{fig:data_spectra}Nonlinearities in neural networks
produce high-frequency power laws. a}~Four example images from a
natural image subset of ImageNet \citep{Stringer19_361} imprinted
to the network. \textbf{b}~Four low-frequency modes of the images
that dominate the spectrum (top), and four high-frequency modes at
the end of the spectrum (bottom). \textbf{c}~Temporal envelope of
the neural data used in \citep{Stringer19_361}. The readout time
is indicated by a dashed line. \textbf{d}~Kernel function (\emph{orange})
as obtained through \cref{eq:kernel-from-mft-1} for $g/g_{V}=1$.
\emph{Gray curve} is the affine-linear kernel \cref{eq:effective_kernel}
that produces the same slope $\nu$ of kernel eigenvalues $\text{\ensuremath{\lambda_{n}=1/}}n^{\nu}$,
as do the \emph{dashed-gray} nonlinear kernels that differ only outside
of the support of data $XX^{\protect\T}$. Overlayed are histograms
of $z$-scored image data $XX^{\protect\T}$ (\emph{blue}) and responses
$\Phi_{X}\Phi^{\protect\T}_{X}$ (\emph{orange}). \textbf{e}~Different
power law coefficients $\nu$ of spectra $\text{\ensuremath{\lambda_{n}=1/}}n^{\nu}$
as a function of the ratio of recurrent connectivity strength $g$
and input projection strength $g_{V}$. \emph{Dashed vertical line}
marks the network parameters $g/g_{V}$ used for other panels. \textbf{f}~Eigenspectrum
and asymptotic power law with exponent $\nu$ (legend) of original
data $X$ (\emph{blue}), and the neural response by the discrete network
theory (\emph{orange}, same parameters as in panel \textbf{d}) and
of the experimental recordings $\Phi_{X}$ from \citet{Stringer19_361}
(\emph{green}). Spectrum resulting from the affine linear kernel \cref{eq:effective_kernel}
(\emph{black}, same parameters as in panel \textbf{d}).}
\end{figure}

So far, we have analyzed the impact of chaos in discrete and continuous
networks on generalization ability in terms of the synaptic strength
and the signaling paradigm. To this end, we have leveraged the Gaussian\textendash process
description of the discrete recurrent dynamics in which all stimulus\textendash response
statistics are encoded by a kernel $k(\bar{\bx}\cdot\bar{\bx}{}^{\prime})$
defined on the correlation $\bar{\bx}\cdot\bar{\bx}{}^{\prime}$ between
(normalized) stimulus patterns $\bar{\bx}\coloneqq\tfrac{1}{\sqrt{N}}\bx$.
In particular, we identified a tradeoff between expressivity and stability.

In recordings of brain activity, often only macroscopic observables
are accessible, such as the eigenspectrum of the trial-averaged neural
covariance, , which typically decays as a power law $\lambda_{n}\sim n^{-\nu}$
with $\nu$ often close to unity \citep{Stringer19_361}. This slower-than-exponential
falloff suggests that brain activity carries a rich signal. We here
use our kernel theory to connect this spectrum to the microscopic
properties of the network.

\paragraph{From kernel geometry to eigenspectra: the role of non-analytical
kernels}

The kernel directly constrains population-level eigenspectra. Given
neural responses $\Phi_{X}\in\mathbb{R}^{P\times N}$ to $P$ stimuli
recorded from $N$ neurons, the (trial-averaged) covariance $C_{XX}=\frac{1}{P}\Phi^{\T}_{X}\Phi_{X}$
and the (empirical) kernel matrix $K_{XX}=\frac{1}{N}\Phi_{X}\Phi^{\T}_{X}$
share the same non-zero eigenvalues (up to rank $\min(P,N)$ and a
global scale factor $P/N$). Hence the kernel's spectral decomposition
\begin{equation}
k(\bar{\bx}\cdot\bar{\bx}{}^{\prime})\;=\;\sum_{n}\lambda_{n}\psi_{n}(\bar{\bx})\,\psi_{n}(\bar{\bx}^{\prime})\label{eq:spectral_decomp}
\end{equation}
directly predicts the decay of eigenvalues observable in recordings.

For high-dimensional stimuli (a generic property of natural sensory
inputs), inner products concentrate near zero and the \emph{local}
behavior of the kernel near perfect correlation $\bar{\bx}\cdot\bar{\bx}^{\prime}=1$
determines the tail of the spectrum. This connection is formalized
by the Funk\textendash Hecke theorem for data on the sphere: the decay
rate of $\{\lambda_{n}\}$ is determined by the kernel's smoothness
at $\bar{\bx}\cdot\bar{\bx}^{\prime}=1$ \citep{Funk15Beitragezurtheorie,dutordoir2020}.

The same local roughness that supplies regularization (\cref{subsec:Binary-networks-are-strong-regularizers})
also shapes the eigenspectrum. Chaos produces a kernel with steep
slope at $\bar{\bx}\cdot\bar{\bx}^{\prime}=1$ which controls the
exponent $\nu$ describing the tail of the power spectrum: kernels
that are steeper (or more singular) at $\bar{\bx}\cdot\bar{\bx}^{\prime}\lesssim1$
place more weight on high-frequency eigenmodes and thus yield smaller
$\nu$.

\paragraph{Network parameters control spectral exponents}

Microscopic network parameters determine the local kernel shape and
therefore the spectral exponent. Concretely, we model stimulus drive
through random projections $V_{ij}\sim\mathcal{N}(0,\,g^{2}_{V}/N_{\text{in}})$
and recurrent couplings scaled by $g$. With the time-dependent boxcar
input used to model the repeated, flashed stimuli in \citep{Stringer19_361},
the network response at readout time $t_{R}$ is 
\begin{align}
\Phi_{X}(t_{R})\; & =\bm{\phi}^{J}\!\big(t_{R};\,h_{\mathrm{ext}}(t\le t_{R})\big)\,,\label{eq:network_response}\\
h_{\mathrm{ext}}(t\le t_{R}) & =a(t\le t_{R})\,VX\,,
\end{align}
where $a(t)$ is a boxcar of duration $T=\SI{0.5}{\second}$ and we
set the neuronal time constant to $\tau=\SI{30}{\milli\second}$.
For the initial condition of the network, we consider random activity
whose variance is matched to the strength of the connectivity, mirroring
an equilibrium state. In our simulations the realizations of the neuronal
random update process entering \cref{eq:transmission_law_main} are
independent across stimulus presentations (trials), modeling trial-to-trial
variability. We use a trial-averaging protocol that follows \citep{Stringer19_361},
so that inter-trial variability can be corrected by correlating across
two distinct repeats of each stimulus.

We consider two limits to illustrate how the ratio $g/g_{V}$ controls
$\nu$:
\begin{itemize}
\item \textbf{Feedforward-dominated, $g_{V}\gg g$}: the input drive dominates.
Because the stimuli persist over some duration, the kernel approaches
the static feedforward form $k_{\text{ff}}(\bar{\bx}\cdot\bar{\bx}{}^{\prime})=\tfrac{2}{\pi}\arcsin(\bar{\bx}\cdot\bar{\bx}{}^{\prime})$.
This kernel emphasizes high-frequency modes and produces a relatively
flat spectrum tail (smaller $\nu$).
\item \textbf{Recurrence-dominated, $g\gg g_{V}$}: strong recurrent input
further decorrelates responses to different static inputs, steepening
the kernel near $\bar{\bx}\cdot\bar{\bx}^{\prime}=1$. In the limit
this tends toward a kernel that is close to zero for $\bar{\bx}\cdot\bar{\bx}^{\prime}<1$.
Because such a kernel is well approximated by a linear function, the
input spectrum is mostly preserved, and thus corresponds to a comparatively
larger value of $\nu$ matching the input (larger). 
\end{itemize}
Due to the employed Heaviside activation function in \cref{eq:transmission_law_main},
only the ratio $g/g_{V}$ matters (not absolute scale), and varying
this ratio in the model systematically shifts the measured power-law
exponent $\nu$.

\paragraph*{A simple effective model reproduces $\nu\simeq1$ exponents}

To obtain a qualitative understanding of which properties shape of
the power spectrum, we consider a simple model. When input correlations
$p(\bx\cdot\bx')$ concentrate near zero (the high-dimensional regime
relevant to natural images), the neural spectrum produced by the detailed
network can be closely approximated by a simple effective model. Define
the effective network response as
\begin{equation}
\bm{\phi}^{\mathrm{eff},\alpha}_{\mathbf{x}}\;\coloneqq\;\sqrt{a}\,\bx\;+\;\sqrt{b}\,\bz_{\bx}\;+\;\sqrt{1-a-b}\,\boldsymbol{\xi}^{\alpha},
\end{equation}
with $\|\bx\|\simeq\sqrt{N}$. Here, the component $\sqrt{a}\,\bx$
contains the stimulus. $\bz_{\bx}$ is a variable that depends on
the stimulus $\bx$, but is uncorrelated across different stimuli,
satisfying $\frac{1}{N}\bz_{\bx}\cdot\bz_{\bx'}=\delta_{\bx,\bx'}$.
It models the discontinuous, ``locally rough'' chaotic response
that is \textit{identical} over trials. Finally, $\boldsymbol{\xi}^{\alpha}$
captures trial-by-trial variability with $\tfrac{1}{N}\ev{\bxi^{\alpha}\cdot\bxi^{\beta}}{}=0$
for $\alpha\neq\beta$. Thus, for $\alpha\neq\beta$, this yields
an affine-linear effective kernel 
\begin{align}
k^{\mathrm{eff}}(\bx\cdot\bx') & =\frac{1}{N}\,\bm{\phi}^{\mathrm{eff},\alpha}_{\bx}\!\cdot\!\bm{\phi}^{\mathrm{eff},\beta}_{\bx'}\label{eq:effective_kernel}\\
 & =a\frac{1}{N}\bx\cdot\bx'\;+\;b\,\delta_{\bx,\bx'}.\nonumber 
\end{align}
Because of its affine-linearity, $k^{\mathrm{eff}}$ directly controls
the decay exponent: the deterministic rough component ($b>0$) compresses
the distribution of input correlations and flattens the spectral tail
beyond the ``white'' part of the spectrum, producing sub-critical
power laws. In practice, choosing parameters that reflect the balanced
regime $g/g_{V}\approx1$ reproduces the experimentally observed exponent
$\nu\approx1$. Concretely, the effective-kernel fit in our figures
yields $\nu\approx0.97$, closely matching \citet{Stringer19_361}.
This simple model therefore demonstrates that a locally rough and
therefore non-differentiable neural code is not in contradiction with
the experimentally observed decay exponents.

\paragraph{Connection to computational properties}

The spectral signatures present a complementary perspective on the
computation to the geometric view of the neural code we discussed
in the preceding sections. A power-law spectrum with relatively small
$\nu$ indicates enhanced high-frequency modes, consistent with the
local non-differentiability that supplies implicit regularization
(\cref{subsec:Binary-networks-are-strong-regularizers}). Concretely,
the low-frequency (smooth) components preserve global stimulus relationships
and support interpolation and generalization, while the high-frequency
(rough) components enable high expressivity, and prevent overfitting.
Chaotic discrete dynamics therefore yield a computationally advantageous
code.

In summary, the eigenspectra of neural kernels provide access to the
computational geometry imposed by chaotic dynamics. Non-analytic kernels
induced by discrete signaling produce power-law spectra; the exponent
$\nu$ is set by the balance of recurrent and feedforward drive and
can be quantitatively linked to experimentally observed values, providing
concrete tests for the theory.

\section{Discussion\label{sec:Discussion}}

This study set out to understand how internally generated chaos
in recurrent spiking networks influences the structure and computational
utility of stimulus representations. By combining kernel methods with
two-replica dynamical mean-field theory, we developed a framework
that connects microscopic chaos to reveal the macroscopic geometry
of neural codes. We found that chaotic activity renders representations
locally rough, inducing a sharp drop in similarity between nearby
stimuli. This discontinuity makes the neural manifold non-differentiable
at small scales. However, these distortions remain confined: at larger
scales, the network's mapping preserves global structure. Thus, chaos
introduces local instability while maintaining global organization.

This local roughness can be computationally beneficial, contrary
to previous claims \citep{London10_123}. Yet, unlike noise, these
fluctuations are deterministic: a form of benign overfitting that
can be useful under certain conditions \citep{Belkin19_15849,Bartlett20Benignoverfittinglinear}.
Our results show that chaotic kernels remain expressive over extended
timescales and support reliable readout despite the underlying instability
of the network dynamics. Trying to produce a similarly regularizing
effect in continuously signaling networks comes at the cost of expressivity. 

We also examined the spectral structure of chaotic representations.
The induced kernels span a broad spectrum, enabling computations at
multiple spatial scales. Notably, the eigenspectra exhibit power-law
decay that are compatible with the experimental observations of Stringer
et al. \citep{Stringer19_361}. This finding resolves a long-standing
paradox: how chaotic dynamics can coexist with globally smooth population
codes.

Using the Funk\textendash Hecke formalism, we showed how the spectral
exponent is tunable by network parameters, particularly the strength
of input projections. This provides a mechanistic explanation for
the diversity of spectral exponents seen across brain areas and experimental
conditions. It also offers a direct link between network dynamics
and the geometry of neural codes.

Our approach can be linked to variability of responses \citep{Destexhe99_1531,Mainen95_1503}:
responses of neurons, even if under near-identical experimental conditions,
are not the same. Instead, there is a strong influence of ongoing
activity \citep{Arieli96} with network responses to identical stimuli
showing a large amount of variability. Combined with the influence
from external inputs, the kernel function qualitatively captures this
property: its drop close to the point of perfect input correlation
amounts to the intrinsic variability present in these biological neuronal
systems \citep{Churchland10}.

We have studied discrete neurons as an abstraction of the discrete
signaling in the form of spikes in the brain. Prior works have shown
that the properties of correlation transmission, which essentially
shape the kernel, are closely related between spiking and binary neurons
\citep{Tchumatchenko10_058102,Keup21_021064}. Yet, a direct study
of spiking network models with regard to the roughness of the code
and the resulting benign overfitting may be interesting to obtain
results that can be mapped more directly to biophysical properties
of neuronal networks. Our comparison to experimentally recorded correlation
spectra focused on the qualitative effect of strong chaos, without
attempting to model the experiment in detail. For example we did not
use a sophisticated network architecture, modeled unobserved external
input or the additional averaging effect of calcium imaging on the
recorded activations.

Our work has analyzed static classification and regression tasks.
Yet, the continuously updated neural state in recurrent networks makes
them particularly well suited to handle temporal sequences. As a first
step, the kernel analysis would allow us to determine the optimal
readout time depending on the nature of the task, complementing prior
work on spectral task\textendash model alignment \citep{bordelon2022population}
and random feature methodologies \citep{Rahimi07Randomfeatureslarge}.
 Another possible extension concerns learning recurrent weights.
Our work studies learning for a disordered, static recurrent connectivity,
implying that computation and learning take place in the lazy regime
\citep{Chizat19_neurips,Yang19}, where recurrent weights stay close
to their initialization. While this approach captures aspects of computation,
it lacks the adaptability necessary for developing more useful inductive
biases. Recent advances on feature learning in feedforward networks
\citep{bordelon2022,Fischer24_10761,ringel2025,lauditi2025} now pave
the way to study feature learning in recurrent networks as well. First
steps in this direction have recently been undertaken in terms of
a theory for artificial recurrent networks with continuous signaling
\citep{bauer2026,clark2026}. Understanding how plasticity interacts
with chaos to shape neural representations is particularly promising,
as we expect that learning the internal weights of the network will
be required, analogous to the rich learning regime of deep networks.
We leave these extensions for future work.

In conclusion, we show that chaotic dynamics in recurrent networks
reshape neural representations in structured, nontrivial ways. The
induced kernels are rough at small scales but globally coherent, supporting
robust and expressive computation. This framework bridges the gap
between dynamical systems theory and the statistical structure of
neural codes, offering a new perspective on the constructive role
of chaos in the brain.

\begin{acknowledgments}We are grateful to Shachar Ashkenazi, Alkesh
Yadav, Lars Schutzeichel, and Vasco Portilheiro for valuable discussions.
This work is supported by the Gatsby Charitable Foundation. Partly
funded by the Deutsche Forschungsgemeinschaft (DFG, German Research
Foundation) as part of the SPP 2205 \textendash{} 533396241. This
work is an open access publication funded by the Deutsche Forschungsgemeinschaft
(DFG, German Research Foundation) \textendash{} 491111487. The authors
gratefully acknowledge the computing time granted by the JARA Vergabegremium
and provided on the JARA Partition part of the supercomputer JURECA
at Forschungszentrum Jülich (computation grant JINB33). JB gratefully
acknowledges a HiDA stipend from Helmholtz.\end{acknowledgments}

\bibliography{brain.bib,cited.bib,bib.bib,misc.bib}

\begin{thebibliography}{67}%
\makeatletter
\providecommand \@ifxundefined [1]{%
 \@ifx{#1\undefined}
}%
\providecommand \@ifnum [1]{%
 \ifnum #1\expandafter \@firstoftwo
 \else \expandafter \@secondoftwo
 \fi
}%
\providecommand \@ifx [1]{%
 \ifx #1\expandafter \@firstoftwo
 \else \expandafter \@secondoftwo
 \fi
}%
\providecommand \natexlab [1]{#1}%
\providecommand \enquote  [1]{``#1''}%
\providecommand \bibnamefont  [1]{#1}%
\providecommand \bibfnamefont [1]{#1}%
\providecommand \citenamefont [1]{#1}%
\providecommand \href@noop [0]{\@secondoftwo}%
\providecommand \href [0]{\begingroup \@sanitize@url \@href}%
\providecommand \@href[1]{\@@startlink{#1}\@@href}%
\providecommand \@@href[1]{\endgroup#1\@@endlink}%
\providecommand \@sanitize@url [0]{\catcode `\\12\catcode `\$12\catcode `\&12\catcode `\#12\catcode `\^12\catcode `\_12\catcode `\%12\relax}%
\providecommand \@@startlink[1]{}%
\providecommand \@@endlink[0]{}%
\providecommand \url  [0]{\begingroup\@sanitize@url \@url }%
\providecommand \@url [1]{\endgroup\@href {#1}{\urlprefix }}%
\providecommand \urlprefix  [0]{URL }%
\providecommand \Eprint [0]{\href }%
\providecommand \doibase [0]{https://doi.org/}%
\providecommand \selectlanguage [0]{\@gobble}%
\providecommand \bibinfo  [0]{\@secondoftwo}%
\providecommand \bibfield  [0]{\@secondoftwo}%
\providecommand \translation [1]{[#1]}%
\providecommand \BibitemOpen [0]{}%
\providecommand \bibitemStop [0]{}%
\providecommand \bibitemNoStop [0]{.\EOS\space}%
\providecommand \EOS [0]{\spacefactor3000\relax}%
\providecommand \BibitemShut  [1]{\csname bibitem#1\endcsname}%
\let\auto@bib@innerbib\@empty
\bibitem [{\citenamefont {Van~Vreeswijk}\ and\ \citenamefont {Sompolinsky}(1996)}]{VanVreeswijk96Chaosneuronalnetworks}%
  \BibitemOpen
  \bibfield  {author} {\bibinfo {author} {\bibfnamefont {C.}~\bibnamefont {Van~Vreeswijk}}\ and\ \bibinfo {author} {\bibfnamefont {H.}~\bibnamefont {Sompolinsky}},\ }\bibfield  {title} {\bibinfo {title} {Chaos in neuronal networks with balanced excitatory and inhibitory activity},\ }\href@noop {} {\bibfield  {journal} {\bibinfo  {journal} {Science}\ }\textbf {\bibinfo {volume} {274}},\ \bibinfo {pages} {1724} (\bibinfo {year} {1996})}\BibitemShut {NoStop}%
\bibitem [{\citenamefont {London}\ \emph {et~al.}(2010)\citenamefont {London}, \citenamefont {Roth}, \citenamefont {Beeren}, \citenamefont {H\"{a}usser},\ and\ \citenamefont {Latham}}]{London10_123}%
  \BibitemOpen
  \bibfield  {author} {\bibinfo {author} {\bibfnamefont {M.}~\bibnamefont {London}}, \bibinfo {author} {\bibfnamefont {A.}~\bibnamefont {Roth}}, \bibinfo {author} {\bibfnamefont {L.}~\bibnamefont {Beeren}}, \bibinfo {author} {\bibfnamefont {M.}~\bibnamefont {H\"{a}usser}},\ and\ \bibinfo {author} {\bibfnamefont {P.~E.}\ \bibnamefont {Latham}},\ }\bibfield  {title} {\bibinfo {title} {Sensitivity to perturbations in vivo implies high noise and suggests rate coding in cortex},\ }\href {https://doi.org/10.1038/nature09086} {\bibfield  {journal} {\bibinfo  {journal} {Nature}\ }\textbf {\bibinfo {volume} {466}},\ \bibinfo {pages} {123} (\bibinfo {year} {2010})}\BibitemShut {NoStop}%
\bibitem [{\citenamefont {Kadmon}\ and\ \citenamefont {Sompolinsky}(2015{\natexlab{a}})}]{Kadmon15_041030}%
  \BibitemOpen
  \bibfield  {author} {\bibinfo {author} {\bibfnamefont {J.}~\bibnamefont {Kadmon}}\ and\ \bibinfo {author} {\bibfnamefont {H.}~\bibnamefont {Sompolinsky}},\ }\bibfield  {title} {\bibinfo {title} {Transition to chaos in random neuronal networks},\ }\href {https://doi.org/10.1103/PhysRevX.5.041030} {\ \textbf {\bibinfo {volume} {5}},\ \bibinfo {pages} {041030} (\bibinfo {year} {2015}{\natexlab{a}})}\BibitemShut {NoStop}%
\bibitem [{\citenamefont {Stringer}\ \emph {et~al.}(2019)\citenamefont {Stringer}, \citenamefont {Pachitariu}, \citenamefont {Steinmetz}, \citenamefont {Carandini},\ and\ \citenamefont {Harris}}]{Stringer19_361}%
  \BibitemOpen
  \bibfield  {author} {\bibinfo {author} {\bibfnamefont {C.}~\bibnamefont {Stringer}}, \bibinfo {author} {\bibfnamefont {M.}~\bibnamefont {Pachitariu}}, \bibinfo {author} {\bibfnamefont {N.}~\bibnamefont {Steinmetz}}, \bibinfo {author} {\bibfnamefont {M.}~\bibnamefont {Carandini}},\ and\ \bibinfo {author} {\bibfnamefont {K.~D.}\ \bibnamefont {Harris}},\ }\bibfield  {title} {\bibinfo {title} {High-dimensional geometry of population responses in visual cortex},\ }\href@noop {} {\bibfield  {journal} {\bibinfo  {journal} {Nature}\ }\textbf {\bibinfo {volume} {571}},\ \bibinfo {pages} {361} (\bibinfo {year} {2019})}\BibitemShut {NoStop}%
\bibitem [{\citenamefont {Mu{\~n}oz}\ \emph {et~al.}(2017)\citenamefont {Mu{\~n}oz}, \citenamefont {Tremblay}, \citenamefont {Levenstein},\ and\ \citenamefont {Rudy}}]{Munoz17_954}%
  \BibitemOpen
  \bibfield  {author} {\bibinfo {author} {\bibfnamefont {W.}~\bibnamefont {Mu{\~n}oz}}, \bibinfo {author} {\bibfnamefont {R.}~\bibnamefont {Tremblay}}, \bibinfo {author} {\bibfnamefont {D.}~\bibnamefont {Levenstein}},\ and\ \bibinfo {author} {\bibfnamefont {B.}~\bibnamefont {Rudy}},\ }\bibfield  {title} {\bibinfo {title} {Layer-specific modulation of neocortical dendritic inhibition during active wakefulness},\ }\href@noop {} {\bibfield  {journal} {\bibinfo  {journal} {Science}\ }\textbf {\bibinfo {volume} {355}},\ \bibinfo {pages} {954} (\bibinfo {year} {2017})}\BibitemShut {NoStop}%
\bibitem [{\citenamefont {Maass}\ \emph {et~al.}(2002)\citenamefont {Maass}, \citenamefont {Natschl\"{a}ger},\ and\ \citenamefont {Markram}}]{Maass02_2531}%
  \BibitemOpen
  \bibfield  {author} {\bibinfo {author} {\bibfnamefont {W.}~\bibnamefont {Maass}}, \bibinfo {author} {\bibfnamefont {T.}~\bibnamefont {Natschl\"{a}ger}},\ and\ \bibinfo {author} {\bibfnamefont {H.}~\bibnamefont {Markram}},\ }\bibfield  {title} {\bibinfo {title} {Real-time computing without stable states: a new framework for neural computation based on perturbations},\ }\href@noop {} {\ \textbf {\bibinfo {volume} {14}},\ \bibinfo {pages} {2531} (\bibinfo {year} {2002})}\BibitemShut {NoStop}%
\bibitem [{\citenamefont {Jaeger}\ and\ \citenamefont {Haas}(2004)}]{Jaeger04_87}%
  \BibitemOpen
  \bibfield  {author} {\bibinfo {author} {\bibfnamefont {H.}~\bibnamefont {Jaeger}}\ and\ \bibinfo {author} {\bibfnamefont {H.}~\bibnamefont {Haas}},\ }\bibfield  {title} {\bibinfo {title} {Harnessing nonlinearity: Predicting chaotic systems and saving energy in wireless communication},\ }\href@noop {} {\bibfield  {journal} {\bibinfo  {journal} {Science}\ }\textbf {\bibinfo {volume} {304}},\ \bibinfo {pages} {78} (\bibinfo {year} {2004})}\BibitemShut {NoStop}%
\bibitem [{\citenamefont {Biswas}\ and\ \citenamefont {Fitzgerald}(2022)}]{biswas2022}%
  \BibitemOpen
  \bibfield  {author} {\bibinfo {author} {\bibfnamefont {T.}~\bibnamefont {Biswas}}\ and\ \bibinfo {author} {\bibfnamefont {J.~E.}\ \bibnamefont {Fitzgerald}},\ }\bibfield  {title} {\bibinfo {title} {Geometric framework to predict structure from function in neural networks},\ }\href {https://doi.org/10.1103/PhysRevResearch.4.023255} {\bibfield  {journal} {\bibinfo  {journal} {Physical Review Research}\ }\textbf {\bibinfo {volume} {4}},\ \bibinfo {pages} {023255} (\bibinfo {year} {2022})}\BibitemShut {NoStop}%
\bibitem [{\citenamefont {Poole}\ \emph {et~al.}(2016)\citenamefont {Poole}, \citenamefont {Lahiri}, \citenamefont {Raghu}, \citenamefont {Sohl-Dickstein},\ and\ \citenamefont {Ganguli}}]{Poole16_3360}%
  \BibitemOpen
  \bibfield  {author} {\bibinfo {author} {\bibfnamefont {B.}~\bibnamefont {Poole}}, \bibinfo {author} {\bibfnamefont {S.}~\bibnamefont {Lahiri}}, \bibinfo {author} {\bibfnamefont {M.}~\bibnamefont {Raghu}}, \bibinfo {author} {\bibfnamefont {J.}~\bibnamefont {Sohl-Dickstein}},\ and\ \bibinfo {author} {\bibfnamefont {S.}~\bibnamefont {Ganguli}},\ }\bibfield  {title} {\bibinfo {title} {Exponential expressivity in deep neural networks through transient chaos},\ }in\ \href {https://proceedings.neurips.cc/paper/2016/file/148510031349642de5ca0c544f31b2ef-Paper.pdf} {\emph {\bibinfo {booktitle} {Advances in Neural Information Processing Systems 29}}}\ (\bibinfo {year} {2016})\BibitemShut {NoStop}%
\bibitem [{\citenamefont {Schoenholz}\ \emph {et~al.}(2017)\citenamefont {Schoenholz}, \citenamefont {Gilmer}, \citenamefont {Ganguli},\ and\ \citenamefont {Sohl-Dickstein}}]{Schoenholz17_01232}%
  \BibitemOpen
  \bibfield  {author} {\bibinfo {author} {\bibfnamefont {S.~S.}\ \bibnamefont {Schoenholz}}, \bibinfo {author} {\bibfnamefont {J.}~\bibnamefont {Gilmer}}, \bibinfo {author} {\bibfnamefont {S.}~\bibnamefont {Ganguli}},\ and\ \bibinfo {author} {\bibfnamefont {J.}~\bibnamefont {Sohl-Dickstein}},\ }\bibfield  {title} {\bibinfo {title} {Deep information propagation},\ }\href {https://openreview.net/forum?id=H1W1UN9gg} {\bibfield  {journal} {\bibinfo  {journal} {5th International Conference on Learning Representations, ICLR 2017 - Conference Track Proceedings}\ } (\bibinfo {year} {2017})}\BibitemShut {NoStop}%
\bibitem [{\citenamefont {Yang}(2019{\natexlab{a}})}]{yang2019}%
  \BibitemOpen
  \bibfield  {author} {\bibinfo {author} {\bibfnamefont {G.}~\bibnamefont {Yang}},\ }\bibfield  {title} {\bibinfo {title} {Scaling limits of wide neural networks with weight sharing: {{Gaussian}} process behavior, gradient independence, and neural tangent kernel derivation},\ }\href@noop {} {\bibfield  {journal} {\bibinfo  {journal} {ArXiv e-prints}\ } (\bibinfo {year} {2019}{\natexlab{a}})},\ \Eprint {https://arxiv.org/abs/1902.04760} {1902.04760} \BibitemShut {NoStop}%
\bibitem [{\citenamefont {Segadlo}\ \emph {et~al.}(2022{\natexlab{a}})\citenamefont {Segadlo}, \citenamefont {Epping}, \citenamefont {van Meegen}, \citenamefont {Dahmen}, \citenamefont {Kr{\"a}mer},\ and\ \citenamefont {Helias}}]{Segadlo22_accepted}%
  \BibitemOpen
  \bibfield  {author} {\bibinfo {author} {\bibfnamefont {K.}~\bibnamefont {Segadlo}}, \bibinfo {author} {\bibfnamefont {B.}~\bibnamefont {Epping}}, \bibinfo {author} {\bibfnamefont {A.}~\bibnamefont {van Meegen}}, \bibinfo {author} {\bibfnamefont {D.}~\bibnamefont {Dahmen}}, \bibinfo {author} {\bibfnamefont {M.}~\bibnamefont {Kr{\"a}mer}},\ and\ \bibinfo {author} {\bibfnamefont {M.}~\bibnamefont {Helias}},\ }\bibfield  {title} {\bibinfo {title} {Unified field theoretical approach to deep and recurrent neuronal networks},\ }\href@noop {} {\  (\bibinfo {year} {2022}{\natexlab{a}})},\ \bibinfo {note} {accepted}\BibitemShut {NoStop}%
\bibitem [{\citenamefont {Keup}\ \emph {et~al.}(2021{\natexlab{a}})\citenamefont {Keup}, \citenamefont {K\"{u}hn}, \citenamefont {Dahmen},\ and\ \citenamefont {Helias}}]{Keup21_021064}%
  \BibitemOpen
  \bibfield  {author} {\bibinfo {author} {\bibfnamefont {C.}~\bibnamefont {Keup}}, \bibinfo {author} {\bibfnamefont {T.}~\bibnamefont {K\"{u}hn}}, \bibinfo {author} {\bibfnamefont {D.}~\bibnamefont {Dahmen}},\ and\ \bibinfo {author} {\bibfnamefont {M.}~\bibnamefont {Helias}},\ }\bibfield  {title} {\bibinfo {title} {Transient chaotic dimensionality expansion by recurrent networks},\ }\href {https://doi.org/10.1103/physrevx.11.021064} {\ \textbf {\bibinfo {volume} {11}},\ \bibinfo {pages} {021064} (\bibinfo {year} {2021}{\natexlab{a}})}\BibitemShut {NoStop}%
\bibitem [{\citenamefont {McCulloch}\ and\ \citenamefont {Pitts}(1943)}]{McCulloch43}%
  \BibitemOpen
  \bibfield  {author} {\bibinfo {author} {\bibfnamefont {W.~S.}\ \bibnamefont {McCulloch}}\ and\ \bibinfo {author} {\bibfnamefont {W.}~\bibnamefont {Pitts}},\ }\bibfield  {title} {\bibinfo {title} {A logical calculus of the ideas immanent in neural nets},\ }\href@noop {} {\ \textbf {\bibinfo {volume} {5}},\ \bibinfo {pages} {115} (\bibinfo {year} {1943})}\BibitemShut {NoStop}%
\bibitem [{\citenamefont {Hebb}(1949)}]{Hebb49}%
  \BibitemOpen
  \bibfield  {author} {\bibinfo {author} {\bibfnamefont {D.~O.}\ \bibnamefont {Hebb}},\ }\href {https://doi.org/10.1002/sce.37303405110} {\emph {\bibinfo {title} {The organization of behavior: A neuropsychological theory}}}\ (\bibinfo  {publisher} {John Wiley \& Sons},\ \bibinfo {address} {New York},\ \bibinfo {year} {1949})\BibitemShut {NoStop}%
\bibitem [{\citenamefont {Amit}\ \emph {et~al.}(1985)\citenamefont {Amit}, \citenamefont {Gutfreund},\ and\ \citenamefont {Sompolinsky}}]{Amit1985}%
  \BibitemOpen
  \bibfield  {author} {\bibinfo {author} {\bibfnamefont {D.~J.}\ \bibnamefont {Amit}}, \bibinfo {author} {\bibfnamefont {H.}~\bibnamefont {Gutfreund}},\ and\ \bibinfo {author} {\bibfnamefont {H.}~\bibnamefont {Sompolinsky}},\ }\bibfield  {title} {\bibinfo {title} {Spin-glass models of neural networks},\ }\href@noop {} {\bibfield  {journal} {\bibinfo  {journal} {Physical Review A}\ }\textbf {\bibinfo {volume} {32}},\ \bibinfo {pages} {1007} (\bibinfo {year} {1985})}\BibitemShut {NoStop}%
\bibitem [{\citenamefont {van Vreeswijk}\ and\ \citenamefont {Sompolinsky}(1996)}]{Vreeswijk96_1724}%
  \BibitemOpen
  \bibfield  {author} {\bibinfo {author} {\bibfnamefont {C.}~\bibnamefont {van Vreeswijk}}\ and\ \bibinfo {author} {\bibfnamefont {H.}~\bibnamefont {Sompolinsky}},\ }\bibfield  {title} {\bibinfo {title} {Chaos in neuronal networks with balanced excitatory and inhibitory activity},\ }\href {https://doi.org/10.1126/science.274.5293.1724} {\bibfield  {journal} {\bibinfo  {journal} {Science}\ }\textbf {\bibinfo {volume} {274}},\ \bibinfo {pages} {1724} (\bibinfo {year} {1996})}\BibitemShut {NoStop}%
\bibitem [{\citenamefont {Renart}\ \emph {et~al.}(2010)\citenamefont {Renart}, \citenamefont {{De La Rocha}}, \citenamefont {Bartho}, \citenamefont {Hollender}, \citenamefont {Parga}, \citenamefont {Reyes},\ and\ \citenamefont {Harris}}]{Renart10_587}%
  \BibitemOpen
  \bibfield  {author} {\bibinfo {author} {\bibfnamefont {A.}~\bibnamefont {Renart}}, \bibinfo {author} {\bibfnamefont {J.}~\bibnamefont {{De La Rocha}}}, \bibinfo {author} {\bibfnamefont {P.}~\bibnamefont {Bartho}}, \bibinfo {author} {\bibfnamefont {L.}~\bibnamefont {Hollender}}, \bibinfo {author} {\bibfnamefont {N.}~\bibnamefont {Parga}}, \bibinfo {author} {\bibfnamefont {A.}~\bibnamefont {Reyes}},\ and\ \bibinfo {author} {\bibfnamefont {K.~D.}\ \bibnamefont {Harris}},\ }\bibfield  {title} {\bibinfo {title} {The asynchronous state in cortical circuits},\ }\href {https://doi.org/10.1126/science.1179850} {\bibfield  {journal} {\bibinfo  {journal} {Science}\ }\textbf {\bibinfo {volume} {327}},\ \bibinfo {pages} {587} (\bibinfo {year} {2010})}\BibitemShut {NoStop}%
\bibitem [{\citenamefont {Glauber}(1963)}]{Glauber63_294}%
  \BibitemOpen
  \bibfield  {author} {\bibinfo {author} {\bibfnamefont {R.}~\bibnamefont {Glauber}},\ }\bibfield  {title} {\bibinfo {title} {Time-dependent statistics of the {I}sing model},\ }\href@noop {} {\ \textbf {\bibinfo {volume} {4}},\ \bibinfo {pages} {294} (\bibinfo {year} {1963})}\BibitemShut {NoStop}%
\bibitem [{\citenamefont {Amari}(1977)}]{Amari77}%
  \BibitemOpen
  \bibfield  {author} {\bibinfo {author} {\bibfnamefont {S.-I.}\ \bibnamefont {Amari}},\ }\bibfield  {title} {\bibinfo {title} {Dynamics of pattern formation in lateral-inhibition type neural fields},\ }\href {https://doi.org/10.1007/bf00337259} {\ \textbf {\bibinfo {volume} {27}},\ \bibinfo {pages} {77} (\bibinfo {year} {1977})}\BibitemShut {NoStop}%
\bibitem [{\citenamefont {Sompolinsky}\ \emph {et~al.}(1988)\citenamefont {Sompolinsky}, \citenamefont {Crisanti},\ and\ \citenamefont {Sommers}}]{Sompolinsky88_259}%
  \BibitemOpen
  \bibfield  {author} {\bibinfo {author} {\bibfnamefont {H.}~\bibnamefont {Sompolinsky}}, \bibinfo {author} {\bibfnamefont {A.}~\bibnamefont {Crisanti}},\ and\ \bibinfo {author} {\bibfnamefont {H.~J.}\ \bibnamefont {Sommers}},\ }\bibfield  {title} {\bibinfo {title} {Chaos in random neural networks},\ }\href {https://doi.org/10.1103/PhysRevLett.61.259} {\ \textbf {\bibinfo {volume} {61}},\ \bibinfo {pages} {259} (\bibinfo {year} {1988})}\BibitemShut {NoStop}%
\bibitem [{\citenamefont {Hornik}\ \emph {et~al.}(1989)\citenamefont {Hornik}, \citenamefont {Stinchcombe},\ and\ \citenamefont {White}}]{Hornik1989}%
  \BibitemOpen
  \bibfield  {author} {\bibinfo {author} {\bibfnamefont {K.}~\bibnamefont {Hornik}}, \bibinfo {author} {\bibfnamefont {M.}~\bibnamefont {Stinchcombe}},\ and\ \bibinfo {author} {\bibfnamefont {H.}~\bibnamefont {White}},\ }\bibfield  {title} {\bibinfo {title} {Multilayer feedforward networks are universal approximators},\ }\href {https://doi.org/10.1016/0893-6080(89)90020-8} {\ \textbf {\bibinfo {volume} {2}},\ \bibinfo {pages} {359} (\bibinfo {year} {1989})}\BibitemShut {NoStop}%
\bibitem [{\citenamefont {Hoerl}\ and\ \citenamefont {Kennard}(1970)}]{Hoerl70Ridgeregressionapplications}%
  \BibitemOpen
  \bibfield  {author} {\bibinfo {author} {\bibfnamefont {A.~E.}\ \bibnamefont {Hoerl}}\ and\ \bibinfo {author} {\bibfnamefont {R.~W.}\ \bibnamefont {Kennard}},\ }\bibfield  {title} {\bibinfo {title} {Ridge regression: Applications to nonorthogonal problems},\ }\href@noop {} {\bibfield  {journal} {\bibinfo  {journal} {Technometrics : a journal of statistics for the physical, chemical, and engineering sciences}\ }\textbf {\bibinfo {volume} {12}},\ \bibinfo {pages} {69} (\bibinfo {year} {1970})}\BibitemShut {NoStop}%
\bibitem [{\citenamefont {Sch{\"o}lkopf}\ \emph {et~al.}(2002)\citenamefont {Sch{\"o}lkopf}, \citenamefont {Smola}, \citenamefont {Bach} \emph {et~al.}}]{Schoelkopf02Learningkernelssupport}%
  \BibitemOpen
  \bibfield  {author} {\bibinfo {author} {\bibfnamefont {B.}~\bibnamefont {Sch{\"o}lkopf}}, \bibinfo {author} {\bibfnamefont {A.~J.}\ \bibnamefont {Smola}}, \bibinfo {author} {\bibfnamefont {F.}~\bibnamefont {Bach}}, \emph {et~al.},\ }\href@noop {} {\emph {\bibinfo {title} {Learning with Kernels: Support Vector Machines, Regularization, Optimization, and Beyond}}}\ (\bibinfo  {publisher} {MIT press},\ \bibinfo {year} {2002})\BibitemShut {NoStop}%
\bibitem [{\citenamefont {Neal}(1996)}]{Neal96}%
  \BibitemOpen
  \bibfield  {author} {\bibinfo {author} {\bibfnamefont {R.~M.}\ \bibnamefont {Neal}},\ }\href {https://doi.org/10.1007/978-1-4612-0745-0} {\emph {\bibinfo {title} {Bayesian Learning for Neural Networks}}}\ (\bibinfo  {publisher} {Springer New York},\ \bibinfo {year} {1996})\BibitemShut {NoStop}%
\bibitem [{\citenamefont {Lee}\ \emph {et~al.}(2019)\citenamefont {Lee}, \citenamefont {Xiao}, \citenamefont {Schoenholz}, \citenamefont {Bahri}, \citenamefont {Novak}, \citenamefont {{Sohl-Dickstein}},\ and\ \citenamefont {Pennington}}]{lee2019wide}%
  \BibitemOpen
  \bibfield  {author} {\bibinfo {author} {\bibfnamefont {J.}~\bibnamefont {Lee}}, \bibinfo {author} {\bibfnamefont {L.}~\bibnamefont {Xiao}}, \bibinfo {author} {\bibfnamefont {S.}~\bibnamefont {Schoenholz}}, \bibinfo {author} {\bibfnamefont {Y.}~\bibnamefont {Bahri}}, \bibinfo {author} {\bibfnamefont {R.}~\bibnamefont {Novak}}, \bibinfo {author} {\bibfnamefont {J.}~\bibnamefont {{Sohl-Dickstein}}},\ and\ \bibinfo {author} {\bibfnamefont {J.}~\bibnamefont {Pennington}},\ }\bibfield  {title} {\bibinfo {title} {Wide neural networks of any depth evolve as linear models under gradient descent},\ }\href@noop {} {\bibfield  {journal} {\bibinfo  {journal} {Advances in neural information processing systems}\ }\textbf {\bibinfo {volume} {32}},\ \bibinfo {pages} {8572} (\bibinfo {year} {2019})}\BibitemShut {NoStop}%
\bibitem [{\citenamefont {Yang}(2019{\natexlab{b}})}]{Yang19}%
  \BibitemOpen
  \bibfield  {author} {\bibinfo {author} {\bibfnamefont {G.}~\bibnamefont {Yang}},\ }\bibfield  {title} {\bibinfo {title} {Wide feedforward or recurrent neural networks of any architecture are gaussian processes}\ }(\bibinfo  {publisher} {Curran Associates, Inc.},\ \bibinfo {year} {2019})\BibitemShut {NoStop}%
\bibitem [{\citenamefont {Segadlo}\ \emph {et~al.}(2022{\natexlab{b}})\citenamefont {Segadlo}, \citenamefont {Epping}, \citenamefont {{van Meegen}}, \citenamefont {Dahmen}, \citenamefont {Kr{\"a}mer},\ and\ \citenamefont {Helias}}]{segadlo2022}%
  \BibitemOpen
  \bibfield  {author} {\bibinfo {author} {\bibfnamefont {K.}~\bibnamefont {Segadlo}}, \bibinfo {author} {\bibfnamefont {B.}~\bibnamefont {Epping}}, \bibinfo {author} {\bibfnamefont {A.}~\bibnamefont {{van Meegen}}}, \bibinfo {author} {\bibfnamefont {D.}~\bibnamefont {Dahmen}}, \bibinfo {author} {\bibfnamefont {M.}~\bibnamefont {Kr{\"a}mer}},\ and\ \bibinfo {author} {\bibfnamefont {M.}~\bibnamefont {Helias}},\ }\bibfield  {title} {\bibinfo {title} {Unified {{Field Theory}} for {{Deep}} and {{Recurrent Neural Networks}}},\ }\href@noop {} {\bibfield  {journal} {\bibinfo  {journal} {arXiv:2112.05589 [cond-mat, stat]}\ } (\bibinfo {year} {2022}{\natexlab{b}})},\ \Eprint {https://arxiv.org/abs/2112.05589} {arXiv:2112.05589 [cond-mat, stat]} \BibitemShut {NoStop}%
\bibitem [{\citenamefont {Rasmussen}\ and\ \citenamefont {Williams}(2006)}]{WilliamsRasmussen06}%
  \BibitemOpen
  \bibfield  {author} {\bibinfo {author} {\bibfnamefont {C.}~\bibnamefont {Rasmussen}}\ and\ \bibinfo {author} {\bibfnamefont {C.}~\bibnamefont {Williams}},\ }\href@noop {} {\emph {\bibinfo {title} {Gaussian Processes for Machine Learning}}},\ Adaptive Computation and Machine Learning\ (\bibinfo  {publisher} {MIT Press},\ \bibinfo {address} {Cambridge, MA, USA},\ \bibinfo {year} {2006})\ p.\ \bibinfo {pages} {248}\BibitemShut {NoStop}%
\bibitem [{\citenamefont {Cohen}\ \emph {et~al.}(2021)\citenamefont {Cohen}, \citenamefont {Malka},\ and\ \citenamefont {Ringel}}]{Cohen21_023034}%
  \BibitemOpen
  \bibfield  {author} {\bibinfo {author} {\bibfnamefont {O.}~\bibnamefont {Cohen}}, \bibinfo {author} {\bibfnamefont {O.}~\bibnamefont {Malka}},\ and\ \bibinfo {author} {\bibfnamefont {Z.}~\bibnamefont {Ringel}},\ }\bibfield  {title} {\bibinfo {title} {Learning curves for overparametrized deep neural networks: A field theory perspective},\ }\href {https://doi.org/10.1103/PhysRevResearch.3.023034} {\ \textbf {\bibinfo {volume} {3}},\ \bibinfo {pages} {023034} (\bibinfo {year} {2021})}\BibitemShut {NoStop}%
\bibitem [{\citenamefont {Cybenko}(1989)}]{Cybenko1989}%
  \BibitemOpen
  \bibfield  {author} {\bibinfo {author} {\bibfnamefont {G.}~\bibnamefont {Cybenko}},\ }\bibfield  {title} {\bibinfo {title} {Approximation by superpositions of a sigmoidal function},\ }\href {https://doi.org/10.1007/bf02551274} {\ \textbf {\bibinfo {volume} {2}},\ \bibinfo {pages} {303} (\bibinfo {year} {1989})}\BibitemShut {NoStop}%
\bibitem [{\citenamefont {Barron}(1994)}]{Barron1994}%
  \BibitemOpen
  \bibfield  {author} {\bibinfo {author} {\bibfnamefont {A.~R.}\ \bibnamefont {Barron}},\ }\bibfield  {title} {\bibinfo {title} {Approximation and estimation bounds for artificial neural networks},\ }\href {https://doi.org/10.1007/bf00993164} {\ \textbf {\bibinfo {volume} {14}},\ \bibinfo {pages} {115} (\bibinfo {year} {1994})}\BibitemShut {NoStop}%
\bibitem [{\citenamefont {Hume}(1896)}]{Hume1896}%
  \BibitemOpen
  \bibfield  {author} {\bibinfo {author} {\bibfnamefont {D.}~\bibnamefont {Hume}},\ }\href@noop {} {\emph {\bibinfo {title} {A Treatise of Human Nature}}}\ (\bibinfo  {publisher} {Clarendon Press},\ \bibinfo {year} {1896})\BibitemShut {NoStop}%
\bibitem [{\citenamefont {Lee}\ \emph {et~al.}(2017)\citenamefont {Lee}, \citenamefont {Bahri}, \citenamefont {Novak}, \citenamefont {Schoenholz}, \citenamefont {Pennington},\ and\ \citenamefont {Sohl-Dickstein}}]{Lee17_00165}%
  \BibitemOpen
  \bibfield  {author} {\bibinfo {author} {\bibfnamefont {J.}~\bibnamefont {Lee}}, \bibinfo {author} {\bibfnamefont {Y.}~\bibnamefont {Bahri}}, \bibinfo {author} {\bibfnamefont {R.}~\bibnamefont {Novak}}, \bibinfo {author} {\bibfnamefont {S.~S.}\ \bibnamefont {Schoenholz}}, \bibinfo {author} {\bibfnamefont {J.}~\bibnamefont {Pennington}},\ and\ \bibinfo {author} {\bibfnamefont {J.}~\bibnamefont {Sohl-Dickstein}},\ }\bibfield  {title} {\bibinfo {title} {Deep neural networks as gaussian processes},\ }\href@noop {} {\ ,\ \bibinfo {pages} {1711.00165} (\bibinfo {year} {2017})},\ \Eprint {https://arxiv.org/abs/arXiv:1711.00165} {arXiv:1711.00165} \BibitemShut {NoStop}%
\bibitem [{\citenamefont {Williams}\ and\ \citenamefont {Rasmussen}(2006)}]{Williams06}%
  \BibitemOpen
  \bibfield  {author} {\bibinfo {author} {\bibfnamefont {C.~K.}\ \bibnamefont {Williams}}\ and\ \bibinfo {author} {\bibfnamefont {C.~E.}\ \bibnamefont {Rasmussen}},\ }\href@noop {} {\emph {\bibinfo {title} {Gaussian Processes for Machine Learning}}},\ \bibinfo {edition} {1st}\ ed.\ (\bibinfo  {publisher} {MIT Press},\ \bibinfo {address} {Cambridge},\ \bibinfo {year} {2006})\BibitemShut {NoStop}%
\bibitem [{\citenamefont {Le~Cun}\ \emph {et~al.}(1991)\citenamefont {Le~Cun}, \citenamefont {Kanter},\ and\ \citenamefont {Solla}}]{LeCun91_2396}%
  \BibitemOpen
  \bibfield  {author} {\bibinfo {author} {\bibfnamefont {Y.}~\bibnamefont {Le~Cun}}, \bibinfo {author} {\bibfnamefont {I.}~\bibnamefont {Kanter}},\ and\ \bibinfo {author} {\bibfnamefont {S.~A.}\ \bibnamefont {Solla}},\ }\bibfield  {title} {\bibinfo {title} {Eigenvalues of covariance matrices: Application to neural-network learning},\ }\href {https://doi.org/10.1103/PhysRevLett.66.2396} {\ \textbf {\bibinfo {volume} {66}},\ \bibinfo {pages} {2396} (\bibinfo {year} {1991})}\BibitemShut {NoStop}%
\bibitem [{\citenamefont {Canatar}\ \emph {et~al.}(2021)\citenamefont {Canatar}, \citenamefont {Bordelon},\ and\ \citenamefont {Pehlevan}}]{canatar2021}%
  \BibitemOpen
  \bibfield  {author} {\bibinfo {author} {\bibfnamefont {A.}~\bibnamefont {Canatar}}, \bibinfo {author} {\bibfnamefont {B.}~\bibnamefont {Bordelon}},\ and\ \bibinfo {author} {\bibfnamefont {C.}~\bibnamefont {Pehlevan}},\ }\bibfield  {title} {\bibinfo {title} {Spectral {{Bias}} and {{Task-Model Alignment Explain Generalization}} in {{Kernel Regression}} and {{Infinitely Wide Neural Networks}}},\ }\href {https://doi.org/10.1038/s41467-021-23103-1} {\bibfield  {journal} {\bibinfo  {journal} {Nature Communications}\ }\textbf {\bibinfo {volume} {12}},\ \bibinfo {pages} {2914} (\bibinfo {year} {2021})},\ \Eprint {https://arxiv.org/abs/2006.13198} {arXiv:2006.13198} \BibitemShut {NoStop}%
\bibitem [{\citenamefont {Dutordoir}\ \emph {et~al.}(2020)\citenamefont {Dutordoir}, \citenamefont {Durrande},\ and\ \citenamefont {Hensman}}]{dutordoir2020}%
  \BibitemOpen
  \bibfield  {author} {\bibinfo {author} {\bibfnamefont {V.}~\bibnamefont {Dutordoir}}, \bibinfo {author} {\bibfnamefont {N.}~\bibnamefont {Durrande}},\ and\ \bibinfo {author} {\bibfnamefont {J.}~\bibnamefont {Hensman}},\ }\bibfield  {title} {\bibinfo {title} {Sparse {{Gaussian}} processes with spherical harmonic features},\ }in\ \href@noop {} {\emph {\bibinfo {booktitle} {International {{Conference}} on {{Machine Learning}}}}}\ (\bibinfo  {publisher} {PMLR},\ \bibinfo {year} {2020})\ pp.\ \bibinfo {pages} {2793--2802}\BibitemShut {NoStop}%
\bibitem [{\citenamefont {Helias}\ and\ \citenamefont {Dahmen}(2019)}]{Helias19_10416}%
  \BibitemOpen
  \bibfield  {author} {\bibinfo {author} {\bibfnamefont {M.}~\bibnamefont {Helias}}\ and\ \bibinfo {author} {\bibfnamefont {D.}~\bibnamefont {Dahmen}},\ }\bibfield  {title} {\bibinfo {title} {Statistical field theory for neural networks},\ }\href@noop {} {\  (\bibinfo {year} {2019})},\ \bibinfo {note} {1901.10416 [cond-mat.dis-nn]}\BibitemShut {NoStop}%
\bibitem [{\citenamefont {Keup}\ \emph {et~al.}(2021{\natexlab{b}})\citenamefont {Keup}, \citenamefont {K{\"u}hn}, \citenamefont {Dahmen},\ and\ \citenamefont {Helias}}]{keup2021}%
  \BibitemOpen
  \bibfield  {author} {\bibinfo {author} {\bibfnamefont {C.}~\bibnamefont {Keup}}, \bibinfo {author} {\bibfnamefont {T.}~\bibnamefont {K{\"u}hn}}, \bibinfo {author} {\bibfnamefont {D.}~\bibnamefont {Dahmen}},\ and\ \bibinfo {author} {\bibfnamefont {M.}~\bibnamefont {Helias}},\ }\bibfield  {title} {\bibinfo {title} {Transient {{Chaotic Dimensionality Expansion}} by {{Recurrent Networks}}},\ }\href {https://doi.org/10.1103/PhysRevX.11.021064} {\bibfield  {journal} {\bibinfo  {journal} {Physical Review X}\ }\textbf {\bibinfo {volume} {11}},\ \bibinfo {pages} {021064} (\bibinfo {year} {2021}{\natexlab{b}})}\BibitemShut {NoStop}%
\bibitem [{\citenamefont {Bertschinger}\ and\ \citenamefont {Natschl{\"a}ger}(2004)}]{Bertschinger04_1413}%
  \BibitemOpen
  \bibfield  {author} {\bibinfo {author} {\bibfnamefont {N.}~\bibnamefont {Bertschinger}}\ and\ \bibinfo {author} {\bibfnamefont {T.}~\bibnamefont {Natschl{\"a}ger}},\ }\bibfield  {title} {\bibinfo {title} {Real-time computation at the edge of chaos in recurrent neural networks},\ }\href@noop {} {\ \textbf {\bibinfo {volume} {16}},\ \bibinfo {pages} {1413} (\bibinfo {year} {2004})}\BibitemShut {NoStop}%
\bibitem [{\citenamefont {Toyoizumi}\ and\ \citenamefont {Abbott}(2011)}]{Toyoizumi11_051908}%
  \BibitemOpen
  \bibfield  {author} {\bibinfo {author} {\bibfnamefont {T.}~\bibnamefont {Toyoizumi}}\ and\ \bibinfo {author} {\bibfnamefont {L.~F.}\ \bibnamefont {Abbott}},\ }\bibfield  {title} {\bibinfo {title} {Beyond the edge of chaos: Amplification and temporal integration by recurrent networks in the chaotic regime},\ }\href@noop {} {\ \textbf {\bibinfo {volume} {84}},\ \bibinfo {pages} {051908} (\bibinfo {year} {2011})}\BibitemShut {NoStop}%
\bibitem [{\citenamefont {Jaeger}(2001)}]{Jaeger01_echo}%
  \BibitemOpen
  \bibfield  {author} {\bibinfo {author} {\bibfnamefont {H.}~\bibnamefont {Jaeger}},\ }\href@noop {} {\emph {\bibinfo {title} {The ``echo state'' approach to analysing and training recurrent neural networks}}},\ \bibinfo {type} {Tech. Rep.}\ \bibinfo {number} {GMD Report 148}\ (\bibinfo  {institution} {German National Research Center for Information Technology},\ \bibinfo {address} {St. Augustin, Germany},\ \bibinfo {year} {2001})\BibitemShut {NoStop}%
\bibitem [{\citenamefont {Bordelon}\ and\ \citenamefont {Pehlevan}(2022{\natexlab{a}})}]{bordelon2022population}%
  \BibitemOpen
  \bibfield  {author} {\bibinfo {author} {\bibfnamefont {B.}~\bibnamefont {Bordelon}}\ and\ \bibinfo {author} {\bibfnamefont {C.}~\bibnamefont {Pehlevan}},\ }\bibfield  {title} {\bibinfo {title} {Population codes enable learning from few examples by shaping inductive bias},\ }\href@noop {} {\bibfield  {journal} {\bibinfo  {journal} {Elife}\ }\textbf {\bibinfo {volume} {11}},\ \bibinfo {pages} {e78606} (\bibinfo {year} {2022}{\natexlab{a}})}\BibitemShut {NoStop}%
\bibitem [{\citenamefont {Bartlett}\ \emph {et~al.}(2020)\citenamefont {Bartlett}, \citenamefont {Long}, \citenamefont {Lugosi},\ and\ \citenamefont {Tsigler}}]{Bartlett20Benignoverfittinglinear}%
  \BibitemOpen
  \bibfield  {author} {\bibinfo {author} {\bibfnamefont {P.~L.}\ \bibnamefont {Bartlett}}, \bibinfo {author} {\bibfnamefont {P.~M.}\ \bibnamefont {Long}}, \bibinfo {author} {\bibfnamefont {G.}~\bibnamefont {Lugosi}},\ and\ \bibinfo {author} {\bibfnamefont {A.}~\bibnamefont {Tsigler}},\ }\bibfield  {title} {\bibinfo {title} {Benign overfitting in linear regression},\ }\href@noop {} {\bibfield  {journal} {\bibinfo  {journal} {Proceedings of the National Academy of Sciences}\ }\textbf {\bibinfo {volume} {117}},\ \bibinfo {pages} {30063} (\bibinfo {year} {2020})}\BibitemShut {NoStop}%
\bibitem [{\citenamefont {Schuecker}\ \emph {et~al.}(2018)\citenamefont {Schuecker}, \citenamefont {Goedeke},\ and\ \citenamefont {Helias}}]{Schuecker18_041029}%
  \BibitemOpen
  \bibfield  {author} {\bibinfo {author} {\bibfnamefont {J.}~\bibnamefont {Schuecker}}, \bibinfo {author} {\bibfnamefont {S.}~\bibnamefont {Goedeke}},\ and\ \bibinfo {author} {\bibfnamefont {M.}~\bibnamefont {Helias}},\ }\bibfield  {title} {\bibinfo {title} {Optimal sequence memory in driven random networks},\ }\href {https://doi.org/10.1103/PhysRevX.8.041029} {\ \textbf {\bibinfo {volume} {8}},\ \bibinfo {pages} {041029} (\bibinfo {year} {2018})}\BibitemShut {NoStop}%
\bibitem [{\citenamefont {Funk}(1915)}]{Funk15Beitragezurtheorie}%
  \BibitemOpen
  \bibfield  {author} {\bibinfo {author} {\bibfnamefont {P.}~\bibnamefont {Funk}},\ }\bibfield  {title} {\bibinfo {title} {Beitr\"age zur theorie der kugelfunktionen},\ }\href@noop {} {\bibfield  {journal} {\bibinfo  {journal} {Mathematische Annalen}\ }\textbf {\bibinfo {volume} {77}},\ \bibinfo {pages} {136} (\bibinfo {year} {1915})}\BibitemShut {NoStop}%
\bibitem [{\citenamefont {Belkin}\ \emph {et~al.}(2019)\citenamefont {Belkin}, \citenamefont {Hsu}, \citenamefont {Ma},\ and\ \citenamefont {Mandal}}]{Belkin19_15849}%
  \BibitemOpen
  \bibfield  {author} {\bibinfo {author} {\bibfnamefont {M.}~\bibnamefont {Belkin}}, \bibinfo {author} {\bibfnamefont {D.}~\bibnamefont {Hsu}}, \bibinfo {author} {\bibfnamefont {S.}~\bibnamefont {Ma}},\ and\ \bibinfo {author} {\bibfnamefont {S.}~\bibnamefont {Mandal}},\ }\bibfield  {title} {\bibinfo {title} {Reconciling modern machine-learning practice and the classical bias{\textendash}variance trade-off},\ }\href {https://doi.org/10.1073/pnas.1903070116} {\bibfield  {journal} {\bibinfo  {journal} {Proceedings of the National Academy of Sciences}\ }\textbf {\bibinfo {volume} {116}},\ \bibinfo {pages} {15849} (\bibinfo {year} {2019})}\BibitemShut {NoStop}%
\bibitem [{\citenamefont {Destexhe}\ and\ \citenamefont {Par\'{e}}(1999)}]{Destexhe99_1531}%
  \BibitemOpen
  \bibfield  {author} {\bibinfo {author} {\bibfnamefont {A.}~\bibnamefont {Destexhe}}\ and\ \bibinfo {author} {\bibfnamefont {D.}~\bibnamefont {Par\'{e}}},\ }\bibfield  {title} {\bibinfo {title} {Impact of network activity on the integrative properties of neocortical pyramidal neurons in vivo},\ }\href@noop {} {\ \textbf {\bibinfo {volume} {81}},\ \bibinfo {pages} {1531} (\bibinfo {year} {1999})}\BibitemShut {NoStop}%
\bibitem [{\citenamefont {Mainen}\ and\ \citenamefont {Sejnowski}(1995)}]{Mainen95_1503}%
  \BibitemOpen
  \bibfield  {author} {\bibinfo {author} {\bibfnamefont {Z.~F.}\ \bibnamefont {Mainen}}\ and\ \bibinfo {author} {\bibfnamefont {T.~J.}\ \bibnamefont {Sejnowski}},\ }\bibfield  {title} {\bibinfo {title} {Reliability of spike timing in neocortical neurons},\ }\href@noop {} {\bibfield  {journal} {\bibinfo  {journal} {Science}\ }\textbf {\bibinfo {volume} {268}},\ \bibinfo {pages} {1503} (\bibinfo {year} {1995})}\BibitemShut {NoStop}%
\bibitem [{\citenamefont {Arieli}\ \emph {et~al.}(1996)\citenamefont {Arieli}, \citenamefont {Sterkin}, \citenamefont {Grinvald},\ and\ \citenamefont {Aertsen}}]{Arieli96}%
  \BibitemOpen
  \bibfield  {author} {\bibinfo {author} {\bibfnamefont {A.}~\bibnamefont {Arieli}}, \bibinfo {author} {\bibfnamefont {A.}~\bibnamefont {Sterkin}}, \bibinfo {author} {\bibfnamefont {A.}~\bibnamefont {Grinvald}},\ and\ \bibinfo {author} {\bibfnamefont {A.}~\bibnamefont {Aertsen}},\ }\bibfield  {title} {\bibinfo {title} {Dynamics of ongoing activity: explanation of the large variability in evoked cortical responses},\ }\href@noop {} {\bibfield  {journal} {\bibinfo  {journal} {Science}\ }\textbf {\bibinfo {volume} {273}},\ \bibinfo {pages} {1868} (\bibinfo {year} {1996})}\BibitemShut {NoStop}%
\bibitem [{\citenamefont {Churchland}\ \emph {et~al.}(2010)\citenamefont {Churchland}, \citenamefont {Yu}, \citenamefont {Cunningham}, \citenamefont {Sugrue}, \citenamefont {Cohen}, \citenamefont {Corrado}, \citenamefont {Newsome}, \citenamefont {Clark}, \citenamefont {Hosseini}, \citenamefont {Scott}, \citenamefont {Bradley}, \citenamefont {Smith}, \citenamefont {Kohn}, \citenamefont {Movshon}, \citenamefont {Armstrong}, \citenamefont {Moore}, \citenamefont {Chang}, \citenamefont {Snyder}, \citenamefont {Lisberger}, \citenamefont {Priebe}, \citenamefont {Finn}, \citenamefont {Ferster}, \citenamefont {Ryu}, \citenamefont {Santhanam}, \citenamefont {Sahani},\ and\ \citenamefont {Shenoy}}]{Churchland10}%
  \BibitemOpen
  \bibfield  {author} {\bibinfo {author} {\bibfnamefont {M.~M.}\ \bibnamefont {Churchland}}, \bibinfo {author} {\bibfnamefont {B.~M.}\ \bibnamefont {Yu}}, \bibinfo {author} {\bibfnamefont {J.~P.}\ \bibnamefont {Cunningham}}, \bibinfo {author} {\bibfnamefont {L.~P.}\ \bibnamefont {Sugrue}}, \bibinfo {author} {\bibfnamefont {M.~R.}\ \bibnamefont {Cohen}}, \bibinfo {author} {\bibfnamefont {G.~S.}\ \bibnamefont {Corrado}}, \bibinfo {author} {\bibfnamefont {W.~T.}\ \bibnamefont {Newsome}}, \bibinfo {author} {\bibfnamefont {A.~M.}\ \bibnamefont {Clark}}, \bibinfo {author} {\bibfnamefont {P.}~\bibnamefont {Hosseini}}, \bibinfo {author} {\bibfnamefont {B.~B.}\ \bibnamefont {Scott}}, \bibinfo {author} {\bibfnamefont {D.~C.}\ \bibnamefont {Bradley}}, \bibinfo {author} {\bibfnamefont {M.~A.}\ \bibnamefont {Smith}}, \bibinfo {author} {\bibfnamefont {A.}~\bibnamefont {Kohn}}, \bibinfo {author} {\bibfnamefont {J.~A.}\ \bibnamefont {Movshon}}, \bibinfo {author} {\bibfnamefont {K.~M.}\ \bibnamefont
  {Armstrong}}, \bibinfo {author} {\bibfnamefont {T.}~\bibnamefont {Moore}}, \bibinfo {author} {\bibfnamefont {S.~W.}\ \bibnamefont {Chang}}, \bibinfo {author} {\bibfnamefont {L.~H.}\ \bibnamefont {Snyder}}, \bibinfo {author} {\bibfnamefont {S.~G.}\ \bibnamefont {Lisberger}}, \bibinfo {author} {\bibfnamefont {N.~J.}\ \bibnamefont {Priebe}}, \bibinfo {author} {\bibfnamefont {I.~M.}\ \bibnamefont {Finn}}, \bibinfo {author} {\bibfnamefont {D.}~\bibnamefont {Ferster}}, \bibinfo {author} {\bibfnamefont {S.~I.}\ \bibnamefont {Ryu}}, \bibinfo {author} {\bibfnamefont {G.}~\bibnamefont {Santhanam}}, \bibinfo {author} {\bibfnamefont {M.}~\bibnamefont {Sahani}},\ and\ \bibinfo {author} {\bibfnamefont {K.~V.}\ \bibnamefont {Shenoy}},\ }\bibfield  {title} {\bibinfo {title} {Stimulus onset quenches neural variability: a widespread cortical phenomenon},\ }\href {https://doi.org/10.1038/nn.2501} {\ \textbf {\bibinfo {volume} {13}},\ \bibinfo {pages} {369} (\bibinfo {year} {2010})}\BibitemShut
  {NoStop}%
\bibitem [{\citenamefont {Tchumatchenko}\ \emph {et~al.}(2010)\citenamefont {Tchumatchenko}, \citenamefont {Malyshev}, \citenamefont {Geisel}, \citenamefont {Volgushev},\ and\ \citenamefont {Wolf}}]{Tchumatchenko10_058102}%
  \BibitemOpen
  \bibfield  {author} {\bibinfo {author} {\bibfnamefont {T.}~\bibnamefont {Tchumatchenko}}, \bibinfo {author} {\bibfnamefont {A.}~\bibnamefont {Malyshev}}, \bibinfo {author} {\bibfnamefont {T.}~\bibnamefont {Geisel}}, \bibinfo {author} {\bibfnamefont {M.}~\bibnamefont {Volgushev}},\ and\ \bibinfo {author} {\bibfnamefont {F.}~\bibnamefont {Wolf}},\ }\bibfield  {title} {\bibinfo {title} {Correlations and synchrony in threshold neuron models},\ }\href@noop {} {\ \textbf {\bibinfo {volume} {104}},\ \bibinfo {pages} {058102} (\bibinfo {year} {2010})}\BibitemShut {NoStop}%
\bibitem [{\citenamefont {Rahimi}\ and\ \citenamefont {Recht}(2007)}]{Rahimi07Randomfeatureslarge}%
  \BibitemOpen
  \bibfield  {author} {\bibinfo {author} {\bibfnamefont {A.}~\bibnamefont {Rahimi}}\ and\ \bibinfo {author} {\bibfnamefont {B.}~\bibnamefont {Recht}},\ }\bibfield  {title} {\bibinfo {title} {Random features for large-scale kernel machines},\ }\href@noop {} {\bibfield  {journal} {\bibinfo  {journal} {Advances in neural information processing systems}\ }\textbf {\bibinfo {volume} {20}} (\bibinfo {year} {2007})}\BibitemShut {NoStop}%
\bibitem [{\citenamefont {Chizat}\ \emph {et~al.}(2019)\citenamefont {Chizat}, \citenamefont {Oyallon},\ and\ \citenamefont {Bach}}]{Chizat19_neurips}%
  \BibitemOpen
  \bibfield  {author} {\bibinfo {author} {\bibfnamefont {L.}~\bibnamefont {Chizat}}, \bibinfo {author} {\bibfnamefont {E.}~\bibnamefont {Oyallon}},\ and\ \bibinfo {author} {\bibfnamefont {F.}~\bibnamefont {Bach}},\ }\bibfield  {title} {\bibinfo {title} {On lazy training in differentiable programming}\ }(\bibinfo {year} {2019})\BibitemShut {NoStop}%
\bibitem [{\citenamefont {Bordelon}\ and\ \citenamefont {Pehlevan}(2022{\natexlab{b}})}]{bordelon2022}%
  \BibitemOpen
  \bibfield  {author} {\bibinfo {author} {\bibfnamefont {B.}~\bibnamefont {Bordelon}}\ and\ \bibinfo {author} {\bibfnamefont {C.}~\bibnamefont {Pehlevan}},\ }\bibfield  {title} {\bibinfo {title} {Population codes enable learning from few examples by shaping inductive bias},\ }\href@noop {} {\bibfield  {journal} {\bibinfo  {journal} {bioRxiv : the preprint server for biology}\ ,\ \bibinfo {pages} {2021}} (\bibinfo {year} {2022}{\natexlab{b}})}\BibitemShut {NoStop}%
\bibitem [{\citenamefont {Fischer}\ \emph {et~al.}(2024)\citenamefont {Fischer}, \citenamefont {Lindner}, \citenamefont {Dahmen}, \citenamefont {Ringel}, \citenamefont {Kr{\"a}mer},\ and\ \citenamefont {Helias}}]{Fischer24_10761}%
  \BibitemOpen
  \bibfield  {author} {\bibinfo {author} {\bibfnamefont {K.}~\bibnamefont {Fischer}}, \bibinfo {author} {\bibfnamefont {J.}~\bibnamefont {Lindner}}, \bibinfo {author} {\bibfnamefont {D.}~\bibnamefont {Dahmen}}, \bibinfo {author} {\bibfnamefont {Z.}~\bibnamefont {Ringel}}, \bibinfo {author} {\bibfnamefont {M.}~\bibnamefont {Kr{\"a}mer}},\ and\ \bibinfo {author} {\bibfnamefont {M.}~\bibnamefont {Helias}},\ }\href@noop {} {\bibinfo {title} {Critical feature learning in deep neural networks}} (\bibinfo {year} {2024}),\ \Eprint {https://arxiv.org/abs/2405.10761} {arXiv:2405.10761 [cond-mat.dis-nn]} \BibitemShut {NoStop}%
\bibitem [{\citenamefont {Ringel}\ \emph {et~al.}(2025)\citenamefont {Ringel}, \citenamefont {Rubin}, \citenamefont {Mor}, \citenamefont {Helias},\ and\ \citenamefont {Seroussi}}]{ringel2025}%
  \BibitemOpen
  \bibfield  {author} {\bibinfo {author} {\bibfnamefont {Z.}~\bibnamefont {Ringel}}, \bibinfo {author} {\bibfnamefont {N.}~\bibnamefont {Rubin}}, \bibinfo {author} {\bibfnamefont {E.}~\bibnamefont {Mor}}, \bibinfo {author} {\bibfnamefont {M.}~\bibnamefont {Helias}},\ and\ \bibinfo {author} {\bibfnamefont {I.}~\bibnamefont {Seroussi}},\ }\href {https://doi.org/10.48550/arXiv.2502.18553} {\bibinfo {title} {Applications of {{Statistical Field Theory}} in {{Deep Learning}}}} (\bibinfo {year} {2025}),\ \Eprint {https://arxiv.org/abs/2502.18553} {arXiv:2502.18553 [stat]} \BibitemShut {NoStop}%
\bibitem [{\citenamefont {Lauditi}\ \emph {et~al.}(2025)\citenamefont {Lauditi}, \citenamefont {Bordelon},\ and\ \citenamefont {Pehlevan}}]{lauditi2025}%
  \BibitemOpen
  \bibfield  {author} {\bibinfo {author} {\bibfnamefont {C.}~\bibnamefont {Lauditi}}, \bibinfo {author} {\bibfnamefont {B.}~\bibnamefont {Bordelon}},\ and\ \bibinfo {author} {\bibfnamefont {C.}~\bibnamefont {Pehlevan}},\ }\href {https://doi.org/10.48550/arXiv.2502.07998} {\bibinfo {title} {Adaptive kernel predictors from feature-learning infinite limits of neural networks}} (\bibinfo {year} {2025}),\ \Eprint {https://arxiv.org/abs/2502.07998} {arXiv:2502.07998 [cs]} \BibitemShut {NoStop}%
\bibitem [{\citenamefont {Bauer}\ \emph {et~al.}(2026)\citenamefont {Bauer}, \citenamefont {Fischer}, \citenamefont {Helias},\ and\ \citenamefont {Palmigiano}}]{bauer2026}%
  \BibitemOpen
  \bibfield  {author} {\bibinfo {author} {\bibfnamefont {J.~P.}\ \bibnamefont {Bauer}}, \bibinfo {author} {\bibfnamefont {K.}~\bibnamefont {Fischer}}, \bibinfo {author} {\bibfnamefont {M.}~\bibnamefont {Helias}},\ and\ \bibinfo {author} {\bibfnamefont {A.}~\bibnamefont {Palmigiano}},\ }\href {https://doi.org/10.48550/arXiv.2602.15593} {\bibinfo {title} {A unified theory of feature learning in {{RNNs}} and {{DNNs}}}} (\bibinfo {year} {2026}),\ \Eprint {https://arxiv.org/abs/2602.15593} {arXiv:2602.15593 [cs]} \BibitemShut {NoStop}%
\bibitem [{\citenamefont {Clark}\ \emph {et~al.}(2026)\citenamefont {Clark}, \citenamefont {Bordelon}, \citenamefont {{Zavatone-Veth}},\ and\ \citenamefont {Pehlevan}}]{clark2026}%
  \BibitemOpen
  \bibfield  {author} {\bibinfo {author} {\bibfnamefont {D.~G.}\ \bibnamefont {Clark}}, \bibinfo {author} {\bibfnamefont {B.}~\bibnamefont {Bordelon}}, \bibinfo {author} {\bibfnamefont {J.~A.}\ \bibnamefont {{Zavatone-Veth}}},\ and\ \bibinfo {author} {\bibfnamefont {C.}~\bibnamefont {Pehlevan}},\ }\href {https://doi.org/10.64898/2026.03.02.708943} {\bibinfo {title} {Structure, disorder, and dynamics in task-trained recurrent neural circuits}} (\bibinfo {year} {2026})\BibitemShut {NoStop}%
\bibitem [{\citenamefont {Coolen}(2000)}]{Coolen00_arxiv_II}%
  \BibitemOpen
  \bibfield  {author} {\bibinfo {author} {\bibfnamefont {A.~C.~C.}\ \bibnamefont {Coolen}},\ }\bibfield  {title} {\bibinfo {title} {Statistical mechanics of recurrent neural networks ii. dynamics},\ }\href@noop {} {\  (\bibinfo {year} {2000})}\BibitemShut {NoStop}%
\bibitem [{\citenamefont {Kadmon}\ and\ \citenamefont {Sompolinsky}(2015{\natexlab{b}})}]{Kadmon15Transitionchaosrandom}%
  \BibitemOpen
  \bibfield  {author} {\bibinfo {author} {\bibfnamefont {J.}~\bibnamefont {Kadmon}}\ and\ \bibinfo {author} {\bibfnamefont {H.}~\bibnamefont {Sompolinsky}},\ }\bibfield  {title} {\bibinfo {title} {Transition to chaos in random neuronal networks},\ }\href@noop {} {\bibfield  {journal} {\bibinfo  {journal} {Physical Review X}\ }\textbf {\bibinfo {volume} {5}},\ \bibinfo {pages} {041030} (\bibinfo {year} {2015}{\natexlab{b}})}\BibitemShut {NoStop}%
\bibitem [{\citenamefont {Bradbury}\ \emph {et~al.}(2018)\citenamefont {Bradbury}, \citenamefont {Frostig}, \citenamefont {Hawkins}, \citenamefont {Johnson}, \citenamefont {Leary}, \citenamefont {Maclaurin}, \citenamefont {Necula}, \citenamefont {Paszke}, \citenamefont {VanderPlas}, \citenamefont {{Wanderman-Milne}},\ and\ \citenamefont {Zhang}}]{bradbury2018}%
  \BibitemOpen
  \bibfield  {author} {\bibinfo {author} {\bibfnamefont {J.}~\bibnamefont {Bradbury}}, \bibinfo {author} {\bibfnamefont {R.}~\bibnamefont {Frostig}}, \bibinfo {author} {\bibfnamefont {P.}~\bibnamefont {Hawkins}}, \bibinfo {author} {\bibfnamefont {M.~J.}\ \bibnamefont {Johnson}}, \bibinfo {author} {\bibfnamefont {C.}~\bibnamefont {Leary}}, \bibinfo {author} {\bibfnamefont {D.}~\bibnamefont {Maclaurin}}, \bibinfo {author} {\bibfnamefont {G.}~\bibnamefont {Necula}}, \bibinfo {author} {\bibfnamefont {A.}~\bibnamefont {Paszke}}, \bibinfo {author} {\bibfnamefont {J.}~\bibnamefont {VanderPlas}}, \bibinfo {author} {\bibfnamefont {S.}~\bibnamefont {{Wanderman-Milne}}},\ and\ \bibinfo {author} {\bibfnamefont {Q.}~\bibnamefont {Zhang}},\ }\href@noop {} {\bibinfo {title} {{{JAX}}: Composable transformations of {{Python}}+{{NumPy}} programs}} (\bibinfo {year} {2018})\BibitemShut {NoStop}%
\bibitem [{\citenamefont {Koltchinskii}\ and\ \citenamefont {Gin{\'e}}(2000)}]{Koltchinskii00Randommatrixapproximation}%
  \BibitemOpen
  \bibfield  {author} {\bibinfo {author} {\bibfnamefont {V.}~\bibnamefont {Koltchinskii}}\ and\ \bibinfo {author} {\bibfnamefont {E.}~\bibnamefont {Gin{\'e}}},\ }\bibfield  {title} {\bibinfo {title} {Random matrix approximation of spectra of integral operators},\ }\href@noop {} {\bibfield  {journal} {\bibinfo  {journal} {Bernoulli. Official Journal of the Bernoulli Society for Mathematical Statistics and Probability}\ ,\ \bibinfo {pages} {113}} (\bibinfo {year} {2000})}\BibitemShut {NoStop}%
\bibitem [{\citenamefont {Braun}(2005)}]{braun2005}%
  \BibitemOpen
  \bibfield  {author} {\bibinfo {author} {\bibfnamefont {M.~L.}\ \bibnamefont {Braun}},\ }\bibfield  {title} {\bibinfo {title} {Spectral properties of the kernel matrix and their relation to kernel methods in machine learning},\ }\href@noop {} {\  (\bibinfo {year} {2005})}\BibitemShut {NoStop}%
\bibitem [{\citenamefont {Suetin}(2001)}]{Suetin2001}%
  \BibitemOpen
  \bibfield  {author} {\bibinfo {author} {\bibfnamefont {{\relax PK}.}~\bibnamefont {Suetin}},\ }\bibfield  {title} {\bibinfo {title} {Ultraspherical polynomials},\ }\href@noop {} {\bibfield  {journal} {\bibinfo  {journal} {Encyclopaedia of mathematics. Springer, Berlin}\ } (\bibinfo {year} {2001})}\BibitemShut {NoStop}%
\end{thebibliography}%

\onecolumngrid

\appendix

\section{Model-independent mean-field theory for random networks\label{app:DMFT}}

This section presents a self-contained derivation of the model-independent
mean-field theory for networks with Gaussian random connectivity $J_{ij}\stackrel{\text{i.i.d.}}{\sim}\mathcal{N}\left(\frac{\bar{g}}{N},\,\frac{g^{2}}{N}\right)$.
This formalism is the basis to condense the network's microscopic
details into at an effective description in terms of Gaussian statistics
of the neurons. To this end, we first reformulate the neuron models
in terms of transmission functionals in \cref{app:transmission_laws}.
Then, we present the calculation for a single network, and finally
for the two-replica case which forms the basis for the kernel function.

The $N$ neurons have inputs $\b h^{J}(t)=\left(h^{J}_{1}(t),...,h^{J}_{N}(t)\right)$
and outputs $\b{\phi}^{J}(t)=\left(\phi^{J}_{1}(t),...,\phi^{J}_{N}(t)\right)$.
For readability, we here omit the superscript indicating dependence
on the connectivity $J$. The neuronal dynamics is described by the
causal conditional probability functional $p_{i}[\phi_{i}|h_{i}]$.
For deterministic neurons, where $\phi_{i}=f_{i}[h_{i}]$ is some
causal functional of the input, one may set $p_{i}[\phi_{i}|h_{i}]=\delta[\phi_{i}-f_{i}[h_{i}]]$.
We use vectorial notation to denote
\begin{align}
p[\boldsymbol{\phi}|\boldsymbol{h}] & =\prod^{N}_{i=1}p_{i}[\phi_{i}|h_{i}],\label{eq:transmission_law}
\end{align}
because, given their inputs $\{h_{i}\}$, neurons are otherwise pairwise
independent. The probability functional $p[\phi_{i}|h_{i}]$ is assumed
to be strictly causal, which is $\phi_{i}(t)$ is independent of $h_{i}(s\ge t)$;
a more explicit notation would be $p[\phi_{i,+}|h_{i}]$, denoting
that the time-argument of $\phi_{i,+}$ is infinitesimally advanced
compared to the history $h_{i}$.

The joint statistics of input and output is then
\begin{equation}
p[\boldsymbol{\phi}_{+},\boldsymbol{h}]=p[\boldsymbol{\phi}_{+}|\boldsymbol{h}]\,p[\boldsymbol{h}|\boldsymbol{\phi}],\label{eq:joint_prob_input_output_app}
\end{equation}
and the distribution of the inputs $\bh$ is given as the marginalization
over $\bphi$ as
\begin{align}
p[\bh]= & \int\D\bphi_{+}\,p[\bphi_{+},\bh]\label{eq:marginalization_x}\\
= & \int\D\bphi_{+}\,p[\boldsymbol{\phi_{+}}|\boldsymbol{h}]\,p[\boldsymbol{h}|\boldsymbol{\phi}].\nonumber 
\end{align}
The connectivity $\bJ$ couples the outputs $\bphi$ of the neurons
back to the input $\bh$ as
\begin{align*}
\bh(t) & =\bJ\,\bphi(t)+\bm{h}_{\text{ext}},
\end{align*}
where $\bm{h}_{\text{ext}}=\bm{V}\bm{x}$ represents an external input
projected by a Gaussian random matrix $V_{ia}\sim\N\left(0,\,1/M\right)$
with the dimension of the input $M$. So in the marginalization \cref{eq:marginalization_x}
over $\bphi$ we need to set

\begin{align}
p[\bh|\bphi]= & \delta\big[\bh-\bJ\bphi-\bm{V}\bm{x}_{\text{ext}}\big]\label{eq:rho_h}\\
= & \int\D\bhh\,\exp\big(\bhh^{\T}\bh\big)\,\exp\big(-\bhh^{\T}\bJ\bphi\big)\exp\big(-\bhh^{\T}\bm{V}\bm{x}_{\text{ext}}\big),\nonumber 
\end{align}
where the path integral measure is $\int\D\bhh=\prod_{t}\prod_{j}\int^{i\infty}_{-i\infty}\frac{d\hat{h}_{j}(t)}{2\pi i}$
and the inner product is meant as $\bhh^{\T}\bh=\sum^{N}_{i=1}\int^{\infty}_{-\infty}dt\,\hat{h}_{i}(t)h_{i}(t)$.
By connecting the outputs back to the inputs, \cref{eq:joint_prob_input_output_app}
may seem to take a circular structure like $p[\boldsymbol{\phi},\boldsymbol{h}]=p[\boldsymbol{\phi}|\boldsymbol{h}]p[\boldsymbol{h}|\boldsymbol{\phi}]$.
But since the first conditional probability is causal, and the second
couples only equal time points, \cref{eq:joint_prob_input_output_app}
is more explicitly represented as 
\begin{align*}
p[\boldsymbol{\phi_{+}},\boldsymbol{h}] & =\prod_{t}p[\boldsymbol{\phi}(t)|\boldsymbol{h}_{<t}]\,p[\boldsymbol{h}(t)|\boldsymbol{\phi}(t)]
\end{align*}
such that the conditioning is ordered in time. This relation forms
the starting point once the transmission functional $p[\boldsymbol{\phi}_{+}|\boldsymbol{h}]$
is known.

\subsection*{Single replicon\label{app:Single-replicon}}

To calculate effective single-neuron statistics, we now perform the
disorder average $\langle\ldots\rangle_{\bJ}$ of \cref{eq:marginalization_x}.
The only term affected is the last exponential factor in the second
line of \cref{eq:rho_h}, which yields
\begin{align}
 & \Big\langle\exp\big(-\bhh^{\T}\,\bJ\,\bphi\big)\Big\rangle_{\bJ\stackrel{\text{i.i.d.}}{\sim}\mathcal{N}\left(\frac{\bar{g}}{N},\,\frac{g^{2}}{N}\right)}\label{eq:disorder_avgd_conn}\\
= & \exp\Big(-\frac{\bar{g}}{N}\,\sum^{N}_{i=1}\hat{h}^{\T}_{i}\,\sum^{N}_{j=1}\phi_{j}+\frac{g^{2}}{2N}\,\sum^{N}_{i,j=1}\big(\hat{h}^{\T}_{i}\phi_{j}\big)^{2}\Big)\nonumber 
\end{align}
by the factorization of the expectation. Here, we redefined the scalar
product such that the last term in the exponent rewrites explicitly
as
\begin{align*}
\big(\hat{h}^{\T}_{i}\phi_{j}\big)^{2} & =\iint dt\,ds\,\hat{h}_{i}(t)\hat{h}_{i}(s)\,\phi_{j}(t)\phi_{j}(s).
\end{align*}
Similarly, for the external input we obtain

\begin{align}
 & \Big\langle\exp\big(-\bhh^{\T}\bm{V}\bm{x}\big)\Big\rangle_{\bm{V}\stackrel{\text{i.i.d.}}{\sim}\mathcal{N}\left(0,\,\frac{1}{M}\right)}\label{eq:disorder_avgd_conn-1}\\
= & \prod^{N}_{i=1}\exp\Big(\frac{1}{2M}\,\sum^{M}_{a=1}\big(\hat{h}^{\T}_{i}x_{a}\big)^{2}\Big)\,.\nonumber 
\end{align}
This suggests the introduction of the auxiliary recurrent fields $\mathcal{R}(t):=\frac{\bar{g}}{N}\sum_{j}\phi_{j}(t)$
and $\mathcal{Q}(t,s):=\frac{g^{2}}{N}\sum_{j}\phi_{j}(t)\phi_{j}(s)$
as well as a deterministic contribution $Q_{\text{ext}}(t,s):=\frac{1}{M}\sum_{a}x_{a}(t)x_{a}(s)$
to rewrite \cref{eq:disorder_avgd_conn} and \cref{eq:disorder_avgd_conn-1}
as
\begin{align}
\prod_{i}\,\exp\Big(-\hat{h}^{\T}_{i}\mathcal{R}+\frac{1}{2}\,\hat{h}^{\T}_{i}\left(\mathcal{Q}+Q_{\text{ext}}\right)\hat{h}_{i}\Big),\label{eq:factorization}
\end{align}
where the bi-linear form is to be read as 
\begin{align*}
\hat{h}^{\T}_{i}\mathcal{Q}\hat{h}_{i} & =\iint dt\,ds\,\hat{h}^{\T}_{i}(t)\mathcal{Q}(t,s)\hat{h}_{i}(s).
\end{align*}
The appearance of the product sign in \cref{eq:factorization} and
the neuron-independent fields $\mathcal{R}$ and $\mathcal{Q}$ signifies
that the problem becomes completely symmetric with regard to neurons.
Enforcing the definitions of the auxiliary fields by Dirac distributions,
represented in Fourier domain analogous to \cref{eq:rho_h}, yields
another pair of fields $\hat{\mathcal{R}}$ and $\hat{\mathcal{Q}}$
and brings \cref{eq:marginalization_x} into the form
\begin{align}
\big\langle p[\bh]\big\rangle_{\bJ}\stackrel{(\ref{eq:marginalization_x}),(\ref{eq:rho_h})}{=} & \int\D\bphi\,p[\boldsymbol{\phi}|\boldsymbol{h}]\,\big\langle\delta[\bh-\bJ\bphi-\bm{h}_{\text{ext}}]\big\rangle_{\bJ}\label{eq:rho_disorder_avgd}\\
\stackrel{(\ref{eq:disorder_avgd_conn}),(\ref{eq:factorization})}{=} & \int\D\{\mathcal{Q},\mathcal{R},\hat{\mathcal{Q}},\hat{\mathcal{R}}\}\exp\big(-\frac{N}{\bar{g}}\hat{\mathcal{R}}^{\T}\mathcal{R}-\frac{N}{g^{2}}\mathcal{\hat{Q}}^{\T}\mathcal{Q}\big)\nonumber \\
 & \times\prod_{i}\int\D\{\phi_{i},\hat{h}_{i}\}\,p[\phi_{i}|h_{i}]\,\exp\Big(\hat{h}^{\T}_{i}h_{i}-\hat{h}^{\T}_{i}\mathcal{R}+\frac{1}{2}\,\hat{h}^{\T}_{i}\left(\mathcal{Q}+Q_{\text{ext}}\right)\hat{h}_{i}+\hat{\mathcal{R}}^{\T}\phi_{i}+\phi^{\T}_{i}\hat{\mathcal{Q}}\phi_{i}\Big).\nonumber 
\end{align}
We now compute the values of the auxiliary fields that provide the
dominant contribution to the probability mass. To this end consider
\begin{align*}
1 & \equiv\int\D\bh\,\big\langle p[\bh]\big\rangle_{\bJ}\\
 & \stackrel{(\ref{eq:rho_disorder_avgd})}{=}\int\D\{\mathcal{Q},\mathcal{R},\hat{\mathcal{Q}},\hat{\mathcal{R}}\}\exp\big(-\frac{N}{\bar{g}}\hat{\mathcal{R}}^{\T}\mathcal{R}-\frac{N}{g^{2}}\mathcal{\hat{Q}}^{\T}\mathcal{Q}\big)\\
 & \times\prod_{i}\int\D\{\phi_{i},h_{i},\hat{h}_{i}\}\,p[\phi_{i}|h_{i}]\,\exp\Big(\hat{h}^{\T}_{i}h_{i}-\hat{h}^{\T}_{i}\mathcal{R}+\frac{1}{2}\,\hat{h}^{\T}_{i}\left(\mathcal{Q}+Q_{\text{ext}}\right)\hat{h}_{i}+\hat{\mathcal{R}}^{\T}\phi_{i}+\phi^{\T}_{i}\hat{\mathcal{Q}}\phi_{i}\Big).
\end{align*}
The exponent in the second line can be considered an action of a field
theory for the auxiliary fields $\{\mathcal{Q},\mathcal{R},\hat{\mathcal{Q}},\hat{\mathcal{R}}\}$.
The integral in the last line appears to the power of $N$, so that
one may rewrite the full expression as
\begin{align*}
 & \int\D\{\mathcal{Q},\mathcal{R},\hat{\mathcal{Q}},\hat{\mathcal{R}}\}\exp\big(N\,\Omega[\mathcal{R},\mathcal{Q},\hat{\mathcal{R}},\hat{\mathcal{Q}}]\big)\\
\text{with }\Omega[\mathcal{R},\mathcal{Q},\hat{\mathcal{R}},\hat{\mathcal{Q}}] & :=-\frac{\mathcal{R}^{\T}\hat{\mathcal{R}}}{\bar{g}}-\frac{\mathcal{Q}^{\T}\mathcal{\hat{Q}}}{g^{2}}\\
 & +\ln\,\int\D\{\phi,h,\hat{h}\}\,p[\phi|h]\,\exp\Big(\hat{h}^{\T}h-\hat{h}^{\T}\mathcal{R}+\frac{1}{2}\,\hat{h}^{\T}\left(\mathcal{Q}+Q_{\text{ext}}\right)\hat{h}+\hat{\mathcal{R}}^{\T}\phi+\phi^{\T}\hat{\mathcal{Q}}\phi\Big).
\end{align*}
Herein, $\Omega$ appears as an effective measure on the scalar fields
$\phi$ that does not follow Gaussian statistics and hence cannot
easily be marginalized.

\subsubsection*{Mean-field equations}

The appearance of $N$ in the exponent $N\,\Omega[\mathcal{R},\mathcal{Q}]$
suggests to perform the integration over the fields $\{\mathcal{Q},\mathcal{R},\hat{\mathcal{Q}},\hat{\mathcal{R}}\}$
in saddle point approximation, demanding $\frac{\delta\Omega}{\delta\{\mathcal{Q},\mathcal{R},\hat{\mathcal{Q}},\hat{\mathcal{R}}\}}\stackrel{!}{=}0$,
which yields four conditions for the saddle point values $R,Q,\hat{R},\text{\ensuremath{\hat{Q}}}$
of the field
\begin{align*}
R(t) & =\bar{g}\,\langle\phi(t)\rangle_{\Omega(R,\bar{Q})},\\
\hat{R}(t) & =\bar{g}\,\langle\hat{h}(t)\rangle_{\Omega(R,\bar{Q})}\equiv0,\\
Q(t,s) & =g^{2}\,\langle\phi(t)\,\phi(s)\rangle_{\Omega(R,\bar{Q})},\\
\hat{Q}(t,s) & =\frac{g^{2}}{2}\,\langle\hat{h}(t)\,\hat{h}(s)\rangle_{\Omega(R,\bar{Q})}\equiv0.
\end{align*}
Here the expectation value is $\langle\ldots\rangle_{\Omega(R,\bar{Q})}=\int\D\phi\,\ldots\,\int\D\{h,\hat{h}\}\,p[\phi|h]\,\exp\Big(\hat{h}^{\T}h-\hat{h}^{\T}R+\frac{1}{2}\,\hat{h}^{\T}\bar{Q}\hat{h}\Big)$
and $\bar{Q}\coloneqq Q+Q_{\text{ext}}$. The outer derivative of
the logarithm appearing in the expression for $\Omega$ does not contribute,
because the normalization condition of the latter distribution is
unity, since the exponential term is the moment-generating functional
of a one-dimensional Gaussian process $h\sim\N(R,\bar{Q})$ and $p[\phi|h]$
is normalized. We can thus rewrite the average
\begin{align}
\langle\ldots\rangle_{\Omega(R,Q)}= & \int\D\phi\,\ldots\,\langle p[\phi|h]\rangle_{h\sim\N(R,\bar{Q})},\label{eq:def_av_Omega}
\end{align}
where the $h$-fields have cumulants

\begin{align*}
\cum{h(t)}{} & =R(t),\\
\cum{h(t)\,h(s)}{} & =\bar{Q}(t,s).
\end{align*}
The auxiliary fields $\hat{h}$ are zero on expectation, which is
a consequence of the normalization \citep[Section X]{Coolen00_arxiv_II,Helias19_10416}.
At this stage, the transmission laws from \cref{app:transmission_laws}
enter and allow for the self-consistent calculation of the statistics.

\subsection*{Two replica \textendash{} Calculation of the kernel function\label{app:Two-replica}\label{app:kernel_mft}}

We introduced the kernel function in \cref{eq:kernel-def}. For
brevity, in this appendix we set the regularization noise to zero,
$\Lambda=0$ and we consider the case of two input data points, denoted
with superscripts $1$ and $2$, respectively. The kernel then describes
the network-transformed pairwise correlation
\begin{equation}
k_{ts}(\bm{x}^{1}\cdot\bm{x}^{2})\coloneqq\frac{1}{N}\bm{\phi}^{J}_{\bm{x}^{1}}(t)\cdot\bm{\phi}^{J}_{\bm{x}^{2}}(s),\label{eq:kernel_app}
\end{equation}
where $\bm{\phi}^{J}_{\bm{x}}$ denotes the nonlinear feature map
that is implemented by the network. It depends on the synaptic connectivity
$J$, activation function $T$, stimulus propagation times $t$ and
$s$, and the noise realization. Note that in the main text, we only
considered responses at equal times, $k_{t}\coloneqq k_{tt}$. To
calculate the kernel, we hence need to perform a replica calculation
that considers a pair of networks with identical connectivity but
different initial conditions for the neurons caused by the inputs
$\bm{x}$.

The formal derivation of mean-field equations that approximates \cref{eq:kernel_app}
proceeds analogous to \cref{app:Single-replicon}: The corresponding
expression to \cref{eq:marginalization_x} and \cref{eq:rho_h} reads
\begin{align*}
p[\bh^{1},\bh^{2}]= & \int\D\{\bphi^{1},\bphi^{2}\}\,p[\bphi^{1},\bphi^{2}|\bh^{1},\bh^{2}]\,\prod^{2}_{\alpha=1}\delta[\bh^{\alpha}-\bJ\,\bphi^{\alpha}-\bm{h}^{\alpha}_{\text{ext}}].
\end{align*}
Here, the conditional density $p[\bphi^{1},\bphi^{2}|\bh^{1},\bh^{2}]$
is a joint distribution across the two replica, because it must allow
the representation of update processes or stochastic activations of
corresponding neurons which have identical realizations between the
two replica.

The important point is the identical matrix $\bJ$ appearing in the
product of the latter two Dirac distributions, which, after introducing
Fourier representations as in \cref{eq:rho_h} and taking the disorder
average over $\bJ$, analogous to \cref{eq:disorder_avgd_conn}, yields
\begin{align}
 & \Big\langle\exp\big(-\bhh^{1\T}\bJ\bphi^{1}-\hat{\bh}^{2\T}\bJ\bphi^{2}\big)\Big\rangle_{\bJ\stackrel{\text{i.i.d.}}{\sim}\mathcal{N}\left(\frac{\bar{g}}{N},\,\frac{g^{2}}{N}\right)}\label{eq:disorder_pair}\\
= & \prod^{N}_{i=1}\prod^{2}_{\alpha=1}\exp\Big(-\frac{\bar{g}}{N}\,\hat{h}^{\alpha\T}_{i}\,\sum^{N}_{j=1}\phi^{\alpha}_{j}+\frac{g^{2}}{2N}\,\sum^{N}_{j=1}\big(\hat{h}^{\alpha\T}_{i}\phi^{\alpha}_{j}\big)^{2}\Big)\nonumber \\
 & \phantom{\prod^{N}_{i=1}}\times\exp\Big(\frac{g^{2}}{N}\,\sum^{N}_{j=1}\hat{h}^{1\T}_{i}\phi^{1}_{j}\,\hat{h}^{2\T}_{i}\phi^{2}_{j}\Big)\nonumber 
\end{align}
for the recurrence. Similar to \cref{eq:disorder_avgd_conn-1} above,
the contribution from the external input reads
\begin{align}
 & \Big\langle\exp\big(-\bhh^{1\T}\bm{V}\bm{x}^{1}-\hat{\bh}^{2\T}\bm{V}\bm{x}^{2}\big)\Big\rangle_{\bm{V}\stackrel{\text{i.i.d.}}{\sim}\mathcal{N}\left(0,\,\frac{1}{M}\right)}\label{eq:disorder_pair-V}\\
= & \prod^{N}_{i=1}\prod^{2}_{\alpha=1}\exp\Big(\frac{1}{2M}\,\sum^{M}_{a=1}\big(\hat{h}^{\alpha\T}_{i}x^{\alpha}_{a}\big)^{2}\Big)\nonumber \\
 & \phantom{\prod^{N}_{i=1}}\times\exp\Big(\frac{1}{M}\,\sum^{M}_{a=1}\hat{h}^{1\T}_{i}x^{1}_{a}\,\hat{h}^{2\T}_{i}x^{2}_{a}\Big).\nonumber 
\end{align}
The penultimate line in \cref{eq:disorder_pair} is the same contribution
for each replica as in the single system; it is treated in the same
manner by introducing pairs of auxiliary fields $\{\mathcal{R}^{\alpha},\hat{\mathcal{R}}^{\alpha},\mathcal{Q}^{\alpha\alpha},\hat{\mathcal{Q}}^{\alpha\alpha}\}_{\alpha\in\{1,2\}}$.
The last line couples the two replica and can be decoupled similarly
by defining
\begin{align}
\mathcal{Q}^{12}(t,s) & :=\frac{g^{2}}{N}\,\sum^{N}_{j=1}\phi^{1}_{j}(t)\,\phi^{2}_{j}(s).\label{eq:def_Q_12}
\end{align}
This definition is enforced by inserting a $\delta$-constraint, represented
as a Fourier integral with the corresponding conjugate field $\hat{\mathcal{Q}}^{12}(t,s)$.
For the external input, we again introduce the deterministic contribution
$Q^{12}_{\text{ext}}(t,s):=\frac{1}{M}\,\sum^{M}_{a=1}x^{1}_{a}(t)\,x^{2}_{a}(s).$

\subsubsection*{Mean-field equations}

The integral over $\{\mathcal{R}^{\alpha},\hat{\mathcal{R}}^{\alpha},\mathcal{Q}^{\alpha\beta},\hat{\mathcal{Q}}^{\alpha\beta}\}_{\alpha,\beta\in\{1,2\}}$
is then taken in saddle point approximation with the resulting non-trivial
saddle point equations
\begin{align}
R^{\alpha}(t) & =\bar{g}\,\langle\phi^{\alpha}(t)\rangle_{\Omega^{12}(\{R,\bar{Q}\})}\,,\label{eq:saddle_replica}\\
Q^{\alpha\beta}(t,s) & =g^{2}\,\langle\phi^{\alpha}(t)\,\phi^{\beta}(s)\rangle_{\Omega^{12}(\{R^{\alpha},\bar{Q}\})}\,,\nonumber 
\end{align}
The remaining $\hat{}-$fields all vanish $\hat{R}^{\alpha}=\hat{Q}^{\alpha\beta}\equiv0$.
The expectation value in \cref{eq:saddle_replica} is taken with the
measure
\begin{align}
\langle\ldots\rangle_{\Omega^{\alpha\beta}(\{R^{\alpha{}^{\prime}},\bar{Q}^{\alpha{}^{\prime}\beta{}^{\prime}}\}_{\alpha{}^{\prime}\beta{}^{\prime}})} & \coloneqq\int\D\{\phi^{\alpha},\phi^{\beta}\}\,\ldots\,\langle p[\phi^{\alpha},\phi^{\beta}|h^{\alpha},h^{\beta}]\rangle_{(h^{\alpha},h^{\beta})\sim\N^{\alpha\beta}(\{R^{\alpha{}^{\prime}},\bar{Q}^{\alpha{}^{\prime}\beta{}^{\prime}}\}_{\alpha{}^{\prime}\beta{}^{\prime}})},\label{eq:def_exp_replica}
\end{align}
where $(h^{\alpha},h^{\beta})\sim\N^{\alpha\beta}(\{R^{\alpha{}^{\prime}},\bar{Q}^{\alpha{}^{\prime}\beta{}^{\prime}}\}_{\alpha{}^{\prime}\beta{}^{\prime}})=\colon\N^{\alpha\beta}$
is a two-dimensional Gaussian processes with cumulants
\begin{align*}
\cum{h^{\alpha}(t)}{} & =R^{\alpha}(t)\equiv R(t),\\
\cum{h^{\alpha}(t)\,h^{\beta}(s)}{} & =\bar{Q}^{\alpha\beta}(t,s),
\end{align*}
where again $\bar{Q}^{\alpha\beta}(t,s)\coloneqq Q^{\alpha\beta}(t,s)+Q^{\alpha\beta}_{\text{ext}}(t,s)$.
The equation for the mean $\cum{h^{\alpha}(t)}{}$ falls back to the
single replicon statistics, reflecting that the presence of the second
replicon does not affect the within-replicon statistics. In summary,
this generalizes the single-replicon case $\alpha=\beta=1$ to the
case to two different systems $1=\alpha\neq\beta=2$.

\subsubsection*{Kernel function}

Using \cref{eq:kernel_app}, \cref{eq:def_Q_12}, and \cref{eq:saddle_replica},
the kernel follows as
\begin{align}
k_{ts}(\bm{x}^{1}\cdot\bm{x}^{2})\overset{\phantom{N\rightarrow\infty}}{=} & \frac{1}{N}\ev{\bm{\phi}^{J}_{\bm{x}^{1}}(t)\cdot\bm{\phi}^{J}_{\bm{x}^{2}}(s)}J\label{eq:def_Q12}\\
\overset{N\rightarrow\infty}{=} & Q^{12}(t,s)/g^{2}\nonumber \\
\overset{\phantom{N\rightarrow\infty}}{=} & \langle\phi^{1}(t)\,\phi^{2}(s)\rangle_{\Omega^{12}(\{R^{\alpha}(t),\,\bar{Q}^{\alpha\beta}(t,s)\})}.\nonumber 
\end{align}
Here, $\phi^{1}$ and $\phi^{2}$ are two scalar stochastic processes
which can be thought of as effective neurons summarizing each replicon.
In \cref{eq:def_exp_replica}, the details of the neuron model, such
as its transmission law, enter via the transmission functional $p[\phi^{1},\phi^{2}|h^{1},h^{2}]$.
To obtain the two-point correlation \cref{eq:def_Q12}, the joint
propagation through the replica needs to be taken into account. We
here closely follow the presentation in \citet{Keup21_021064}.

The work of \citet{Schuecker18_041029} formulated a PDE for the two-replica
correlation which extended the ODE found by \citet{Sompolinsky88_259}
for the autocorrelation of a single replicon. We here treat it on
equal footing with discrete networks to better compare the approaches.
For readability, we now move time indices to subscripts for these
scalar mean-field quantities where the neuron index has become superfluous.
We aim to cast the correlation of the activations $\phi$ in terms
of the pre-activations $h$ and subsequently average with respect
to the mean-field measure $(h^{\alpha}_{t},h^{\beta}_{s})\sim\N(\{R^{\alpha{}^{\prime}},Q^{\alpha{}^{\prime}\beta{}^{\prime}}\}_{\alpha{}^{\prime}\beta{}^{\prime}})$
as per \cref{eq:def_exp_replica}. To this end, we need the joint
neural transmission function $p[\phi^{1},\phi^{2}|h^{1},h^{2}]$.
We will derive it in the next section by extending the single-replicon
measure $p[\phi\,|\,h]$.

\section{Evaluation of mean-field equations for the network models\label{app:transmission_laws}}

We will now use \cref{eq:def_av_Omega} to evaluate the theory that
we formulated up to here on general grounds on the discrete and continuous
network models. To provide \cref{eq:transmission_law} for the models
in \cref{eq:transmission_dynamics_glauber} and \cref{eq:rate_from_avg},
we here reformulate them in terms of respective single-neuron probabilistic
transmission laws $p[\phi_{i}(t)\,|\,h_{i}]\equiv p[\phi(t)\,|\,h]\:\forall\,i$.

As a starting point, we reformulate the dynamical equations of the
models,

{\addtolength{\jot}{0.0em}

\begin{align}
\phi_{i}(t+dt)= & \left(1-\mathcal{\theta}_{\text{up}}\right)\,\phi_{i}(t)+\mathcal{\theta}_{\text{up}}\,\mathcal{T}_{i}(h_{i}(t))\label{eq:binary}\\
{\scriptstyle \downarrow\text{ average over }} & {\scriptstyle \text{\ensuremath{\theta_{\text{up}}\in\{0,1\}}}}\nonumber \\
\phi_{i}(t+dt)= & \left(1-\frac{dt}{\tau}\right)\,\phi_{i}(t)+\frac{dt}{\tau}\,T(h_{i}(t))+\xi(t).\label{eq:rate}
\end{align}

}Here, $h_{i}(t)=\sum_{j}J_{ij}\phi_{j}(t)$ again is the recurrent
synaptic input, $\mathcal{\theta}_{\text{up}}\in\{0,1\}$ is a Bernoulli
variable with $p(\theta_{\text{up}}=1)=\frac{dt}{\tau}$, and $\mathcal{T}_{i}\in\{-1,1\}$
with $p(\mathcal{T}_{i}=1)=\frac{1}{2}\big(T(h_{i}(t))+1\big)$. This
shows that the models are related by averaging over the stochastic
processes controlling updates $\theta_{\text{up}}$ and the neural
activation $\mathcal{T}$.

Importantly, both transmission laws \cref{eq:binary} and \cref{eq:rate}
are identical over neurons given the inputs $h_{i}$, inducing $p_{i}[\phi_{i}(t)|h_{i}]\equiv p[\phi_{i}(t)|h_{i}]\:\forall\,i$.
This forms the basis for using them after the disorder average in
the mean-field calculation below.

\subsection*{Continuous networks\label{app:Continuous-networks}}

The original formulation of continuous recurrent networks goes back
to \citet{Amari77}, interpreting \cref{eq:rate} as a synaptic input
flux following 
\begin{equation}
\tau\partial_{t}\phi_{i}(t)=T(h_{i}(t))+\xi_{i}(t)-\phi_{i}(t).\label{eq:amari_rate}
\end{equation}
It reflects that the change in the neuron in each time step is proportional
to the synaptic input $h_{i}(t)$ transformed by the activation $T$,
minus the neuronal flux $-\phi_{i}(t)$ which is fed back through
the recurrent circuitry. This equation can be integrated to give the
single-neuron functional $p^{c}_{i}\equiv p^{c}$

\begin{align}
p^{c}[\phi_{t}|h] & =\Big\langle\delta\Big[\phi_{t}-\int^{t}_{-\infty}\frac{dt^{\prime}}{\tau}\,e^{-(t-t^{\prime})/\tau}\,\big(T(h_{t^{\prime}})+\xi_{t^{\prime}}\big)\Big]\Big\rangle_{\xi}.\label{eq:p_cont}
\end{align}

Because \cref{eq:rate} is a deterministic mapping up to a noise term,
we can generalize \cref{eq:p_cont} with the two-replica formalism
to
\begin{align*}
p^{c}[\phi^{\alpha}_{t},\phi^{\beta}_{s}|h^{\alpha},h^{\beta}] & =\Big\langle\delta\Big[\phi^{\alpha}_{t}-\int^{t}_{-\infty}\frac{dt^{\prime}}{\tau}\,e^{-(t-t^{\prime})/\tau}\,\big(T(h^{\alpha}_{t^{\prime}}+\xi^{\alpha}_{t^{\prime}})\big)\Big]\,\delta\Big[\phi^{\beta}_{s}-\int^{s}_{-\infty}\frac{ds^{\prime}}{\tau}\,e^{-(t-t^{\prime})/\tau}\,\big(T(h^{\beta}_{s^{\prime}})+\xi^{\beta}_{s^{\prime}}\big)\Big]\Big\rangle_{\xi}.
\end{align*}
Next, one can use this joint density to compute \cref{eq:def_Q12},
where the integral over $\phi^{\alpha}$ and $\phi^{\beta}$ becomes
trivial, namely
\begin{align*}
Q^{\alpha\beta}_{ts} & =g^{2}\,\langle\phi^{\alpha}_{t}\,\phi^{\beta}_{s}\rangle_{\Omega^{\alpha\beta}.}\\
 & =g^{2}\,\int^{t}_{-\infty}\frac{dt^{\prime}}{\tau}\,e^{-(t-t^{\prime})/\tau}\int^{s}_{-\infty}\frac{ds^{\prime}}{\tau}\,e^{-(t-t^{\prime})/\tau}\,\big\langle\,\big\langle\big(T(h^{\alpha}_{t^{\prime}})+\xi^{\alpha}_{t^{\prime}}\big)\,\big(T(h^{\beta}_{s^{\prime}})+\xi^{\beta}_{s^{\prime}}\big)\big\rangle_{\xi}\,\big\rangle_{\N^{\alpha\beta}}.
\end{align*}
To bring this equation into differential form, we apply $(\tau\partial_{t}+1)\,(\tau\partial_{s}+1)$,
giving
\begin{align*}
(\tau\partial_{t}+1)\,(\tau\partial_{s}+1)\,Q^{\alpha\beta}_{ts} & =g^{2}\,\big\langle\,\big\langle\big(T(h^{\alpha}_{t})+\xi^{\alpha}_{t}\big)\,\big(T(h^{\beta}_{s})+\xi^{\beta}_{s}\big)\big\rangle_{\xi}\big\rangle_{\N^{\alpha\beta}}.
\end{align*}
The inner expectation value can now be taken ($h$ is given by the
outer one and we do not have any statistical dependencies between
the two), so

\begin{align*}
(\tau\partial_{t}+1)\,(\tau\partial_{s}+1)\,Q^{\alpha\beta}_{ts} & =g^{2}\,\big\langle T(h^{\alpha}_{t})\,T(h^{\beta}_{s})\big\rangle_{\N^{\alpha\beta}}+g^{2}\,\ev{\xi^{\alpha}_{t}\xi^{\beta}_{s}}{\xi}.
\end{align*}
For a white noise of some magnitude $\sqrt{D}$ we get $g^{2}\ev{\xi^{\alpha}_{t}\xi^{\beta}_{s}}{}=D\,\delta(t-s)$.

\subsubsection*{Single replicon}

We can now instantiate the replicon indices to be the same $h^{\alpha}=h^{\beta}\equiv h^{1}=h$
to obtain the evolution of the kernel within a single replicon,

\begin{align*}
\left(\tau\partial_{t}+1\right)\left(\tau\partial_{s}+1\right)Q_{ts} & =g^{2}\langle T(h_{t})\,T(h_{s})\rangle_{(h_{t},h_{s})\sim\N(R_{t},R_{s},\bar{Q}_{tt},\bar{Q}_{ss},\bar{Q}_{ts})}+\ev{\xi_{t}\xi_{s}}{}.
\end{align*}
This generalizes the stationary autocorrelation found by \citet{Sompolinsky88_259}
to time-dependent inputs.

\subsubsection*{Two replica}

For the correlation between replica which enters the kernel, we set
$\alpha=1$ and $\beta=2$. This gives
\[
\left(\tau\partial_{t}+1\right)\left(\tau\partial_{s}+1\right)Q^{12}_{ts}=g^{2}\langle T(h^{1}_{t})\,T(h^{2}_{s})\rangle_{(h^{1}_{t},h^{2}_{s})\sim\N^{12}(R_{t},R_{s},\bar{Q}_{tt},\bar{Q}_{ss},\bar{Q}^{12}_{ts})}+\ev{\xi^{1}_{t}\xi^{2}_{s}}{},
\]
where the solution to the first equation enters as the off-diagonal
of the $h^{\alpha\beta}$-covariance matrix in the second equation.
This generalizes the solution of \citet{Schuecker18_041029} to time-dependent
inputs.

\subsection*{Discrete networks\label{app:Discrete-networks}}

Discrete neurons are constrained by 
\begin{equation}
\phi_{i}(t)\,\phi_{i}(t)=1\label{eq:discrete_constraint}
\end{equation}
due to the discrete state $\phi_{i}=\pm1$ of each neuron. To obtain
a continuous network that has an identical mean-field description,
an external driving noise to the network needs to be introduced \citep{Keup21_021064}.

In the discrete networks, the neuronal states are updated at specific
time points according to the Glauber dynamics \cref{eq:transmission_dynamics_glauber}.
After averaging the over the Poisson update process $\theta_{\text{up}}$
in \cref{eq:binary}, the activation probability takes the form $T_{p}(h_{t})=\big(\frac{1}{2}T(h_{t})+1\big)\in\left[0,1\right]$
which leads to the single-neuron functional $p^{d}_{i}\equiv p^{d}$
\begin{align}
p^{d}[\phi_{t}=1|h] & =\int^{t}_{-\infty}\frac{dt^{\prime}}{\tau}\,e^{-(t-t^{\prime})/\tau}\,T_{p}(h_{t^{\prime}}).\label{eq:p_disc}
\end{align}
The update process in the Glauber dynamics constitutes a source of
stochasticity that prevents the simple argument above when extending
to two-point correlators. Assume $t>s>0$. Both systems undergo the
same update process. We first consider the contribution from the case
that $\phi^{\alpha}_{t}=\phi^{\alpha}_{s}=1$. There are hence two
possibilities: Either there is no update in $[s,t]$ (which happens
with probability $e^{-(t-s)/\tau}$) or the last update within $[s,t]$
takes $\phi^{\alpha}_{t}$ to the up-state. We add up these contributions:

\begin{align}
p[\phi^{\alpha}_{t}=1,\phi^{\beta}_{s}=1|h^{\alpha}h^{\beta}]= & p[\text{no update in \ensuremath{[s,t]}},\,\phi^{\alpha}_{s}=1,\phi^{\beta}_{s}=1|h^{\alpha}h^{\beta}]\label{eq:ansatz_corr_updates}\\
 & +p[\text{\ensuremath{\geq1} update in}\ensuremath{[s,t]},\,\phi^{\alpha}_{t}=1,\phi^{\beta}_{s}=1|h^{\alpha}h^{\beta}]\nonumber \\
\nonumber \\= & p[\text{no update in \ensuremath{[s,t]}}|\phi^{\alpha}_{s}=1,\phi^{\beta}_{s}=1,h^{\alpha}h^{\beta}]p[\phi^{\alpha}_{s}=1,\phi^{\beta}_{s}=1|h^{\alpha}h^{\beta}]\nonumber \\
 & +\sum_{t^{\prime}}p[\text{last update at \ensuremath{t^{\prime}\in}}\ensuremath{[s,t]},\text{\ensuremath{\phi^{\alpha}_{t^{\prime}}=1}},\phi^{\beta}_{s}=1|h^{\alpha}h^{\beta}].\nonumber 
\end{align}
The two terms are determined as follows:
\begin{enumerate}
\item At time $s$, both variables are in state $\phi^{\alpha}_{s}=\phi^{\beta}_{s}=1$
and there is no update within $[s,t]$. The probability for this event
is $e^{-(t-s)/\tau}\,p(\phi^{\alpha}_{s}=1,\phi^{\beta}_{s}=1|h^{\alpha}h^{\beta})$.
\item At time $s$, variable $\phi^{\beta}_{s}$ is in state $\phi^{\beta}_{s}=1$
and\textbf{ $\phi^{\alpha}_{s}$} is arbitrary, which happens with
the probability that the last update took $\phi^{\beta}_{s}$ to the
up-state, which is $p(\phi^{\beta}_{s}=1)=\int^{s}_{-\infty}\frac{ds^{\prime}}{\tau}\,e^{-(s-s^{\prime})}\,T_{p}(h^{\beta}_{s^{\prime}})$
and within $[s,t]$ the last update that brought $\phi^{\alpha}$
into state $\phi^{\alpha}_{t}=1$. The probability for the joint occurrence
of this event is
\begin{align*}
p[\text{\ensuremath{\geq1} update in }\ensuremath{[s,t]},\phi^{\alpha}_{t}=1,\phi^{\beta}_{s}=1|h^{\alpha}h^{\beta}] & =\underbrace{\int^{s}_{-\infty}\frac{ds^{\prime}}{\tau}\,e^{-(s-s^{\prime})/\tau}\,T_{p}(h^{\beta}_{s^{\prime}})}_{p(\phi^{\beta}_{s}=1|h^{\alpha}h^{\beta})}\,\underbrace{\int^{t}_{s}\frac{dt^{\prime}}{\tau}\,e^{-(t-t^{\prime})/\tau}\,T_{p}(h^{\alpha}_{t^{\prime}})}_{p(\text{\ensuremath{\geq1} update in}\ensuremath{[s,t]},\,\phi^{\alpha}_{t}=1|h^{\alpha}h^{\beta})}.
\end{align*}
\end{enumerate}
Because $\phi^{\alpha}_{t}=\phi^{\alpha}_{s}=1$, the contribution
to the autocorrelation function is positive. Likewise, we obtain contributions
from the other three cases $(\phi^{\alpha}_{t},\phi^{\beta}_{s})\in\{(1,-1),(-1,1),(-1,-1)\}$,
which contribute with $-1$, $-1$, $+1$, respectively. Together
this yields for case 2
\begin{align*}
 & 1\cdot1\,T^{\alpha}_{p}T^{\beta}_{p}+1\cdot(-1)\,T^{\alpha}_{p}\left(1-T^{\beta}_{p}\right)+(-1)\cdot1\,\left(1-T^{\alpha}_{p}\right)T^{\beta}_{p}+(-1)\cdot(-1)\,\left(1-T^{\alpha}_{p}\right)\left(1-T^{\beta}_{p}\right)\\
= & \left(T^{\alpha}_{p}-\left(1-T^{\alpha}_{p}\right)\right)\left(T^{\beta}_{p}-\left(1-T^{\beta}_{p}\right)\right)=\left(2T^{\alpha}_{p}-1\right)\left(2T^{\beta}_{p}-1\right)\equiv T^{\alpha}T^{\beta},
\end{align*}
recovering the activation function $T(\circ)\in[-1,1]$.

After multiplying both contributions by $g^{2}$ and averaging over
$(h^{\alpha}_{t},\,h^{\beta}_{s})\sim\N^{\alpha\beta}$ we get
\begin{align}
Q^{\alpha\beta}_{ts} & =e^{-(t-s)/\tau}\,Q^{\alpha\beta}_{ss}+g^{2}\int^{s}_{-\infty}\frac{ds^{\prime}}{\tau}\,e^{-(s-s^{\prime})/\tau}\,\int^{t}_{s}\frac{dt^{\prime}}{\tau}\,e^{-(t-t^{\prime})/\tau}\,\big\langle T(h^{\alpha}_{t^{\prime}})\,T(h^{\beta}_{s^{\prime}})\big\rangle_{\N^{\alpha\beta}}.\label{eq:both_events}
\end{align}
Taking a derivative with regard to $t$ one has
\begin{align}
\tau\partial_{t}\,Q^{\alpha\beta}_{ts} & =-Q^{\alpha\beta}_{ts}+g^{2}\int^{s}_{-\infty}\frac{ds^{\prime}}{\tau}\,e^{-(s-s^{\prime})/\tau}\,\big\langle T(h^{\alpha}_{t})\,T(h^{\beta}_{s^{\prime}})\big\rangle_{\N^{\alpha\beta}},\label{eq:diffeq_q12}
\end{align}
with the initial condition given by the equal-time correlation $Q^{\alpha\beta}_{ss}$.
To compute the right hand side, one hence needs the joint statistics
of $h^{\alpha}_{-\infty<s^{\prime}<s}$ and $h^{\beta}_{t}$.

Now introduce the auxiliary variable
\begin{align*}
z^{\alpha\beta}_{ts} & :=g^{2}\int^{s}_{-\infty}\frac{ds^{\prime}}{\tau}\,e^{-(s-s^{\prime})/\tau}\,\big\langle T(h^{\alpha}_{t})\,T(h^{\beta}_{s^{\prime}})\big\rangle_{\N^{\alpha\beta}}
\end{align*}
to write the differential equation \cref{eq:diffeq_q12} as
\begin{align}
(\tau\partial_{t}+1)\,Q^{\alpha\beta}_{ts}= & z^{\alpha\beta}_{ts}.\label{eq:forward_q}
\end{align}
The quantity $z^{\alpha\beta}_{ts}$ itself obeys a differential equation
\begin{align}
(\tau\partial_{s}+1)\,z^{\alpha\beta}_{ts} & =g^{2}\big\langle T(h^{\alpha}_{t})\,T(h^{\beta}_{s})\big\rangle_{\N^{\alpha\beta}}.\label{eq:forward_z}
\end{align}
Finally, using \citep[eq. (A26)]{Keup21_021064}, we can obtain the
diagonal $Q^{\alpha\beta}_{tt}$ that was undetermined by this algorithm
so far

\begin{equation}
\left(\tau\frac{d}{dt}+1\right)Q^{\alpha\beta}_{tt}=g^{2}\big\langle1-2|T_{p}(h^{\alpha}_{t})-T_{p}(h^{\beta}_{t})|\big\rangle_{\N^{\alpha\beta}},\label{eq:ode_diag}
\end{equation}
where the notation $\nicefrac{d}{dt}$ means changing both arguments
simultaneously. For the special case $\alpha=\beta$ and the constraint
$Q^{\alpha\alpha}_{tt}\equiv Q_{tt}=g^{2}\ev{\phi_{t}\phi_{t}}{}=g^{2}$,
this correctly reproduces $\frac{d}{dt}Q_{tt}=0$.

\subsubsection*{Single replicon}

We can now again instantiate the replicon indices $\alpha=\beta=1$
to obtain the evolution within a single replicon,
\begin{align*}
(\tau\partial_{t}+1)\,Q_{ts}= & z_{ts}\\
(\tau\partial_{s}+1)\,z_{ts}= & g^{2}\big\langle T(h_{t})\,T(h_{s})\big\rangle_{(h_{t},h_{s})\sim\N(R_{t},R_{s},\bar{Q}_{tt},\bar{Q}_{ss},\bar{Q}_{ts})}\\
Q_{tt}= & g^{2}.
\end{align*}

\subsubsection*{Two replica}

For the correlation between replica that enters the kernel, we set
$\alpha=1$ and $\beta=2$. We first consider the case where both
replica experience identical draws of the update point process. This
gives
\begin{align*}
(\tau\partial_{t}+1)\,Q^{12}_{ts}= & z^{12}_{ts}\\
(\tau\partial_{s}+1)\,z^{12}_{ts}= & g^{2}\big\langle T(h^{1}_{t})\,T(h^{2}_{s})\big\rangle_{(h^{1}_{t},h^{2}_{s})\sim\N^{12}(R_{t},R_{s},\bar{Q}_{tt},\bar{Q}_{ss},\bar{Q}^{12}_{ts})}\\
\left(\tau\frac{d}{dt}+1\right)Q^{12}_{tt}=g^{2} & \begin{cases}
\ev{1-2|T_{p}(h^{1}_{t})-T_{p}(h^{2}_{t})|}{(h^{1}_{t},h^{2}_{t})\sim\N^{12}(R_{t},R_{t},\bar{Q}_{tt},\bar{Q}_{tt},\bar{Q}^{12}_{tt})} & \text{activations correlated}\\
\ev{T_{p}(h^{1}_{t})T_{p}(h^{2}_{t})}{(h^{1}_{t},h^{2}_{t})\sim\N^{12}(R_{t},R_{t},\bar{Q}_{tt},\bar{Q}_{tt},\bar{Q}^{12}_{tt})} & \text{activations independent}
\end{cases},
\end{align*}
Here, the operator $\frac{d}{dt}$ acts on both time arguments, i.e.
yields $\frac{1}{dt}(Q^{12}_{t+dt,t+dt}-Q^{12}_{tt})$. The third
line is the equal-time evolution for correlated activations that reproduces
equation (A26) from \citet{Keup21_021064}.

For the case that the update processes are drawn independently, the
two-replica functional in \cref{eq:def_exp_replica} factorizes, 
\[
p(\phi^{1}_{t},\phi^{2}_{s}|h^{1},h^{2})=p(\phi^{1}_{t}|h^{1})\,p(\phi^{2}_{s}|h^{2}).
\]
Averaging the update process in each replicon, \cref{eq:p_disc} results
in $\left(\tau\partial_{t}+1\right)\langle\phi^{\alpha}_{t}\rangle_{\mathrm{up}|h}=T(h^{\alpha}_{t})$
so that we get
\[
\left(\tau\partial_{t}+1\right)\left(\tau\partial_{s}+1\right)Q^{12}_{ts}=g^{2}\begin{cases}
\ev{1-2|T_{p}(h^{1}_{t})-T_{p}(h^{2}_{s})|}{(h^{1}_{t},h^{2}_{t})\sim\N^{12}(R^{1}_{t},R^{2}_{t},\bar{Q}_{tt},\bar{Q}_{ss},\bar{Q}^{12}_{ts})} & \text{activations correlated and \ensuremath{t=s,}}\\
\ev{T_{p}(h^{1}_{t})T_{p}(h^{2}_{s})}{(h^{1}_{t},h^{2}_{t})\sim\N^{12}(R^{1}_{t},R^{2}_{s},\bar{Q}_{ts},\bar{Q}_{ss},\bar{Q}^{12}_{ts})} & \text{else}
\end{cases}.
\]
For the case of independent activations, this reproduces the case
of the continuous network with replica-independent noise, as discussed
in \citet{Kadmon15Transitionchaosrandom,Keup21_021064}.

\subsubsection*{Solving the equations}

Both the continuous and discrete case can be solved by integrating
forward in time with an Euler scheme from a suitable initial condition
such as $Q^{\alpha\beta}_{t<0\,0}$. While the single replicon case
can be solved in a standalone manner, the two replica solution depend
on the single replicon solution. However, it is not necessary to calculate
and store the single replicon statistics that enter the equation upfront:
As they only depend on them in a time-local manner, the system can
be solved jointly. We provide a solver in JAX \citep{bradbury2018}
that implements this algorithm in the supplementary material.

\subsection{Limiting cases\label{app:Small-decorrelation-Deltapeak}}

\subsubsection*{Small decorrelation}

Let $c=\frac{2}{\sqrt{\pi}}g^{2}\langle T{}^{\prime}(h)\rangle_{\mathcal{N}(0,\,Q_{0})}.$
Defining a small decorrelation $\epsilon_{t}=Q_{0}-Q^{12}_{tt}$,
\citep{Keup21_021064} finds

\[
\left(\tau\partial_{t}+1\right)\epsilon_{t}=c\sqrt{\epsilon_{t}}
\]
with solution
\begin{align*}
\epsilon_{t} & =\left(c-\left(c-\sqrt{\epsilon_{0}}\right)e^{-t/2\tau}\right)^{2}.
\end{align*}
To analyze the discontinuity, we take the limit $\epsilon_{0}\rightarrow0$,
\begin{align*}
\epsilon_{t} & =\left(c\left(1-e^{-t/2\tau}\right)+\sqrt{\epsilon_{0}}e^{-t/2\tau}\right)^{2}\\
 & \simeq c^{2}\left(1-e^{-t/2\tau}\right)^{2},
\end{align*}
giving a drop $\Delta_{t}/c^{2}=\left(1-e^{-t/2\tau}\right)^{2}$
that is finite for any finite time.%

\subsection*{Microscopic analysis of chaos in networks}

\subsection*{Discrete networks\label{app:microscopic-analysis-of-chaos-discrete}}

It has been noted that discrete networks exhibit a stronger kind of
chaoticity. This is due to the way in which a change in activity affects
the global network state. Suppose that the value of a single unit
gets flipped, $\phi_{i}=1\rightarrow-1$, so $|\delta\phi_{i}|=2$. 

By the recurrent coupling, this will lead to a change in the synaptic
inputs at every downstream neuron $j$ (including $i$)
\[
h_{j}\rightarrow h_{j}+\delta h_{j},
\]
with $\delta h_{i}\simeq\delta h_{j}=J_{ji}\delta\phi_{i}\simeq g/\sqrt{N}$,
where we have used $\ev{|J_{ij}|}{\simeq}g/\sqrt{N}$. The probability
that this changes the activation at each neuron $j$ in the next update
is $\mathcal{C}\delta h_{j}$, where $\mathcal{C}$ is a constant
depending on the state of the activation. To see this, the activation
mechanism can be phrased as a thresholding operation on a randomly
distributed decision variable $r_{i}\sim\mathcal{U}(-1,1)$ at every
neuron that the activation function output $T(h_{j})$ is compared
against.

As there are $N$ units in the system, we hence expect $N_{dt}=p_{\text{act}}p_{\text{up}}N=\mathcal{C}\delta h_{j}\,\frac{dt}{\tau}N$
units to be flipped after a time $dt$ due to the original perturbation
$\delta\phi^{(0)}_{j}$, introducing a set of $N_{dt}$ flips $\left\{ \delta\phi_{k}(dt)\right\} $
with magnitude $\delta\phi_{k}(dt)=2$ due to the discreteness of
the state. 

Now, we make a specific choice for $dt$: It should be sufficiently
small such that only some neurons get updated, but at the same time
sufficiently large such that at least a single neuron is updated,
$N_{dt}=1$. These constraints necessitate the choice $dt(N)=1/\sqrt{N}$.
After feeding this flip back through the recurrent connectivity, we
arrive at 
\begin{align}
d\delta h_{i}(dt) & \simeq\frac{g}{\sqrt{N}}N_{dt}\cdot2\label{eq:sqrt_growth}\\
 & =\frac{g}{\sqrt{N}}\,1\cdot2.\nonumber 
\end{align}
If we compare to an exponential growth ansatz $\frac{d\delta h_{i}(dt)}{dt}=\lambda\,\delta h_{i}e^{\lambda dt}$,
we find
\[
\lambda\simeq\frac{1}{\delta h_{i}}\frac{d\delta h_{i}(dt)}{dt}=\frac{1}{g/\sqrt{N}}\,\frac{g/\sqrt{N}}{1/\sqrt{N}}=\sqrt{N}\,\overset{N\rightarrow\infty}{\longrightarrow}\,\infty,
\]
so that perturbations $\delta h_{i}$ grow with diverging Lyapunov
exponent.

\subsection*{Continuous networks}

For continuous networks, similar considerations can be made. Consider
changing the state of a neuron $i$ by an infinitesimal amount $\delta\phi_{i}\ll2$.
After integrating the dynamics by $dt$, we get
\begin{align*}
\delta h_{j}(dt) & =\left(\frac{g}{\sqrt{N}}\sqrt{N}\right)\,\delta\phi_{i}\,dt\\
 & =\left(\frac{g}{\sqrt{N}}\sqrt{N}\right)\,\mathcal{C}\delta h_{i}\,dt
\end{align*}
at every downstream neuron $j$. Here, $\mathcal{C}$ again is a constant
depending on the working point of the activation function. Notably,
we had to account for subadditivity by multiplying by $\sqrt{N}$
instead of $N$, as all neurons get updated simultaneously. After
another pass through the recurrent connectivity, we end up with 
\[
\delta h_{i}(dt)=\left(\frac{g}{\sqrt{N}}\sqrt{N}\right)\,\mathcal{C}\delta h_{j}dt,
\]
at the initial perturbed neuron $i$. This in total gives
\[
\frac{d\delta h_{i}}{dt}=\left(\frac{g}{\sqrt{N}}\sqrt{N}\right)^{2}\,\mathcal{C}\mathcal{C}{}^{\prime}\delta h_{i}.
\]
Here the Lyapunov exponent is identified with the $\mathcal{O}(1)$-constant
$\mathcal{C}\mathcal{C}{}^{\prime}g^{2}$ and does not diverge.

\section{Kernel trick\label{app:Kernel-trick}}

In this section, we detail the derivation of the optimal readout $\bm{w}^{\star}$
and how it emerges from training examples. Furthermore, introduce
a regularization $\Lambda$ to the loss. 

Let $\Phi_{X}=\left\{ \phi_{x^{\alpha},i}\right\} _{\substack{\alpha=1,\ldots,P\\
i=1\ldots N
}
}\in\mathbb{R}^{P\times N}$ denote the feature matrix. Differentiation of the objective \cref{eq:loss-rr}
gives

\begin{align*}
\bm{w}^{\star} & =\argmin_{\bm{w}}\,\mathcal{L}(\bm{w};Y_{X})\\
 & =\left(\Phi^{\T}_{X}\Phi_{X}+N\Lambda\right)^{-1}\Phi^{\T}_{X}Y_{X}\\
 & =\Phi^{\T}_{X}\,\underbrace{\left(\Phi_{X}\Phi^{\T}_{X}+N\Lambda\right)^{-1}Y_{X}}_{\bm{\alpha}}\,.
\end{align*}
Herein, $C_{ij}\coloneqq\left(\Phi^{\T}_{X}\Phi_{X}\right)_{ij}$
is the correlation matrix across $N$ neurons, and $K_{\alpha\beta}\coloneqq\frac{1}{N}\left(\Phi_{X}\Phi^{T}_{X}\right)_{\alpha\beta}$
is the kernel matrix across $P$ features. Herein, $\bm{\alpha}=(\alpha_{1},\ldots,\alpha_{P})^{\T}$
are coefficients which select a subset of support vectors from $\Phi_{X}$
to make a connection to the SVM literature. Further, we used that
matrix products are associative, so

\begin{align*}
\Phi^{\T}_{X}\left(\Phi_{X}\Phi^{\T}_{X}+N\Lambda\right) & =\left(\Phi^{\T}_{X}\Phi_{X}+N\Lambda\right)\Phi^{\T}_{X}\\
\Rightarrow\left(\Phi^{\T}_{X}\Phi_{X}+N\Lambda\right)^{-1}\Phi^{\T}_{X} & =\Phi^{\T}_{X}\left(\Phi_{X}\Phi^{\T}_{X}+N\Lambda\right){}^{-1}.
\end{align*}
A prediction on a test input $\bm{x}^{*}$ is obtained from projection
onto the readout $\bm{w}^{\star}$:

\begin{align*}
\bar{y}_{\bm{x}^{*}} & =\bphi_{\bx^{*}}\cdot\bm{w}^{\star}\\
 & =\bphi_{\bx^{*}}\left(\Phi^{\T}_{X}\Phi_{X}+N\Lambda\right)^{-1}\Phi^{\T}_{X}Y_{X}\\
 & =\bphi_{\bx^{*}}\Phi^{\T}_{X}\left(\Phi_{X}\Phi^{\T}_{X}+N\Lambda\right)^{-1}Y_{X}\\
 & =k_{\bx^{*}X}\,k^{-1}_{XX}\,Y_{X}.
\end{align*}

\section{Reservoir computing yields Gaussian prior\label{app:RC-yields-GP}}

In this section we show that reservoir computing in the limit of
large $N$ is equivalent to Bayesian inference. The network output
given the parameters $J$, $\bw$, and $t_{R}$ is

\begin{align*}
y_{\bm{x}}(t_{R}) & =\bm{w}\cdot\bm{\phi}^{J}_{\bm{x}}(t_{R})+\eta_{\bm{x}},
\end{align*}
where $\eta_{\bm{x}}\stackrel{\text{i.i.d. over \ensuremath{\bm{x}}}}{\sim}\N(0,\,\Lambda)$
is a Gaussian readout noise that is drawn independently for each input
$\bm{x}$.

The prior distribution of the output under the Gaussian priors $w_{i}\overset{\text{i.i.d.}}{\sim}\N(0,\,1/N)$
and $J_{ij}\overset{\text{i.i.d.}}{\sim}\N(0,\,g^{2}/N)$ is given
by
\begin{align}
p(\{y_{\bm{x}}\}_{\bm{x}\in X}) & =\Big\langle\prod_{\bm{x}\in X}\delta\big[y_{\bm{x}}-\bm{w}\cdot\bm{\phi}^{J}_{\bm{x}}(t_{R})-\eta_{\bm{x}}\big]\Big\rangle_{\bw,J,\eta_{\bm{x}}}.\label{eq:prior_y_app}
\end{align}

Conditioned on $J$, the expectation over $\eta_{\bm{x}}$ and $\bw$
in \cref{eq:prior_y_app} again yields a Gaussian process for $y_{\bm{x}}$
as a sum of centered Gaussians, 
\begin{align*}
p(\{y_{\bm{x}}\}_{\bm{x}\in X}) & =\Big\langle\N\big(\{y_{\bm{x}}\}_{\bm{x}\in X}\,|\,0,\,\{N^{-1}\bm{\phi}^{J}_{\bm{x}}(t_{R})\cdot\bm{\phi}^{J}_{\bm{x^{\prime}}}(t_{R})+\delta_{\bm{x}\bm{x}^{\prime}}\Lambda\}_{\bm{x},\bm{x}^{\prime}\in X}\big)\Big\rangle_{J},
\end{align*}
where the kernel of the Gaussian process for a given value of $J$
is $k^{J}_{t_{R}}(\bm{x},\bm{x}^{\prime})=N^{-1}\bm{\phi}^{J}_{\bm{x}}(t_{R})\cdot\bm{\phi}^{J}_{\bm{x^{\prime}}}(t_{R})+\delta_{\bm{x}\bm{x}^{\prime}}\Lambda$.
As we show in \cref{app:kernel_mft}, the inner product $N^{-1}\bm{\phi}^{J}_{\bm{x}}(t_{R})\cdot\bm{\phi}^{J}_{\bm{x^{\prime}}}(t_{R})$
in the large $N$-limit is self-averaging and concentrates to \cref{eq:def_Q12}
\begin{align*}
N^{-1}\bm{\phi}^{J}_{\bm{x}}(t_{R})\cdot\bm{\phi}^{J}_{\bm{x^{\prime}}}(t_{R}) & \stackrel{N\to\infty}{\to}Q^{(\bm{x}\bm{x}^{\prime})}_{t_{R}t_{R}}/g^{2}=\langle\phi_{\bm{x}}(t_{R})\,\phi_{\bm{x}^{\prime}}(t_{R})\rangle_{\Omega(\{R^{\alpha}_{t_{R}},\,Q^{\alpha\beta}_{t_{R}t_{R}}\}_{\alpha,\beta\in\{\bm{x},\bm{x}^{\prime}\}})},
\end{align*}
which hence becomes independent of the realization of $J$ so that
we obtain for the network prior the Gaussian process

\begin{align*}
\{y_{\bm{x}}\}_{\bm{x}\in X} & \stackrel{N\to\infty}{\sim}\N(0,\,k_{t}(XX^{T})),\\
k_{t}(\bm{x}\cdot\bm{x}^{\prime}) & =\langle\phi_{\bm{x}}(t_{R})\,\phi_{\bm{x}^{\prime}}(t_{R})\rangle_{\Omega(\{R^{\alpha}_{t_{R}},\,Q^{\alpha\beta}_{t_{R}t_{R}}\}_{\alpha,\beta\in\{\bm{x},\bm{x}^{\prime}\}})}+\delta_{\bm{x}\bm{x}^{\prime}}\Lambda.
\end{align*}
A complementary view on the prior \cref{eq:prior_y_app} can be obtained
by first taking the expectation with regard to $\eta$, which yields

\begin{align*}
p(\{y_{\bm{x}}\}_{\bm{x}\in X}) & =\big\langle\N\big(y_{\bm{x}}\,|\,\bm{w}\cdot\bm{\phi}^{J}_{\bm{x}}(t_{R}),\Lambda\big)\big\rangle_{\bw,J}\\
 & \propto\big\langle\exp\big(-\sum_{\bx}\,\|y_{\bm{x}}-\bm{w}\cdot\bm{\phi}^{J}_{\bm{x}}(t_{R})\|^{2}/(2\Lambda)\big)\big\rangle_{\bw,J},
\end{align*}
showing that the prior can be considered introducing a likelihood
$p(y|\bm{w},J)\propto\exp\left(-\|y-\bm{w}\cdot\bm{\phi}^{J}(t_{R})\|^{2}/(2\Lambda)\right)$,
where $\|\ldots\|^{2}$ denotes the $L_{2}$-norm over patterns $\bx$.

Note that if the prior is Gaussian, \cref{eq:gp-post} corresponds
to finding
\[
\bar{y}_{\bm{x}^{*}}=\argmax_{y_{\bm{x}^{*}}}\,p(y_{\bm{x}^{*}},Y_{X})=\argmax_{y^{*}}\,\max_{w,J}\,p(y,Y|w,J)\,p(w,J).
\]

The Gaussian process prediction hence is identical to finding a global
optimum of $J$ and $w$ , a consequence of the self-averaging nature
of large networks. From this form, we can now relate the Tikhonov
regularization in \cref{eq:loss-rr} to a noise, $\Lambda\equiv\ev{\eta^{2}_{\bx}}{}$.
This reveals that an additive constant in the kernel can be interpreted
as a Bayesian prior on the reliability of the data.

\section{Local extrapolation shape of kernels\label{app:Local-extrapolation-shape}}

The analysis of the kernel spectrum provides a general view on learning
in terms of eigenfunctions. In analyzing the kernels that correspond
to chaotic networks, we are especially interested in the short-range
extrapolation. This is controlled by the fast modes of the kernel.
In this regime, we can understand the extrapolation described by the
networks also directly in the space of inputs by neglecting the influence
of distant training examples. To see this, we approximate \cref{eq:y_pred_GP}
for the case that the test point $\bx^{*}$ is close to a nearest
neighbor training point $\bx^{\text{nn}}\in X$ but $\mathcal{O}(\epsilon)$-distant
from all others $X\backslash\bx^{\text{nn}}$. Without loss of generality
consider a kernel with unit diagonal $k_{\bx\bx}=1$:

Partitioning the training set $X=(x^{\text{nn}},\,X\backslash\bx^{\text{nn}})$
into the nearest neighbor to a test site and more distant points,
we get

\begin{align*}
\bar{y}_{\bx^{*}} & =k_{\bx^{*}X}\,\left(\begin{array}{cc}
1 & \bm{\epsilon}^{T}\\
\bm{\epsilon} & K_{\backslash\bx^{\text{nn}}}
\end{array}\right)^{-1}Y_{X}\\
 & =k_{\bx^{*}X}\,\left(\left(\begin{array}{cc}
1 & \bm{0}^{T}\\
\bm{0} & K^{-1}_{\backslash\bx^{\text{nn}}}
\end{array}\right)+\mathcal{O}(\epsilon)\right)Y_{X}\\
 & =\left(\begin{array}{c}
k_{\bx^{*}\bx^{\text{nn}}}\\
\bm{\epsilon}
\end{array}\right)^{T}\left(\begin{array}{cc}
1 & \bm{0}^{T}\\
\bm{0} & K^{-1}_{\backslash\bx^{\text{nn}}}
\end{array}\right)Y_{X}\:+\:\mathcal{O}(\epsilon)\\
 & =k_{\bx^{*}\bx^{\text{nn}}}Y_{\bx^{\text{nn}}}\:+\:\mathcal{O}(\epsilon).
\end{align*}
Here, we abbreviated $K_{\backslash\bx^{\text{nn}}}\coloneqq K_{X\backslash\bx^{\text{nn}}\,X\backslash\bx^{\text{nn}}}$.
In the second line, we used an identity on the inverse of a $2\times2$
block matrix. In particular, this entails that for only distant training
points, the extrapolation from a training pattern $\bx^{\text{nn}}$
takes the form $k_{\bx^{*}\bx^{\text{nn}}}Y_{\bx^{\text{nn}}}$.

\section{Spectral analysis of kernel functions}

\subsection*{}

\subsubsection*{}

\subsubsection*{Connection to finite-size kernel matrices\label{app:operator-and-matrix-spectra}}

\begin{figure}[!tbp]
\centering{}\includegraphics[width=1\columnwidth]{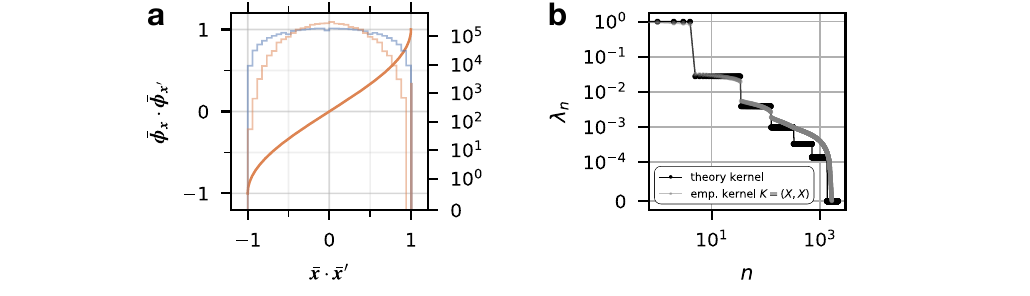}\caption{\textbf{\label{fig:funk-hecke}Correspondence of finite kernel matrices
and kernel operators. a}~Generic kernel function $\tfrac{2}{\pi}\,\arcsin(\protect\bx\cdot\protect\bx^{\prime})$
(\emph{orange}) applied to $d=5$ dimensional Gaussian i.i.d. data
$X$ (\emph{blue histogram}), producing \emph{orange histogram}. \textbf{b}~Eigenvalue
spectra produced by Funk-Hecke formula \cref{eq:funk-hecke-short}
(\emph{black}), but repeated according to their multiplicity $\#m(l,d)$,
i.e. $n$ enumerates the multi-index $(lm),m=1\ldots\#m(l,d)$. \emph{Gray}
dots show the spectrum of the finite-size matrix $K_{XX}=k(XX^{T})$
produced from passing the data $X$ through the kernel function.}
\end{figure}

The Funk-Hecke formula \citep{Funk15Beitragezurtheorie,dutordoir2020}
(see \cref{eq:funk-hecke-short}) allows us to derive the eigenvalues
$\lambda_{l}$ from a kernel operator function $k(u=\bx\cdot\bx^{\prime})$
if data $X$ is uniform on a space of dimensionality $d$. The resulting
eigenspectrum has countably infinite number of eigenvalues, and leading
eigenvalues are well separated. We here demonstrate how this relates
to the contiguous empirical eigenspectra of finite-size matrices $K_{XX}\in\mathbb{R}^{P\times P}$
discussed in the main text and reported by \citep{Stringer19_361}as
their size grows large.

Such finite-size evaluations of the kernel function will indeed approximate
the true eigenvalues \citep{Koltchinskii00Randommatrixapproximation,braun2005},
but will contain each according to its multiplicity $\#m(l,d)=\binom{d+l-1}{l}-\binom{d+l-3}{l-2}\approx d^{l}/l!$
\citep{canatar2021}, where $m$ is the multiplicity of eigenvalue
$\lambda_{l}$.

\subsubsection*{Linear kernels have degenerate spectra\label{app:Linear-kernels-have-degenerate-spectra}}

Networks in the regular regime exhibit linear kernel functions, $k(u=\bm{x}\cdot\bm{x}{}^{\prime})\propto u$.
In this section, we show that this entails a degenerate spectrum with
just a single eigenmode. Consequentially, the kernel generalizes poorly
to other modes. 

If the kernel is linear, we can express it in terms of the first Gegenbauer
polynomial \citep{Suetin2001}
\[
k(u)\propto C^{a}_{1}(u)=2a\,u.
\]
For uniform data distribution, the Funk-Hecke formula \citep[Eq.  (10)]{dutordoir2020}
then yields

\begin{align}
\lambda_{l} & \propto\int^{1}_{-1}du\,k(u)C^{a}_{l}(u)(1-u^{2})^{a-1/2}\label{eq:funk-hecke-short}\\
 & =\int^{1}_{-1}du\,C^{a}_{1}(u)C^{a}_{l}(u)(1-u^{2})^{a-1/2}\nonumber \\
 & =\delta_{1l},\nonumber 
\end{align}
where we used orthogonality of the Gegenbauer polynomials with respect
to the measure $(1-u^{2})^{a-1/2}$.

\subsection*{}

\section{Details on comparison to experimental spectra\label{app:stringer-details}}

In the experiment, a fixed set of stimuli $X\in\mathbb{R}^{P\times N_{\text{in}}}$
is presented to animals at two distinct points in time, which we denote
$1=\alpha\neq\beta=2$ in accordance with the setup in \cref{app:stringer-details}.
We model the resulting responses as 
\begin{align*}
\bm{\phi}^{1}_{x}(t_{R}) & =\bm{\phi}^{J}[t_{R};\,h_{\text{ext}}(t)=a(t)Vx;\,\zeta^{1}_{x}]\in\mathbb{R}^{N},\\
\bm{\phi}^{2}_{x}(t_{R}) & =\bm{\phi}^{J}[t_{R};\,h_{\text{ext}}(t)=a(t)Vx;\,\zeta^{2}_{x}]\in\mathbb{R}^{N}.
\end{align*}
where $\bphi^{J}[t_{R};\,h_{\text{ext}}(t);\,\zeta]$ is the discrete
network forward mapping as described by \cref{eq:transmission_dynamics_glauber}
and $\zeta$ stands for the inherent variability in the network from
the Glauber update process that is independent of the input. In particular,
we assume that this process is completely independent between the
time points of experiment and stimuli (including identical stimuli
$x=x^{\prime}$ that),
\[
\text{MI}[\zeta^{1}_{x};\zeta^{2}_{x^{\prime}}]=0\quad\forall\,x,x^{\prime}\in X.
\]
In the experiment, this could also represent unobserved, uncorrelated
inputs to the area of the neural recording. The matrix of responses
are then formed by stacking these vectors into matrices, $\Phi^{1}_{X}(t_{R})\coloneqq\text{Mat}\bigl(\bm{\phi}^{1}_{x}(t_{R})\bigr){}_{x}$,
$\Phi^{2}_{X}(t_{R})\coloneqq\text{Mat}\bigl(\bm{\phi}(t_{R})\bigr){}_{x}$.

According with the mean-field theory developed in \cref{app:transmission_laws},
the kernel is still a valid (i.e. positive-definite) covariance
\[
K_{XX}=\ev{\bm{\phi}^{1}_{X}\bm{\phi}^{2T}_{X}}{J,\zeta}\in\mathbb{R}^{P\times P}
\]

so that the spectrum may safely be calculated. Note that different
from the kernel with $k(\bx\cdot\bx)=1$ which would give the spectrum
of a normal empirical covariance matrix of a dataset where each measurement
appears once, here the kernel does not need to equal $1$ on equal
inputs; $K_{XX}$ is the matrix that (up to $\min(P,N)$) has the
same spectrum as the matrix $C_{XX}\in\mathbb{R}^{N\times N}$ that
the cross-validated-PCA (cvPCA) in \citep{Stringer19_361} is estimating.
\end{document}